\documentclass[10pt,journal,compsoc]{IEEEtran}
\usepackage[ruled,linesnumbered,vlined]{algorithm2e}
\usepackage{bm,subfigure,graphicx,amsmath,amssymb,xcolor,cite}
\usepackage{stfloats}
\usepackage{ragged2e}
\usepackage{amsmath,amsthm}
\usepackage{cuted}
\usepackage{graphicx} 
\usepackage{diagbox}
\usepackage{hyperref}
\usepackage{array}
\usepackage{multirow}

\graphicspath{{figures/}}

\allowdisplaybreaks

\begin{document}

\title{A Truthful Auction for Graph Job Allocation in Vehicular Cloud-assisted Networks}

\author{Zhibin Gao*, Minghui LiWang*, \IEEEmembership{Member, IEEE}, Seyyedali Hosseinalipour, \IEEEmembership{Student Member, IEEE}, Huaiyu~Dai, \IEEEmembership{Fellow, IEEE}, Xianbin Wang, \IEEEmembership{Fellow, IEEE}

\thanks{Zhibin Gao (gaozhibin@xmu.edu.cn) is with the school of Informatics, Xiamen University, Fujian, China. Minghui LiWang (corresponding author, mliwang@uwo.ca) and Xianbin Wang (xianbin.wang@uwo.ca) are with the department of Electrical and Computer Engineering, University of Western Ontario, Ontario, Canada. Seyyedali Hosseinalipour (shossei3@ncsu.edu) and Huaiyu Dai (hdai@ncsu.edu) are with the department of Electrical and Computer Engineering, North Carolina State University, NC, USA. $*$: These authors contributed equally to this work.}

\vspace{0.3cm}

\begin{abstract}
\justifying
Vehicular cloud computing has emerged as a promising solution to fulfill users' demands on processing computation-intensive applications in modern driving environments. Such applications are commonly represented by graphs consisting of components and edges. However, encouraging vehicles to share resources poses significant challenges owing to users' selfishness. In this paper, an auction-based graph job allocation problem is studied in vehicular cloud-assisted networks considering resource reutilization. Our goal is to map each buyer (component) to a feasible seller (virtual machine) while maximizing the buyers' utility-of-service, which concerns the execution time and commission cost. First, we formulate the auction-based graph job allocation as an integer programming (IP) problem. Then, a Vickrey-Clarke-Groves based payment rule is proposed which satisfies the desired economical properties, truthfulness and individual rationality. We face two challenges: 1) the above-mentioned IP problem is NP-hard; 2) one constraint associated with the IP problem poses addressing the subgraph isomorphism problem. Thus, obtaining the optimal solution is practically infeasible in large-scale networks. Motivated by which, we develop a structure-preserved matching algorithm by maximizing the utility-of-service-gain, and the corresponding payment rule which offers economical properties and low computation complexity. Extensive simulations demonstrate that the proposed algorithm outperforms the benchmark methods considering various problem sizes.
\end{abstract}

\begin{IEEEkeywords}
Vehicular cloud-assisted networks, truthful auction, graph job allocation, subgraph isomorphism.
\end{IEEEkeywords}}

\maketitle


\section{Introduction}

\IEEEPARstart{W}{ith} the rapid growth of connected devices in Internet of things (IoT) networks, especially smart cars, the Internet of vehicles (IoV) has become an emerging paradigm that offers safety, convenience, and entertainment for both drivers and passengers. Furthermore, technological advances in computing processors and sensing devices has offered innovative solutions for vehicular applications with resource hungry features. Such applications include face/behavior identification, simultaneous localization and mapping, autonomous driving, and advanced driver assistants, all of which require massive computational resources~\cite{1,2}. Notably, graph-based representation is utilized to characterize the non-negligible internal structures of such applications. In such scenarios, each application (the ``application'' is interchangeable with ``job'' or ``graph job'' throughout) is modeled as a graph, where the vertices (components) represent either data sources or data processing units, while the edges describe the dependency between the vertices~\cite{3}.

However, the limitation of a single smart vehicle caused by resource and capability constraints of on-board equipment, may hinder the fulfillment of graph job execution requirements. One approach is to process such applications on remote data centers or mobile edge computing (MEC) servers~\cite{4,5,6}. Nonetheless, the former may incur enormous costs due to bulk data transfer and latency, while the limited resource as well as signal coverage of MEC servers may cause unsatisfactory offloading performance. Thus, vehicular cloud computing (VCC)~\cite{2,7} has been proposed to provide flexible and effective computing services for the users (called job owners, JOs). In VCC platform, vehicles with surplus resources act as servers (called service providers, SPs) to form vehicular clouds (VCs) via vehicle-to-vehicle (V2V) communications, which enables the processing of jobs in parallel. We summarize two major challenges facing the allocation of graph jobs over VCs:

\noindent 1) Obtaining the feasible mappings between the components of the jobs and the SPs while satisfying the heterogeneous computational demands of the components requires solving the subgraph isomorphism problem, which is known to be NP-complete.

\noindent 2) Avoiding behaviors that sellers may misreport their bids while motivating the selfish SPs to lease their idle resources is a major challenge, since reporting untruthful bid may potentially lead to a higher profit for the SPs. 

Addressing the aforementioned challenges is the main motivation of this paper. We propose innovative policies for a marketplace considering resource reutilization. The proposed framework comprises three participants: sellers (service providers), buyers (service requestors), and the broker that manages the acution while offering the desired economical properties of truthfulness and individual rationality~\cite{8}. We investigate an auction mechanism for graph job allocation, where each JO has a job modeled as an undirected weighted graph of components (buyers) with different demands for execution time. A virtual machine (VM)-based~\cite{9} representation is utilized to quantize available resources of SPs. Also, the VC is abstracted to an undirected weighted graph consisting of VMs (sellers) that can be reutilized after release, where sellers can provide heterogeneous computational capabilities. To maximize the sum of the utility-of-service, an optimal and a sub-optimal algorithm together with the relevant payment rules are proposed to map each buyer to a feasible seller. The mapping ensures that the requirements of the graph job execution are met, and also provides truthfulness as well as individual rationality as two key desired market properties.

\subsection{Related work}
There existing works devoted to studying the computation-intensive job allocation that can be roughly divided into two categories: a) allocation of bit stream-represented jobs without considering the inherent dependencies, such as~\cite{5,10,11,6,12,13}; and b) allocation of graph-represented jobs upon existence of inner dependencies among the components, which is also the main focus of this paper, such as~\cite{3,14,15,16,17,18,19}. Moreover, the environment in which the graph job allocation is studied can be divided into three types based on the dynamism of the network topology: static, semi-static, and dynamic. By assuming the static topologies for both the servers and the users, authors in~\cite{14} proposed a randomized job scheduling algorithm that stabilizes a system with job arrivals/departures and facilitates a smooth trade-off between the minimization of the average execution cost and the servers' queue length. In semi-static environment where the topologies of either the computing servers or the service requestors are fixed, a novel framework for energy-efficient graph job allocation in geo-distributed cloud networks was introduced in~\cite{3}, where solutions are obtained for data center networks considering various scales. A lyapunov optimization-based dynamic offloading approach is developed in~\cite{15} to satisfy the constraints on energy conservation and application execution time. The scheduling of parallel jobs composed of a set of independent tasks was studied in~\cite{16} by mainly considering energy consumption and job completion time. In~\cite{17}, a VC-based computation offloading mechanism was introduced which enabled tasks being executed in different vehicles to minimize the overall response time while enhancing the capacity of the edge clouds. The graph job allocation problem in dynamic network environments have rarely been studied. In such networks, the mobility of servers and users as well as interdependency of the components pose major challenges to the design of applicable allocation methods. We were among the few working on addressing such challenges. A randomized job allocation mechanism based on hierarchical tree decomposition was proposed in our previous work~\cite{18}, which efficiently solves the allocation problem considering the trade-off between the task completion time and data exchange cost among the SPs. In~\cite{19}, we investigated the multi-graph-task offloading problem while considering the potential competition among components caused by the concurrency of multiple tasks.

\begin{figure*}[b!]
\centering
\subfigure[]{\includegraphics[width=9cm, height=7cm]{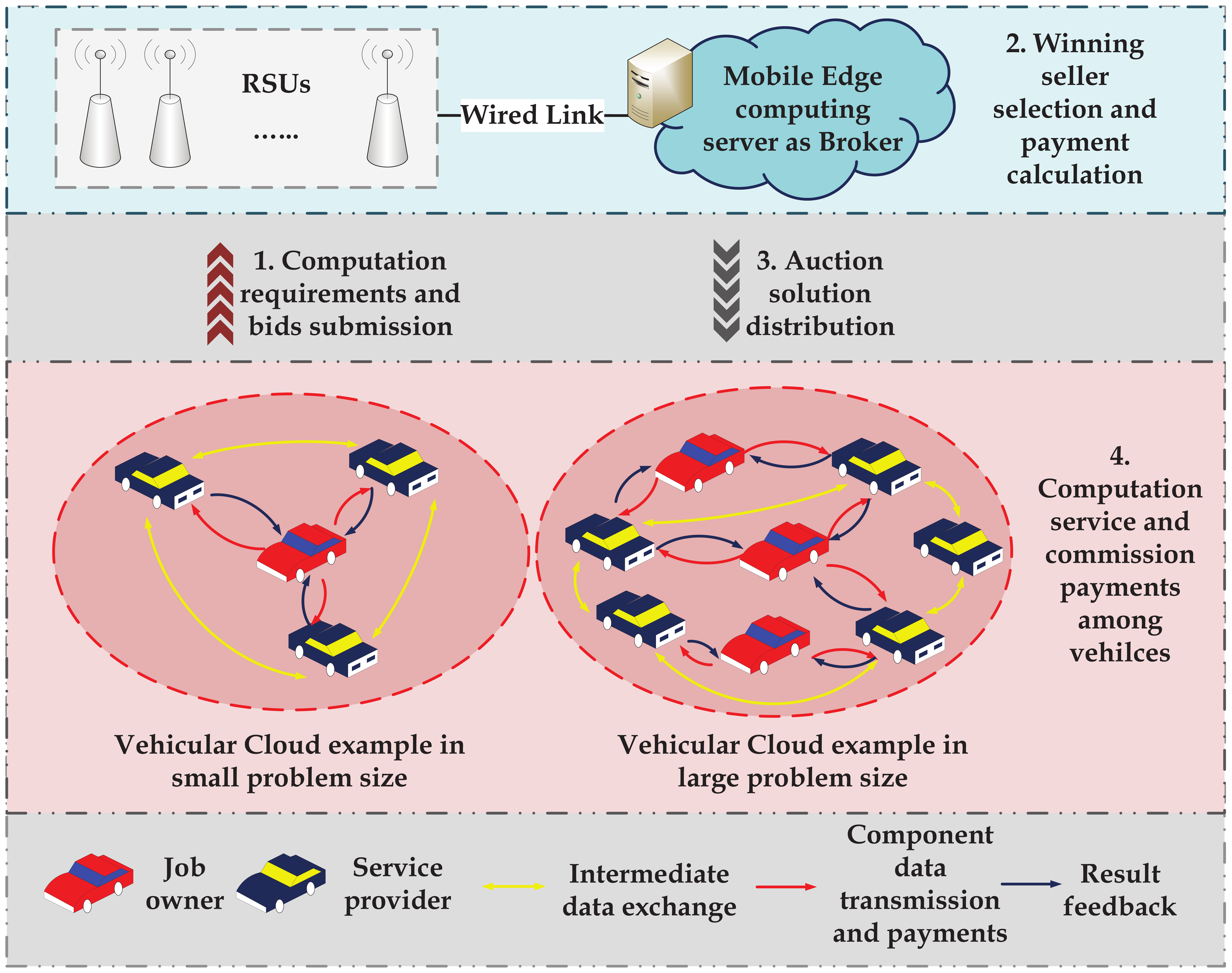}}
\subfigure[]{\includegraphics[width=9cm, height=7cm]{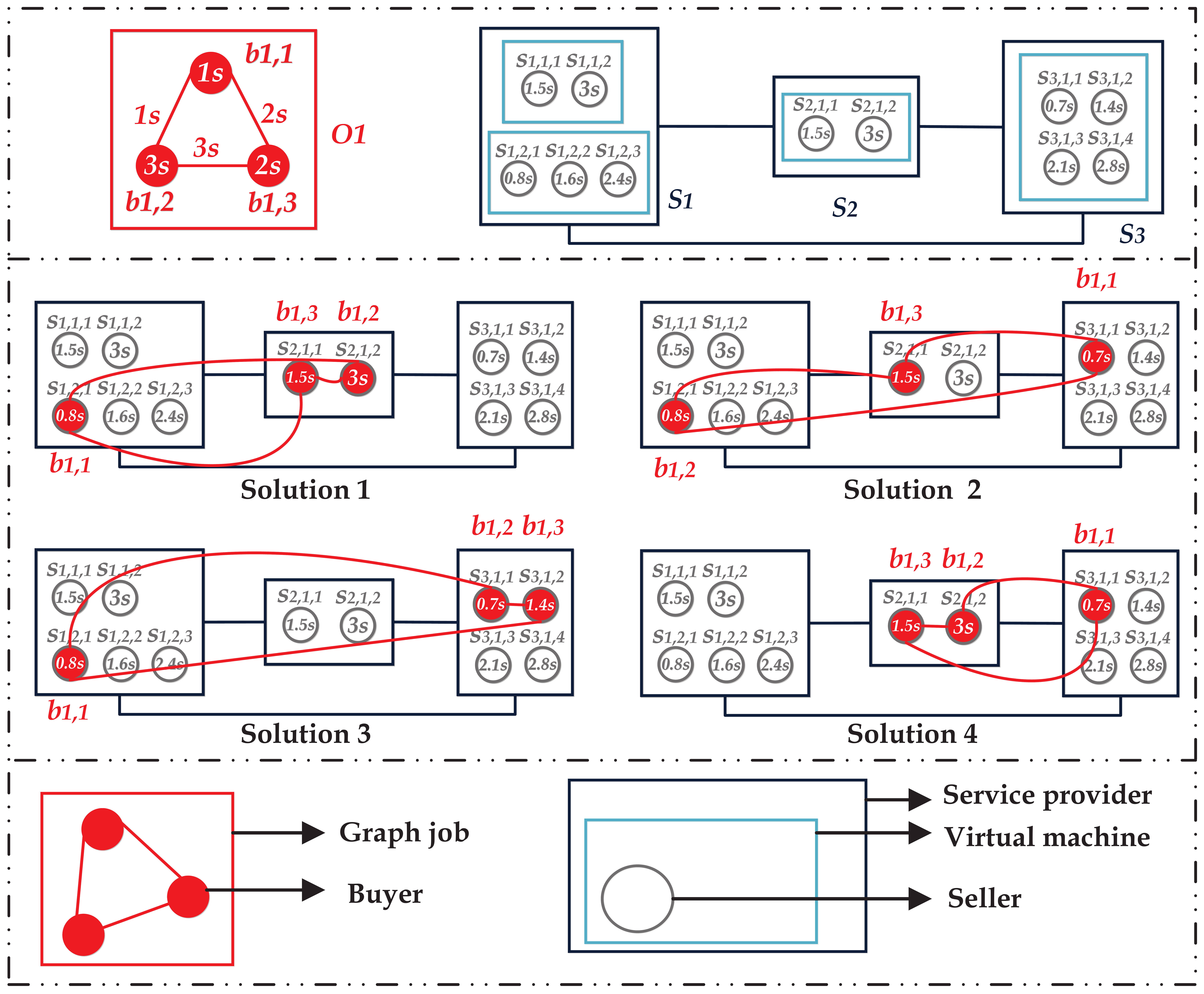}}
\caption{(a) The framework of the proposed auction for VC-assisted graph job allocation; (b) Example of possible solutions of graph job allocation.}
\end{figure*}

However, economical factors and incentive issues related to the graph job allocation haven't been fully addressed in the above-mentioned works. Auctions have been widely applied and regarded as a popular trading form for effective price discovery and resource sharing among the users. For the semi-static network environments~(e.g., \cite{20,21,22,23,24}), authors in~\cite{20} proposed two double auction schemes with dynamic pricing to determine the matched pairs between industrial-IoT mobile devices and edge servers. In~\cite{21}, a stochastic offloading approach in mobile cloud computing through online auction was presented while achieving desirable economic properties such as truthfulness and individual rationality. A truthful double auction mechanism was proposed in~\cite{22} to bridge users' task requirements and providers' resources in two-sided cloud markets. In~\cite{23}, the authors modeled the VM resource allocation problem among edge clouds and mobile users as an n-to-one weighted bipartite graph matching problem with 0-1 knapsack constraints, and then designed a greedy approximation algorithm to solve the problem. Authors in~\cite{24} focused on vehicular fog computing (VFC) environment based on smart parking, where a VFC-aware parking reservation auction was proposed to guide the on-the-move vehicles to the available parking places with less effort and exploite the computing services of the parked vehicles to assist the delay-sensitive tasks. In the dynamic environment, authors in \cite{7} proposed different truthful auction models for homogeneous and heterogeneous task models (e.g., tasks with the same and different resource requirements). In~\cite{25}, a distributed auction model was investigated to facilitate the resource trading between the owner of the tasks and the mobile devices participating in task execution. However, the above mentioned auction-based works maily focused on the bit stream-represented jobs, and thus the inherent interdependencies among the components of the jobs has not been fully considered.

\subsection{Novelty and Contribution}
In existing literature concerned with the semi-static or dynamic network environment, addressing the job allocation problem under graph-based representation while applying auction theory are rarely investigated. Moreover, the cocept of resource reutilization allowing the usage of a VM after its release is neglected in most of the above-mentioned works, which offers more options to buyers. 
To the best of our knowledge, this paper is among the first which proposes a truthful auction model for the graph job allocation problem over VCs while considering resource reutilization. The main contributions of this paper can be summarized as follows:

\begin{itemize}
\item We establish a novel VC-assisted graph job allocation marketplace which enables multiple computation-intensive graph jobs to be mapped (offloaded) to the SPs, while considering the reutilization of resources. 

\item We propose an auction-based graph job allocation mechanism, where the components of jobs are acted as buyers with various demands. Sellers are defined by considering VM reutilization with heterogeneous capabilities. To achieve the maximum total utility-of-service of the buyers, we formulate the graph job allocation as an integer programming (IP) problem under limited opportunistic communications, which is NP-hard. Moreover, one of the constraints related to preserving the graph job structures requires addressing the subgraph isomorphism problem. We tailor a Vickrey–Clarke–Groves (VCG)-based payment rule to protect the benefits of the sellers, prevent misreporting behaviors, and achieve individual rationality.

\item To tackle the IP problem for obtaing a feasible mapping between the buyers and the sellers, we first develop an optimal algorithm to find the graph job allocation solution while guaranteeing truthfulness and individual rationality from different perspectives of the SPs, the VMs and the sellers. We show that this method suffers from a high computational complexity, which makes it ineffective when encountering a large number of jobs and SPs or upon having complicated job and network topology. 

\item To obtain a near optimal solution for the proposed IP problem in polynomial time, we propose a structure-preserved matching algorithm based on maximizing the utility-of-service-gain and its corresponding payment rule. We demonstrate that the low computational complexity of this method makes it a good fit for large and fast-changing IoV networks. Moreover, we prove that the proposed method enjoys desirable economic properties of truthfulness and individual rationality. 

\item Based on thorough numerical analysis and comparative evaluations, we demonstrate that the performance of the proposed low complexity structure-preserved matching algorithm can approach that of the optimal algorithm, while outperforming baseline methods considering various problem sizes.
\end{itemize}

The rest of this paper is organized as follows.  System model and design targets are introduced in Section~2. We formulate the IP problem and propose an adequate VCG-based payment rule in Section~3. This section also contains the proofs of the desired economical properties and the related optimal graph job allocation algorithm. In Section~4, a structure-preserved matching algorithm based on utility-of-service-gain maximization is proposed. The conomical properties are also proved to be true in the same section. The performance evaluation through comprehensive simulations is conducted in Section~5 before drawing the conclusion in Section~6.

\section{System Model and Design Targets}

In the following, we mainly study the problem in one VC for tractable analysis,  since the proposed auction-based graph job allocation mechanism
is universal across all VCs. Consider a VC containsing $|\bm{O}|$ JOs and $|\bm{S}|$ SPs and several RSUs alongside the road. Note that the data size of resulting feedback of the job execution is much smaller than that of the application data~\cite{4}, alternatively, results can be transferred back to the JO via a multi-hop V2V routing path or uploaded to RSUs for future delivery in case the V2V connection is disrupted, and thus neglected in this paper. The framework of the proposed auction for VC-assisted graph job allocation is shown in Fig. 1(a), and the main notations are summarized in \textbf{TABLE 1}. 
The related models in this paper are introduced hereafter.

\subsection{Opportunistic contact model}

\begin{table}[!t]
\renewcommand\arraystretch{1.15}
\begin{center}
     \caption{Major notations}
     \setlength{\tabcolsep}{0.2mm}{
\begin{tabular}{|c|c|}
\hline
Notation& 
Explanation\\
\hline
$\bm{O}, \bm{S}$ & UAV set, SP set \\
\hline
$\bm{G^{O_n}}, \bm{V^{O_n}}, \bm{E^{O_n}}, \bm{W^{O_n}}$ & graph job, set of components, set of edges, \\
& and set of weights of $O_n$  \\
\hline
$\bm{G^{s}}, \bm{V^{s}}, \bm{E^{s}}, \bm{W^{s}}$ & graph structure, set of sellers, set of edges, \\
& and set of weights  of VC\\
\hline
$n$, $x$ & the index of JO and buyer, respectively\\
\hline
$m$, $y$, $n$ & the index of SP, VM and seller, respectively\\
\hline
$\bm{\mathcal{B}}$, $\bm{\mathcal{D}}$ & the bidding matrix, the demand matrix\\
\hline
$\bm{\mathcal{K}}$, $\bm{\mathcal{K^*}}$ & the binary matrix, the optimal solution\\
\hline
$b_{n,x}$ & buyer (the $x^{th}$ component of $\bm{G^{O_n}}$)\\
\hline
$t_{n,x}$ & the tolerable execution time of $b_{n,x}$\\
\hline
$S_{m}, v_{m,y}$ & SP, the $y^{th}$ VM of $S_{m}$\\
\hline
$ s_{m,y,r}$ & seller (the $r^{th}$ seller related to $v_{m,y}$ of $S_{m}$ )\\
\hline
$c_{m,y,r}$ & the computational capability of seller $ s_{m,y,r}$\\
\hline
$p_{m,y,r}, q_{m,y,r}$ & the bid and true price of seller $ s_{m,y,r}$\\
\hline
$g^{n,x}_{m,y,r}$ & the time saved by mapping $b_{n,x}$ to $s_{m,y,r}$\\
\hline
$\kappa^{n,x}_{m,y,r}$ & the binary indicater that represents \\
& the mapping between $b_{n,x}$ and $s_{m,y,r}$\\
\hline
$\bm{\mathcal{R}_{O_n}}$ & the set of sellers coveraged by $O_n$\\
\hline
$\bm{L_{n,x}}, \bm{L_{Bro}}$ & the preference list of $b_{n,x}$ and the broker\\
\hline
$\widetilde{s_{i, j, k}}$ & the seller located behind $s_{i, j, k}$ in \\
&buyer's preference list\\
\hline
$\widetilde{g^{n,x}_{i,j,k}}$, $\widetilde{p_{i,j,k}}$ & the execution time saved when mapping \\ &$b_{n,x}$ to $\widetilde{s_{i, j, k}}$, the bid of $\widetilde{s_{i, j, k}}$\\
\hline
\end{tabular}}
\end{center}
\end{table}
A contact event between vehicles $n $ and $m$ occurs during $\tau \in ({\tau}_{1}, {\tau}_{2})$ when $\| L_n ({\tau}_{1})-L_m({\tau}_{1})\|>R, \| L_n(\tau)-L_m(\tau) \| \leq R$ and $ \| L_n ({\tau}_{2})-L_m({\tau}_{2})\| >R$, where $L_n(\tau)$ and $L_m(\tau)$ denote the respective locations, $\| \cdot \|$ represents the Euclidean distance, and $R$ indicates the communication radius of the vehicles. Generally, the contact duration among the vehicles obeys an exponential distribution~\cite{26,27} with parameter ${\lambda}_{n, m}$; thus, the probability of the contact duration $\Delta {\tau}_{n, m}$ between two vehicles $n $ and $m$ being larger than $T$ is given by $prob(\Delta {\tau}_{n, m}>T|{\lambda}_{n, m})={\rm e}^{-T{\lambda}_{n, m}}$. Correspondingly, the larger the value of $prob(\Delta {\tau}_{n, m}>T|{\lambda}_{n, m})$ is, the more insurance can be achieved to protect data interaction among the respective vehicles. 
\subsection{Modeling the buyers}

Consider a set of JOs $\bm{O}$, where each JO $O_n\in \bm{O}$ owns a graph job ${\bm{G}}^{\bm{O_n}}=\left({\bm{V}}^{\bm{O_n}}, {\bm{E}}^{\bm{O_n}}, {\bm{W}}^{\bm{O_n}}\right)$ that contains a set of components ${\bm{V}}^{\bm{O_n}} =\left\{b_{n, x}|n\in \left\{1,2,\dots, | \bm{O}| \right\}, x\in \left\{1,2,\dots, |{\bm{V}}^{\bm{O_n}}|\right\}\right\}$ where each component $b_{n,x}$ is associated with a tolerable execution time $t_{n, x}$ (seconds); and a set of edges ${\bm{E}}^{\bm{O_n}}=\left\{{e}^{O_n}_{xx'}|x, x'\in \left\{1,2,\dots, |{\bm{V}}^{\bm{O_n}}|\right\}, x\neq x'\right\}$ with associated weight ${\bm{W}}^{\bm{O_n}}=\left\{{\omega}^{O_n}_{xx'}|x, x'\in \left\{1,2,\dots, |{\bm{V}}^{\bm{O_n}}|\right\}, x\neq x'\right\}$. In this model, the edges represent data flows among the components and the weight ${\omega}^{O_n}_{xx'}$ of edge ${e}^{O_n}_{xx'}$ indicates the requested connect duration for intermediate data exchange between components $b_{n, x}$ and $b_{n, x'}$, which is considered to be equal to or lower than the smaller tolerable execution time of the two. Also, the contact duration of the SPs that handle components $b_{n, x}$ and $b_{n, x'}$ should be ideally equal to or larger than ${\omega}^{O_n}_{xx'}$. A job ${\bm{G}}^{\bm{O_n}}$ represents how the computation is supposed to be splitted among components in ${\bm{V}}^{\bm{O_n}}$. In this paper, each component $b_{n, x}\in {\bm{V}}^{\bm{O_n}}$ is seen as a \textbf{\textit{buyer}} and mapped to an available seller in the related VC, by paying a certain commission for the computing service.

\subsection{Modeling the sellers}

We use a VM-based presentation to describe the available resources of the SPs. In this case, each SP $S_m\in \bm{S}$ has a collection of VMs ${\bm{VM_{m}}}=\left\{v_{m,1}, \dots, v_{m,|{\bm{VM_{m}}}|}\right\}$ that are fully connected (namely, can communicate continuously with each other), where $ v_{m,y}$ denotes the $y{\rm th}$ VM on $S_m$. Moreover, $v_{m,y}$ can provide the computational capability of $c_{m,y}$ (seconds) that is the execution time of processing one component of a graph job. The VM reutilization factor enables each VM $v_{m,y}$ to serve more than one buyer after every $c_{m,y}$ (seconds); namely, to cover multiple buyers, $v_{m,y}$ can process a buyer after the completion of the previous buyers. Thus, each VM $v_{m, y}$ is modeled as a collection of virtual sellers ${\bm{s}}^{{\bm{v}}_{\bm{m},\bm{y}}}=\left\{s_{m,y,r}|r\in \left\{1,2,\dots,  \widetilde{{r}_{m,y}}\right\}\right\}$, where $s_{m, y, r}$ refers to the $r^{\text{th}}$ virtual seller derived from VM  $v_{m,y}$, that can provide the computational capability $c_{m,y,r}=r\times c_{m,y}$ for processing one buyer. Notably, the value of ${\widetilde{r_{m,y}}}\times c_{m,y}$ should be equal to or less than the maximum demand among buyers. For notational simplicity, we use “\textbf{\textit{seller}}” as the substitution of ``virtual seller'' for indicating each $s_{m,y,r}$.

Fig. 1(b) shows a scenario of a graph job with three components (3 buyers), and a VC containing three SPs (SP1 has 2 VMs/5 sellers, SP2 has 1 VM/2 sellers, SP3 has 1 VM/4 sellers). Notably, the maximum tolerable execution time among buyers is 3~seconds. Take VM $v_{1,2}$ (the second VM on the first SP) which provides $c_{1,2}=0.8$ seconds for processing one component as an example. Considering VM reutilization ($\widetilde{{r}_{1,2}}=3$), $v_{1,2}$ can be divided into three sellers $s_{1,2,1}$, $s_{1,2,2}$, $s_{1,2,3}$ with computational capabilities of $c_{1,2,1}=1\times0.8=0.8$ seconds, $c_{1,2,2}=2\times 0.8=1.6$~seconds and $c_{1,2,3}=3\times0.8=2.4$~seconds, respectively. Fig. 1(b) also depicts several feasible solutions of mapping the buyers to the sellers. For instance, in Solution 1, buyers $b_{1,1}$, $b_{1,2}$, and $b_{1,3}$ are mapped to sellers $s_{1,2,1}$, $s_{2,1,1}$, and $s_{2,1,2}$, respectively.

\subsection{Modeling the VC}

In a VC, we assume that each JO can communicate with at least one SP via one-hop V2V communication. It is assumed that the JOs are under pressure from insufficient local resources so that they prefer to integrate resources from the SPs in the related VC; thus, we do not consider the VMs on the JOs as available resources. Consequently, a VC is represented as graph ${\bm{G}}^{\bm{s}} =\left({\bm{V}}^{\bm{s}}, {\bm{E}}^{\bm{s}}, {\bm{W}}^{\bm{s}}\right)$ containing a set of sellers ${\bm{V}}^{\bm{s}}=\left\{s_{m, y, r}|S_m\in\bm{S}, {{v_{m,y} \in {\bm{VM_m}}, s}_{m,y,r} \in \bm{s}}^{{\bm{v}}_{\bm{m}, \bm{y}}}\right\}$; edge set $\bm{E^s}=\left\{{\rm e}^{m, y, r}_{m',y',r'}|s_{m,y,r}, s_{m',y',r'}\in \bm{V^s}, s_{m,y,r}\neq s_{m',y',r'}\right\}$, where each edge ${e}^{m,y,r}_{m',y',r'}$\linebreak indicates that seller $s_{m,y,r}$ can communicate with $s_{m',y',r'}$; and weight set $\bm{W^s}=\left\{{\lambda}^{m, y, r}_{m', y', r'}|s_{m,y,r}, s_{m',y',r'}\in \bm{V^s}, s_{m,y,r}\neq s_{m',y',r'}\right\}$ related to edges. Notably, sellers of the same SP are fully connected which brings ${\lambda}^{m, y, r}_{m', y', r'}=0$ when $m=m'$; otherwise, ${\lambda}^{m, y, r}_{m', y', r'}={\lambda}_{m, m'}$, which refers to the exponential distribution parameter of the contact duration between the vehicles.

\subsection{The broker}

A broker acts as an intermediate agent or auctioneer who hosts and directs auction processes. In this paper, the edge computing server can conduct auctions as a broker~\cite{7}.

\subsection{The economic model}

In our proposed auction-based framework for graph job allocation, each seller $s_{m, y, r}$ has a bid $p_{m, y, r}$ which denotes the reported price for processing one component and offers a computational capability $c_{m, y, r}$. Also, it has a true valuation $q_{m, y, r}$ \footnote{The ``true price'' and ``true valuation'' are utilized interchangeably for the rest of this paper.} that is unknown to the broker, the buyers, and the other SPs. 
The true price $q_{m,y,r}$ of seller $s_{m,y,r}$ is defined as a monotone decreasing function of $c_{m,y,r}$ as (1), 
\begin{align}
\label{eq1}
q_{m, y, r}={\mathcal{U}}^S \left(c_{m, y, r}\right).
\end{align}
${\mathcal{U}}^S$ denotes a monotone decreasing function, where a lower execution time $c_{m,y,r}$ leads to a higher price owing to a more powerful computational capability that $s_{m,y,r}$ can provide. To prevent misreporting behavior where $p_{m, y, r}\neq q_{m, y, r}$, the mechanism proposed in this paper enjoys truthfulness that incentivizes sellers to provide the bidding information as the true valuation (i.e., $p_{m, y, r}=q_{m, y, r}$) by ensuring no additional benefits and possible risks brought by misreporting. 

The gross utility $g^{n,x}_{m,y,r}$ of buyer $b_{n, x}$ is defined as the time saved by enjoying the computing service of seller $s_{m,y,r}$, defined as (2).
\begin{align}
\label{eq2}
g^{n,x}_{m,y,r}=t_{n, x}-c_{m, y, r}
\end{align}
The gross utility that each seller offers to buyers is non-identical, where the larger value of $g^{n, x}_{m, y, r}$ represents that buyer $b_{n, x}$ can save more time on execution. 

Based on (1) and (2), the utility-of-service (UoS) of each buyer $b_{n, x}$ is defined as the utility each buyer can get after paying for the computing service, which can be calculated as ${\alpha}_ng^{n,x}_{m,y,r}-p_{m,y,r}$, where ${\alpha}_n$ denotes the sensitivity factor between price and execution time of $O_n$. The VM reutilization can effectively handle various buyers with different preferences related to various prices and execution times the the sellers can provide. Thus, a buyer with a lower sensitivity on execution time can wait for the release of an applicable VM by enjoying a discounted price. 

\subsection{Design targets}

In this paper, truthfulness and individual rationality are considered as key design targets~\cite{25,28}, which are formally defined below.
\medskip

\noindent
\textbf{Definition 1 (Truthfulness):}
An auction is truthful if every seller keeps his bid equal to his true valuation. Sellers in a truthful auction marketplace are not willing to take any risks that may have bad effects on their utilities.

\medskip
\noindent
\textbf{Definition 2 (Individual rationality):}
An auction is individually rational if all the bidders are guaranteed to receive non-negative utilities. 

Our proposed framework includes the following main steps (see Fig. 1(a)): first, the buyers publish their demands (e.g., the tolerable execution time) and the sellers provide their bids. Then, applicable allocation solutions can be determined by the broker and transfered to all the participants. Note that any feasible allocation should consider the structural characteristics of the jobs and the VC. Aiming to maximize the total UoS of buyers while protecting the sellers' benefits, each buyer is matched to a seller under the constraints of opportunistic communications and available resources.

\section{Problem Formulation and The Optimal Algorithm}

This section first presents the problem formulation, and the corresponding payment rule. Then, an optimal algorithm is proposed for distributing graph jobs over VCs.

\subsection{Problem formulation}
Let the binary indicator ${\kappa}^{n,x}_{m,y,r}$ represents the assignment of buyer $b_{n,x}$ to seller $s_{m,y,r}$, where ${\kappa}^{n,x}_{m,y,r}=1$ when buyer $b_{n,x}$ is mapped to $s_{m,y,r}$, and ${\kappa}^{n,x}_{m,y,r}=0$, otherwise. For notational simplicity, let $\bm{\mathcal{K}}={\left[{\kappa}^{n, x}_{m, y, r}\right]}_{{s_{m,y,r} \in \bm{V^s}}, {1\leq n\leq|\bm{O}|,1\leq x\leq|{\bm{V}}^{\bm{o_n}}|}}$ denote the corresponding binary matrix. Given the bidding matrix of sellers $\bm{\mathcal{B}}={\left[p_{m,y,r}\right]}_{{{s_{m,y,r} \in \bm{V^s}}}}$ and the demand matrix of buyers $\bm{\mathcal{D}}={\left[t_{n,x}\right]}_{1\leq n\leq|\bm{O}|,1\leq x\leq|{\bm{V}}^{\bm{o_n}}|}$, we fomulate the graph job allocation as an integer programming problem $\bm{P}$ given in (4), aiming to maximize the total UoS of buyers ${\mathcal{F(\bm{\mathcal{K}})}}$ shown in (3). 


\begin{align}
\label{eq3}
{\mathcal{F(\bm{\mathcal{K}})}}=\sum^{|\bm{S}|}_{m=1}\sum^{|\bm{O}|}_{n=1}\sum^{|{\bm{VM_m}}|}_{y=1}\sum^{|{\bm{V}}^{\bm{o_n}}|}_{x=1}\sum^{{\widetilde{r_{m,y}}}}_{r=1}\left(\alpha_n g^{n,x}_{m,y,r}-p_{m,y,r}\right){\kappa}^{n,x}_{m,y,r}\tag{3}
\end{align}

\begin{strip}
\hrulefill
\begin{align}
\label{eq4}
\bm{P}:~\bm{\mathcal{K}^*}=\mathop{\arg \max}_{\bm{\mathcal{K}}}{\mathcal{F}\left(\bm{\mathcal{K}}\right)}\tag{4}
\end{align}
s.t.
\begin{align}
& {\kappa}^{n, x}_{m, y, r}\triangleq 0,\text{ if }S_m{\notin \mathcal{R}}_{\bm{o}_{n}}\text{ or }~ t_{n, x}<c_{m, y,r}\text{ or }\left(\alpha_ng^{n, x}_{m, y, r}-p_{m, y, r}\right)\leq 0, \tag{C1}\\
& {\rm e}^{-{\lambda}_{mm'}\times{\omega}^{O_n}_{xx'}}\geq \varepsilon , \forall {\rm e}^{O_n}_{xx'}\in {\bm{E}}^{\bm{o}_{n}}, \text{ if } m\neq m' \text{ and }~{\kappa}^{n, x}_{m, y, r}\cdot {\kappa}^{n, x'}_{m', y', r'}=1, \tag{C2}\\
& \sum^{|\bm{S}|}_{m=1}{\sum^{|{\bm{VM_m}}|}_{y=1}{\sum^{{\widetilde{r_{m,y}}}}_{r=1}{{\kappa}^{n, x}_{m, y, r}}}}=1,~\forall b_{n, x},1\leq n\leq | \bm{O}|,1\leq x\leq | {\bm{V}}^{\bm{o_n}}|, \tag{C3}\\
& \sum^{|\bm{O}|}_{n=1}{\sum^{|{\bm{V}}^{\bm{o_n}}|}_{x=1}{{\kappa}^{n, x}_{m, y, r}}}\leq 1, \forall s_{m, y, r},1\leq m\leq | \bm{S}|,1\leq y\leq | {\bm{VM_m}}|,1\leq r\leq {\widetilde{r_{m,y}}},\tag{C4}
\end{align}
\hrulefill
\end{strip}

\noindent
Notation $S_m\in {\mathcal{R}}_{\bm{O_n}}$ indicates that $S_m$ is located within the communication range of $O_n$. In problem $\bm{P}$, constraint (C1) ensures one-hop V2V job allocation from JOs to SPs where buyer $b_{n, x}$ is mapped to a seller that meets the requirement on the tolerable execution time $t_{n, x}$ and positive UoS, and thus guarantees the successful execution of components. Constraint (C2) represents a probabilistic constraint which ensures that if two connected buyers $b_{n, x}$ and $b_{n, x'}$ are assigned to two sellers belonging to two different SPs $S_m$ and $S_{m'}$, the probability of the contact duration between $S_m$ and $S_{m'}$ being larger than ${\omega}^{o_n}_{xx'}$ must be greater than the threshold $\varepsilon \in (0,1]$. Also, 
(C2) relies on addressing the subgraph isomorphism problem, to preserve the graph job structures during auction. Constraint (C3) and (C4) guarantee the one-to-one mapping between the buyers and the sellers. Concretely, they guarantee that each buyer can only be mapped to one seller, and each seller can process no more than one buyer.

\subsection{The VCG-based payment rule}

Having the above defined IP problem, we use a VCG-based~\cite{28} payment rule given in (5), 
where the payment of a winning seller $s_{i, j, k}$ depends on the damage it causes to other participants, which is calculated as the difference between the optimal value of the objective function with and without seller $s_{i, j, k}$.
\begin{align}
\label{eq5}
{\mathcal{P}}_{i, j, k}& =\mathcal{F} (\bm{\mathcal{K}^*})-{\mathcal{F}}_{{\bm{V}}^{\bm{s}}\backslash \{s_{i, j, k}\}}(\bm{\Theta^*})\notag \\
& \quad~ +p_{i, j, k}\times{{\kappa}^{n, x~*}_{i, j, k}}, \tag{5}
\end{align}

\noindent
where $\mathcal{F}(\bm{\mathcal{K}}^*)$ denotes the maximum value of function (3). Also, ${\mathcal{F}}_{{\bm{V}}^{\bm{s}}\backslash\{s_{i, j, k}\}}(\bm{\Theta^*})$ indicates the maximum value of (3) without the participation of seller $s_{i, j, k}$, where the relevant optimal solution is represented by ${ \bm{\Theta^*}={\left[{\theta}^{n, x~*}_{m, y, r}\right]}_{s_{m,y,r}\neq s_{i,j,k}}}$.

\subsection{Analysis of truthfulness and individual rationality}

\noindent
\textbf{Proposition 1 (Truthfulness):}
Given the payment rule in (5), every seller $s_{i, j, k}\in {\bm{V}}^{\bm{s}}$ in the proposed auction is willing to have the bid equal to his true valuation ($p_{i, j, k}=q_{i, j, k}$).
\begin{proof}
\noindent
Under a given bidding matrix of other sellers in the related VC, consider two cases for seller $s_{i, j, k}$ as follows,

\noindent
\textbf{Case 1}: $s_{i, j, k}$ has his bid $p_{i, j, k}$ equal to the true valuation $q_{i, j, k}$ ($p_{i, j, k}=q_{i, j, k}$), and the corresponding utility $u_{i, j, k}$ is calculated as below based on the payment rule shown in (5),

\begin{align}
\label{eq6}
u_{i, j, k}& ={\mathcal{P}}_{i, j, k}-q_{i, j, k}\times{{\kappa}^{n, x~*}_{i, j, k}}\notag \\
& =\mathcal{F} (\bm{\mathcal{K}^*})-{\mathcal{F}}_{{\bm{V}}^{\bm{s}}\backslash \{s_{i, j, k}\}} (\bm{\Theta^*}).\tag{6}
\end{align}

\noindent
\textbf{Case 2}: $s_{i, j, k}$ misreports his bid denoted as $p'_{i, j, k}$ ($p'_{i, j, k}\neq q_{i, j, k}$), and the corresponding net utility is calculated in (7), in which $\bm{\mathcal{K}^{*\prime}}={\left[{{\kappa}^{n, x~*\prime}_{m, y, r}}\right]}_{s_{m,y,r}\in \bm{V^s}, 1 \leq n \leq |\bm{O}|, 1 \leq X \leq |\bm{V^{O_n}}|}$ denotes the graph job allocation solution when  $s_{i, j, k}$ misreports his bid, and ${{\kappa}^{n, x~*\prime}_{i, j, k}}$ indicates the relevant assignment of $s_{i, j, k}$.
\begin{align}
\label{eq7}
u'_{i, j, k}& ={\mathcal{P}}'_{i, j, k}-q_{i, j, k}\times{{\kappa}^{n, x~*\prime}_{i, j, k}}\notag \\
& =\mathcal{F} (\bm{\mathcal{K}^{*\prime}})-{\mathcal{F}}_{{\bm{V}}^{\bm{s}}\backslash \{s_{i, j, k}\}} ({\bm{\Theta^{*}}})\notag \\
&\quad~ +p'_{i, j, k}\times{{\kappa}^{n, x~*\prime}_{i, j, k}}-q_{i, j, k}\times{{\kappa}^{n, x~*\prime}_{i, j, k}}\tag{7}
\end{align}

The difference between the utilities of the above mentioned two cases is calculated in (8).
\begin{align}
\label{eq7}
u_{i, j, k}-u'_{i, j, k}& =\mathcal{F} (\bm{\mathcal{K}^*})-\mathcal{F} (\bm{\mathcal{K}^{*\prime}})\notag \\
& \quad~ -p'_{i, j, k}\times{{\kappa}^{n, x~*\prime}_{i, j, k}}+q_{i, j, k}\times{{\kappa}^{n, x~*\prime}_{i, j, k}}.\tag{8}
\end{align}
According to (3), we have 
\begin{align}
\label{eq9}
\mathcal{F} (\bm{\mathcal{K}^*})& =\sum^{|\bm{S}|}_{m=1}\sum^{|\bm{O}|}_{n=1}\sum^{|{\bm{VM_m}}|}_{y=1}\sum^{|{\bm{V}}^{\bm{o_n}}|}_{x=1}\sum^{{\widetilde{r_{m,y}}}}_{r=1}\notag\\
&\quad~ {{\left(\alpha_ng^{n, x}_{m, y, r}-p_{m, y, r}\right)\times{{\kappa}^{n, x~*}_{m, y, r}}}},\tag{9}
\end{align}
and 

\begin{align}
\label{eq10}
& \mathcal{F} ({\bm{\mathcal{K}^{*\prime}}})=\sum^{|\bm{S}|}_{m=1}\sum^{|\bm{O}|}_{n=1}\sum^{|{\bm{VM_m}}|}_{y=1}\sum^{|{\bm{V}}^{\bm{o_n}}|}_{x=1}\sum^{{\widetilde{r_{m,y}}}}_{r=1}~{{\alpha}_ng^{n, x}_{m, y, r}\times{{\kappa}^{n, x~*\prime}_{m, y, r}}}\notag \\ 
&-\sum^{|\bm{S}|}_{m=1}\sum^{|\bm{O}|}_{n=1}\sum^{|{\bm{VM_\bm{m}}}|}_{y=1}\sum^{|{\bm{V}}^{\bm{o_n}}|}_{x=1}\sum^{{\widetilde{r_{m,y}}}}_{r=1}~{p_{m, y, r}\times{{\kappa}^{n, x~*\prime}_{m, y, r}}}\notag \\
&-{p'_{i, j, k}\times{{\kappa}^{n, x~*\prime}_{i, j, k}}+q_{i, j, k}\times{{\kappa}^{n, x~*\prime}_{i, j, k}}}.\tag{10}
\end{align}
Correspondingly, (8) can be calculated as (11).

\begin{strip}
\hrulefill
\vspace{-0.8cm}
\begin{align*}
\label{eq11}
&  \notag \\
& \notag \\
&u_{i, j, k}-u'_{i, j, k}=\sum^{|\bm{S}|}_{m=1}\sum^{|\bm{O}|}_{n=1}\sum^{|{\bm{VM_m}}|}_{y=1}\sum^{|{\bm{V}}^{\bm{o_n}}|}_{x=1}\sum^{{\widetilde{r_{m,y}}}}_{r=1}\left(\alpha_ng^{n, x}_{m, y, r}-p_{m, y, r}\right)\times{{\kappa}^{n, x~*}_{m, y, r}}-\sum^{|\bm{S}|}_{m=1}{\sum^{|\bm{O}|}_{n=1}{\sum^{|{\bm{VM_m}}|}_{y=1}{\sum^{|{\bm{V}}^{\bm{o_n}}|}_{x=1}{\sum^{{\widetilde{r_{m,y}}}}_{r=1}{{\alpha}_ng^{n, x}_{m, y, r}\times{{\kappa}^{n, x~*\prime}_{m, y, r}}}}}}}+ \notag \\
&\sum^{|\bm{S}|}_{m=1}{\sum^{|\bm{O}|}_{n=1}{\sum^{|{\bm{VM_m}}|}_{y=1}{\sum^{|{\bm{V}}^{\bm{o_n}}|}_{x=1}{\sum^{{\widetilde{r_{m,y}}}}_{r=1}{p_{m, y, r}\times{{\kappa}^{n, x~*\prime}_{m, y, r}}}}}}}+p'_{i, j, k}\times{{\kappa}^{n, x~*\prime}_{i, j, k}}-q_{i, j, k}\times{{\kappa}^{n, x~*\prime}_{i, j, k}}- p'_{i, j, k}\times{{\kappa}^{n, x~*\prime}_{i, j, k}}+q_{i, j, k}\times{{\kappa}^{n, x~*\prime}_{i, j, k}}\notag \\
&=\sum^{|\bm{S}|}_{m=1}{\sum^{|\bm{O}|}_{n=1}{\sum^{|{\bm{VM_m}}|}_{y=1}{\sum^{|{\bm{V}}^{\bm{o_n}}|}_{x=1}{\sum^{{\widetilde{r_{m,y}}}}_{r=1}{{\left(\alpha_ng^{n, x}_{m, y, r}-p_{m, y, r}\right)\times{{\kappa}^{n, x~*}_{m, y, r}}}}}}}}- \sum^{|\bm{S}|}_{m=1}\sum^{|\bm{O}|}_{n=1}{\sum^{|{\bm{VM_m}}|}_{y=1}{\sum^{|{\bm{V}}^{\bm{o_n}}|}_{x=1}{\sum^{{\widetilde{r_{m,y}}}}_{r=1}{{\left(\alpha_ng^{n, x}_{m,y,r}-p_{m,y,r}\right)\times{{\kappa}^{n, x~*\prime}_{m, y, r}}}}}}}.\tag{11}
\end{align*}
\hrulefill
\end{strip}

\noindent Apparently in (11), solution ${{\kappa}^{n, x~*\prime}_{m, y, r}}$ is included in the solution space of the graph job allocation problem. Owing to that ${{\kappa}^{n, x~*}_{m, y, r}}$ represents the optimal solution which is always equal to or better than other solutions, we have $u_{i, j, k}-u'_{i, j, k}\geq 0$ and sellers do not obtain better utilities by misreporting in the auction. In other words, sellers will always report their true valuations.
\end{proof}

\begin{algorithm*}
\caption{The optimal algorithm for the auction-based graph job allocation}
\SetKwInOut{Input}{Input}\SetKwInOut{Output}{Output}
\Input{graph jobs ${\bm{G}}^{\bm{O_n}}=\{ {\bm{V^{O_n}}}, {\bm{E^{O_n}}}, {\bm{W^{O_n}}}\}$, VC graph ${\bm{G}}^{\bm{s}}=\{ {\bm{V^s}}, {\bm{E^s}}, {\bm{W^s}}\}$, $\bm{\mathcal{B}}$, $\bm{\mathcal{D}}$}

\Output{the optimal solution ${\bm{{\mathcal{K}}^{*}}}$, the related payments ${\mathcal{P}}_{i, j, k}$ of the winning sellers}

//~Stage 1: The buyer-seller pair selection procedure

Initialization: ${\bm{B}}^{\bm{*}}\leftarrow \{\bm{V}^{\bm{O}_{\bm{1}}}\cup \bm{V}^{\bm{O}_{\bm{2}}}\dots \cup \bm{V}^{\bm{O}_{|\bm{O}|}}\}$, $\bm{MatchT}\leftarrow []$, ${\bm{\mathcal{K}}}^{\bm{*}}\leftarrow []$, $\mathrm{b}\leftarrow length ({\bm{B^*}})$, $\mathrm{s}\leftarrow length (\bm{V}^{\bm{s}})$,

${\bm{B}}^{\bm{buyer}}\leftarrow \{{\bm{B}}^{\bm{buyer}}_{\bm{i}}| i\in \{1, 2, \dots, \mathrm{b}!\}, | {\bm{B}}^{\bm{buyer}}_{\bm{i}}|=\mathrm{b}\}$; \% each sequence ${\bm{B}}^{\bm{buyer}}_{\bm{i}}$ denotes a permutation of the elements in ${\bm{B^*}}$

${\bm{S^*}}\leftarrow \{{\bm{S}}^{\bm{seller}}_{\bm{j}}| j\in \{1, 2, \dots, \mathcal{C}(\mathrm{s},\mathrm{b})\}, | {\bm{S}}^{\bm{seller}}_{\bm{j}}|=\mathrm{b}, {\bm{S}}^{\bm{seller}}_{\bm{j}}\bm{\subseteq} \bm{V}^{\bm{s}}\}$, \% each sequence ${\bm{S}}^{\bm{seller}}_{\bm{j}}$ denotes a permutation of $\mathrm{b}$ elements in $\bm{V}^{\bm{s}}$ 

\For{$i=1$ to $\mathrm{b}!$}{
\For{$j=1$ to $ \mathcal{C}(\mathrm{s}, \mathrm{b})$}{
\If{a buyer in ${\bm{B}}^{\bm{buyer}}_{\bm{i}}$ can be mapped to a seller in ${\bm{S}}^{\bm{seller}}_{\bm{j}}$ while meeting constraints (C1) and (C2)}{

${\bm{\mathcal{K}}}_{\bm{ij}}\leftarrow \{{\bm{B}}^{\bm{buyer}}_{\bm{i}}, {\bm{S}}^{\bm{seller}}_{\bm{j}}\}$, $\bm{MatchT}\leftarrow \bm{MatchT}\cup {\bm{\mathcal{K}}}_{\bm{ij}}$,

\uElse{${\bm{\mathcal{K}}}_{\bm{ij}}\leftarrow []$,}
}}}
${\bm{\mathcal{K^*}}}\leftarrow $ the solution with maximum value of (3) in $\bm{MatchT}$, \% the optimal matching between the buyers and sellers

//~Stage 2: Payment calculation for winning sellers

\For{$s_{i, j, k}\in \bm{V}^{\bm{s}}$ and ${\kappa}^{n, x}_{i, j, k}=1$ }{
calculate ${\mathcal{P}}_{i, j, k}$ according to the payment rule (5),
}
\textbf{end algorithm}
\end{algorithm*}

\noindent
\textbf{Proposition 2 (Individual rationality of the sellers):}
Given the payment rule in (5), every seller $s_{i, j, k}\in {\bm{V}}^{\bm{s}}$ in the proposed auction are individual rational, in other words, sellers will not be in worse positions after participation.
\begin{proof}
\noindent
Given that the auction is truthful and the payment rule defined in (5), the utility of a winning seller $s_{i, j, k}$ is calculated as (6), where ${{\kappa}^{n, x~*}_{m, y, r}}$ and ${{\theta}^{n, x~*}_{m, y, r}}$ indicate the optimal allocation of cases with and without $s_{i, j, k}$. For notational simplicity, ${{\bm{V}}^{\bm{s}'}\triangleq\bm{V^s}}\backslash \{s_{i, j, k}\}$ denotes the sellers set without $s_{i, j, k}$, we consider the following two cases:

\noindent
\textbf{Case 1}: $\mathcal{F} (\bm{\mathcal{K}^*})\geq {\mathcal{F}}_{{\bm{V}}^{\bm{s}'}}({\bm{\Theta^*}}$). According to (5), the utility $u_{i, j, k}$ of seller $s_{i, j, k}$ is greater than or equal to zero. Consequently, $s_{i, j, k}$ will get non-negative benefit after participation.

\noindent
\textbf{Case 2}: $\mathcal{F} (\bm{\mathcal{K}^*})< {\mathcal{F}}_{{\bm{V}}^{\bm{s}'}}({\bm{\Theta^*}}$). Given that ${\bm{V}}^{\bm{s}'}\bm{\subseteq} {\bm{V}}^{\bm{s}}$, the solution space over set ${\bm{V}}^{\bm{s}'}$ is a subset of that over ${\bm{V}}^{\bm{s}}$, and thus the value of $\mathcal{F} (\bm{\mathcal{K}^*})$ is always equal to or larger than ${\mathcal{F}}_{{\bm{V}}^{\bm{s}'}}({\bm{\Theta^*}}$). Consequently, \textbf{Case 2} will never happen.

In conclusion, all sellers in the auction have the property of individual rationality.
\end{proof}

\noindent
\textbf{Proposition 3 (Individual rationality of VMs):}
All VMs are willing to take part in the proposed auction scheme.
\begin{proof}
\noindent
A VM is divided into multiple sellers with different capabilities owing to the resource reutilization factor considered in this paper. According to (6), the utility of a VM $v_{i, j}$ participates in the auction is thus calculated as:

\begin{align}
\label{eq12}
u_{i, j}& =\sum^{{\widetilde{r_{i,j}}}}_{k=1}{{\mathcal{P}}_{i, j, k}}-\sum^{{\widetilde{r_{i,j}}}}_{k=1}{q_{i, j, k}\times{{\kappa}^{n, x~*}_{i, j, k}}}\notag \\
&=\mathcal{F} (\bm{\mathcal{K}^*})\times \sum^{{\widetilde{r_{i,j}}}}_{k=1}{{\kappa}^{n, x~*}_{i, j, k}}-\sum^{{\widetilde{r_{i,j}}}}_{k=1}{{\mathcal{F}}_{{\bm{V}}^{\bm{s}}\backslash \{s_{i, j, k}\}} ({\bm{\Theta^*}})}\notag \\
&+\sum^{{\widetilde{r_{i,j}}}}_{k=1}{p_{i, j, k}\times{{\kappa}^{n, x~*}_{i, j, k}}} -\sum^{{\widetilde{r_{i,j}}}}_{k=1}{q_{i, j, k}\times{{\kappa}^{n, x~*}_{i, j, k}}}\tag{12}
\end{align}

Given that all the sellers are truthful, we have:
\begin{align}
\label{eq13}
u_{i, j}=\mathcal{F} (\bm{\mathcal{K}^*})\times \sum^{{\widetilde{r_{i,j}}}}_{k=1}{{\kappa}^{n, x~*}_{i, j, k}}-\sum^{{\widetilde{r_{i,j}}}}_{k=1}{{\mathcal{F}}_{{\bm{V}}^{\bm{s}}\backslash \{s_{i, j, k}\}} ({\bm{\Theta^*}})}.\tag{13}
\end{align}

Notably, when VM $v_{i, j}$ does not take part in the auction, all sellers $s_{i, j, k}\in {\bm{s}}^{{\bm{v}}_{\bm{i}, \bm{j}}}$ will accordingly exit the auction, which brings ${\mathcal{F}}_{{\bm{V}}^{\bm{s}}\backslash \{s_{i, j, k}\}} ({\bm{\Theta^*}})\triangleq {\mathcal{F}}_{{\bm{V}}^{\bm{s}}\backslash {\bm{s}}^{{\bm{v}}_{\bm{i}, \bm{j}}}}$ $ ({\bm{\Theta^{**}}})$. ${\bm{V}}^{\bm{s}}\backslash {\bm{s}}^{{\bm{v}}_{\bm{i}, \bm{j}}}$ denotes the set of sellers when VM $v_{i, j}$ does not take part in the auction (all sellers of VM $v_{i, j}$ are excluded) and $\bm{\Theta^{**}}={[\theta^{n,x~**}_{m,y,r}]}_{v_{m,y}\neq v_{i,j}}$ is the corresponding optimal solution. Thus, (13) can be rewrote as:

\begin{align}
\label{eq14}
u_{i, j}=\left(\mathcal{F} (\bm{\mathcal{K}^*})-\mathcal{F}_{{\bm{V}}^{\bm{s}}\backslash \bm{ s^{v_{i, j}}}} (\bm{{\Theta^{**}}})\right)\times \sum^{{\widetilde{r_{i,j}}}}_{k=1}{{\kappa}^{n, x~*}_{i, j, k}}.\tag{14}
\end{align}

Since ${\bm{\mathcal{K}^*}}$ stands for the optimal solution over set ${\bm{V}}^{\bm{s}}$, we have $\mathcal{F} ({\bm{\mathcal{K}^*}})\geq {\mathcal{F}}_{{\bm{V}}^{\bm{s}}\backslash {\bm{s}}^{{\bm{v}}_{\bm{i}, \bm{j}}}} ({\bm{\Theta^{**}}})$ which brings $u_{i, j}\geq 0.$ Thus, all the VMs are willing to participate in the proposed auction scheme. 
\end{proof}

\noindent
\textbf{Proposition 4 (Individual rationality of  SPs):}
All SPs are willing to take part in the proposed auction.

\begin{proof}

The utility $u_i$ of SP $S_i$ can be calculated as:
\begin{align}
\label{eq15}
u_i& =\sum^{|{\bm{VM}}_{\bm{i}}|}_{j=1}{u_{i, j}}. \tag{15}
\end{align}

According to \textbf{Proposition 3}, we have $u_i\geq 0$. Thus, all SPs in this auction are individual rational.
\end{proof}

\subsection{The optimal graph job allocation}

Our proposed optimal algorithm that matches each buyer to a feasible seller, proceeds in two stages, the pseudo-code of which is given in \textbf{Algorithm 1}. In stage 1, the best graph job allocation solution is obtained by exhaustive search and comparisons of the values of (3). Then, in stage 2, the payment of each winning seller is calculated according to (5), which guarantees the truthfulness. Specifically in \textbf{Algorithm 1}, lines 2-4 stands for the initialization procedure; in lines 5-10 we search for the possible mapping set $\bm{MatchT}$, wherein each mapping ${\bm{\mathcal{K}}_{\bm{ij}}}$ denotes the assignments from the buyer set to a feasible seller set; line 11 selects the best mapping with the maximum value of the total UoS of buyers given by (3); lines 13-14 in stage 2 we calculate the payments for the winning sellers.

\textbf{Algorithm 1} solves the NP-hard problem given in (4). Moreover, the constraint (C2) of (4) relies on addressing the subgraph isomorphism problem, which is also challenging to be solved~\cite{3,18}. Specially, obtaining the optimal solution requires computation complexity of $\mathcal{O} \left(\mathrm{b}!\times \mathcal{C} \left(\mathrm{s}, \mathrm{b}\right)\right)$, where $\mathrm{b}=\sum^{|\bm{O}|}_{n=1}{|{\bm{V}}^{\bm{O_n}}|}$ and $\mathrm{s}=|{\bm{V}}^{\bm{s}}|$ denotes the total number of buyers and sellers in the related VC, respectively. Symbol ``!'' is used as the factorial notation and $ \mathcal{C}(\mathrm{s}, \mathrm{b})$ stands for the ${\mathrm{s}-choose-\mathrm{b}}$ operation. Due to that jobs are modeled as undirected graphs where components (buyers) do not have a particular execution sequence. Moreover, the calculation of the payments of the winning sellers leads to the computational complexity of $\mathcal{O} \left(\mathrm{b}\times \mathrm{b}!\times \mathcal{C} \left(\mathrm{s}, \mathrm{b}\right)\right)$. Thus, a broker can hardly obtain the solution for larger-scale and real-life networks with the inreasing number of buyers and sellers. In consequence, a structure-preserved matching algorithm is proposed in the next section to solve the problem in polynomial time complexity, while achieving the maximization of the utility-of-service-gain (UoSG).

\section{The Structure-Preserved Matching Algorithm based on UoSG Maximization}

We propose a structure-preserved matching algorithm based on UoSG maximization (hereafter, we also use ``MaxUoSG'' for notational simplicity) as an efficient way to solve the graph job allocation problem in VC-assisted networks. Specifically, UoSG is defined as the profit a VC can obtain after matching one pair of buyer and seller. We later demonstrate that our proposed method enjoys truthfulness and individual rationality properties. 

\subsection{The proposed structure-preserved matching based on UoSG maximization}

For notational simplicity, we define $v^{n,x}_{m,y,r}\triangleq \left(\alpha_ng^{n, x}_{m, y, r}-p_{m, y, r}\right)$. We also define the preference lists from the perspectives of both buyers and the broker to identify the buyer-seller pairs that may bring more gains to the total UoS of the buyers.
Notably, if seller $s_{m,y,r}$ is out of the communication coverage of buyer $b_{n, x}$ or it makes the value of $\left(\alpha_ng^{n, x}_{m, y, r}-p_{m, y, r}\right)$ less than zero, it will not be included in both buyer's and broker's preference lists so as to prevent non-positive UoS. For $b_{n, x}\in {\bm{V}}^{\bm{O_n}}$, the preference list $\bm{L_{n,x}}$ is defined as (16) sorted by the value of $v^{n, x}_{m, y, r}$ in a non-ascending order.

\begin{align}
\label{eq16}
\bm{L_{n,x}}&:~\left(1, b_{n, x}, s_{m, y, r}, v^{n, x}_{m, y, r}\right){\succ}_{b_{n, x}}\dots \notag \\
&{\succ}_{b_{n, x}}\left({{\widetilde{|\bm{ {\mathcal{R}}_{{b}_{n, x}}}|}+1}},b_{n, x}, {s}_{b_{n, x}}, v_{b_{n, x}}\right),\tag{16}
\end{align}

\noindent where $\widetilde{|\bm{{\mathcal{R}}_{{\bm{b}}_{{n}, \bm{x}}}}|}$ denotes a number that meets $ 0<\widetilde{|\bm{\mathcal{R}_{b_{n,x}}}|}\leq |\bm{{\mathcal{R}_{\bm{b}_{n,x}}}}|$. Each item in the above list is represented by a tetrad that contains index in the list, a buyer, a seller and the corresponding $v^{n, x}_{m, y, r}$, symbol ${\succ}_{b_{n,x}}$ represents the preference relation, where the seller on the left-side of ``${\succ}_{b_{n,x}}$'' can bring more benefit to buyer $b_{n, x}$ as compared to all the sellers on the right-side. To ensure the property of truthfulness, a virtual seller $s_{b_{n, x}}$ is added to the end of the list $\bm{L_{n,x}}$ as a critical indicator with the critical UoS $v_{b_{n, x}}\triangleq \left({\alpha}_n\widetilde{g^{n, x}_{i, j, k}}-\widetilde{p_{i, j, k}}\right)$ 
be slightly lower than the values of all UoS in the list. The relevant proofs of economical properties will be given in Section 4.3.

The broker's preference list given by (17) is regarded as a union set of all $\bm{L_{n,x}}$ excluding the virtual sellers, sorted by non-ascending order of the value of $v^{n, x}_{m, y, r}$, where the mapping between a buyer and a seller on the left-side of ``${\succ}_{broker}$'' can bring higher benefit to the buyers as compared to those on the right-side. An example of preference lists of buyers and the broker is shown in Fig.~2.

\begin{align}
\label{eq17}
\bm{L_{Bro}}&:~\left(1,b_{n, x}, s_{m, y, r}, v^{n, x}_{m, y, r}\right) {\succ}_{broker}~\notag \\
&\left(2,b_{n', x'}, s_{m', y', r'}, v^{n', x'}_{m', y', r'}\right){\succ}_{broker} \cdots \tag{17}
\end{align}

The proposed MaxUoSG contains the job allocation stage and the payment decision stage. The preference list $\bm{L_{Bro}}$ will be mainly referred to in the former stage aiming to maximize the total UoS. Notably, this might sacrifice some buyers' benefits. For instance, $\left(1, b_{n, x}, s_{i, j, k}, v^{n, x}_{i, j, k}\right)$ and $\left(1, b_{n', x'}, s_{i, j, k}, v^{n', x'}_{i, j, k}\right)$ located at the top of $\bm{L_{n,x}}$ and $\bm{L}_{\bm{n',x'}}$ respectively; however, the broker is inclined to match $b_{n, x}$ to $s_{i, j, k}$ since $\left(\alpha_ng^{n, x}_{i, j, k}-p_{i, j, k}\right)>\left(\alpha_{n'}g^{n', x'}_{i, j, k}-p_{i, j, k}\right)$. Then, lists of the buyers $\bm{L_{n,x}}$ are used to calculate the payments for winning sellers.
\begin{figure}[h!t]
\centerline{\includegraphics[width=0.9\linewidth]{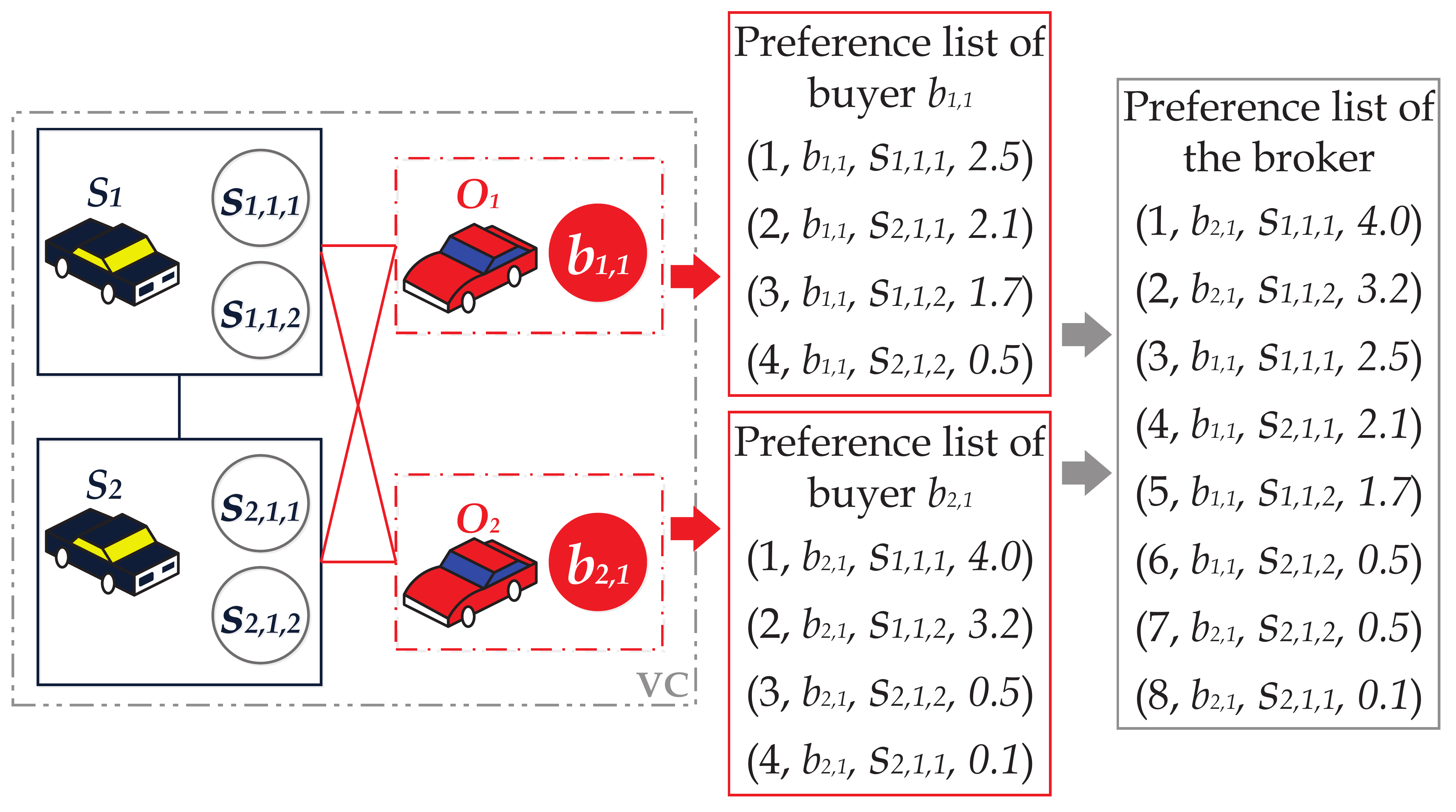}}
\caption{Example of the preference lists of buyers and the broker.}
\end{figure}

\subsection{The related payment rule}

For notational simplicity, let $\widetilde{s_{i, j, k}}$ be the first seller located behind $s_{i, j, k}$ in list $\bm{L_{n,x}}$. We define the payment of a winning seller $s_{i, j, k}$ as 
\begin{align}
\label{eq18}
{\mathcal{P}}^{\#}_{i, j, k}= \left({\alpha}_ng^{n, x}_{i, j, k}-\left({\alpha}_n\widetilde{g^{n, x}_{i, j, k}}-\widetilde{p_{i, j, k}}\right)\right)\times{{\kappa}^{n, x~\#}_{i, j, k}},\tag{18}
\end{align}
where $\widetilde{g^{n, x}_{i, j, k}}=t_{n, x}-\widetilde{c_{i, j, k}}$ denotes the benefit of buyer $b_{n, x}$ from enjoying the computing service of seller $\widetilde{s_{i, j, k}}$. Symbol $\widetilde{c_{i,j,k}}$ and $\widetilde{p_{i, j, k}}$ represents the execution time and the bid of $\widetilde{s_{i, j, k}}$, respectively. In (18), ${{\kappa}^{n, x~\#}_{i, j, k}}$ indicates the obtained solution via the proposed MaxUoSG algorithm. 

\begin{algorithm*}
\caption{The structure-preserved matching algorithm based on UoSG maximization}
\SetKwInOut{Input}{Input}\SetKwInOut{Output}{Output}
\Input{graph jobs ${\bm{G}}^{\bm{O_n}}=\{ {\bm{V^{O_n}}}, {\bm{E^{O_n}}}, {\bm{W^{O_n}}}\}$, VC graph ${\bm{G}}^{\bm{s}}=\{ {\bm{V^s}}, {\bm{E^s}}, {\bm{W^s}}\}$, $\bm{\mathcal{B}}$, 
$\bm{\mathcal{D}}$, $\bm{L_{n,x}}$, $\bm{L_{Bro}}$}

\Output{the sub-optimal solution $\bm{{\mathcal{K}}^{sub}}$, the related payments ${\mathcal{P}}^{\#}_{i, j, k}$ of the winning sellers}

//~Stage 1: The buyer-seller pair selection procedure

Initialization: ${\bm{B^*}}\leftarrow \{\bm{V}^{\bm{O}_{\bm{1}}}\cup \bm{V}^{\bm{O}_{\bm{2}}}\dots \cup \bm{V}^{\bm{O}_{|\bm{O}|}}\}$, ${\bm{S^*}}\leftarrow \bm{V}^{\bm{s}}$, $\mathrm{b}\leftarrow length ({\bm{B^*}})$, $\mathrm{s}\leftarrow length ({\bm{S^*}})$, $\mathrm{L}\leftarrow length (\bm{L_{Bro}})$,

${\bm{\mathcal{K}^{sub}}}\leftarrow []$, $\bm{MatchT}\leftarrow []$, $ indTry\leftarrow 0$, $ indMatch \leftarrow 1$,

$\bm{MatchT}\leftarrow$ the first item in $\bm{L_{Bro}}$, \% add the first item of $\bm{LBro}$ to $\bm{MatchT}$

\While{the index of the last item in $\bm{MatchT}\le \mathrm{L}$~and~$indTry \le \mathrm{L}$}{

\For{$ indTry=(indMacth+1)$~to~$\mathrm{L}$}{
\If{the buyer-seller pair of index $indTry$ in $\bm{L_{Bro}}$ is not in $\bm{MatchT}$}{

$\bm{MTemp} \leftarrow$ the item with index $indTry$ in $\bm{L_{Bro}}$,

 $\bm{TempB}\leftarrow$ the buyer in $\bm{MTemp}$, \% save the relevant buyer 

 $\bm{TempS}\leftarrow$ the seller in $\bm{MTemp}$, \% save the relevant seller

\If{match $\bm{TempB}$ to $\bm{TempS}$ can meet constraints (C1) and (C2), and prevent the interdependency with buyers and sellers in $\bm{MatchT}$ }{

$\bm{MatchT}\leftarrow \bm{MTemp}\cup \bm{MatchT}$, \% add the new buyer-seller pair to $\bm{MatchT}$

$indMatch \leftarrow$ the index of the last item in $\bm{MatchT}$, \% update $ indMatch $

}
}
}

\If{the number of buyers in $\bm{MatchT}==\mathrm{b}$}{
go to stage 2, \% if all the buyers are successfully mapped to the sellers, go to stage 2 (line 26)}
\If{the number of buyers in $\bm{MatchT}< \mathrm{b}$~and~$indMatch < \mathrm{L}$~and~$indTry==\mathrm{L}$}{
\If{the number of buyers in $\bm{MatchT}==1$}{
$ indMatch \leftarrow$ the index of the first item in $\bm{MatchT}+1$, \% restart the matching procedure from the second item in $\bm{L_{Bro}}$ 

the first item in $\bm{MatchT}\leftarrow$ the item with ${indMatch}$ in $\bm{L_{Bro}}$,

\uElse{the last item in $\bm{MatchT} \leftarrow[]$, \% delete the last successful matching pair in $\bm{MatchT}$ and keep searching from the next item of which in $\bm{L_{Bro}}$}
}

}
\If{the number of buyers in $\bm{MatchT}< \mathrm{b}$~and~$indMatch==\mathrm{L}$~and~$indTry==\mathrm{L}$}{go to line 28,}
}

${\bm{\mathcal{K}^{sub}}}\leftarrow$ all buyer-seller pairs in $\bm{MatchT}$, \% a feasible matching between the buyers and the sellers

//~Stage 2: Payment calculation for winning sellers

\For{$s_{i, j, k}\in \bm{V}^{\bm{s}}$ and ${\kappa}^{n, x}_{i, j, k}=1$ }{
calculate ${\mathcal{P}}^{\#}_{i, j, k}$ according to the payment rule in (18),
}

\textbf{end algorithm}

\end{algorithm*}

The pseudo-code of the proposed MaxUoSG algorithm is given in \textbf{Algorithm 2}. The notation $\bm{MatchT}$ stands for the match table that stores all the successfully matched buyer-seller pairs, and $\bm{\mathcal{K}^{sub}}$ denotes the final graph job allocation solution. In stage 1 (lines 1-24), a mapping between the buyers and the winning sellers are obtained according to $\bm{L_{Bro}}$, while taking into account the graph job structures. In stage 2 (lines 25-27), the payment of each winning seller is calculated based on the buyer's preference list. More precisely, lines 2-4 correspond to the initialization procedure. Lines 5-13 search for feasible buyer-seller pairs from the top of list $\bm{L_{Bro}}$, while meeting the structure-preservation and one-to-one matching constraints. For the former constraint, the structure of each graph job has to be preserved, while the later constraint requires that the buyers and sellers that are in current $\bm{MatchT}$ should be ignored during the matching procedure. Lines 14-15 stand for the successful matching where all the buyers are mapped to the feasible sellers. Lines 16-19 correspond to the case in which no feasible buyer-seller pairs can be chosen after the first item of $\bm{L_{Bro}}$, restart the matching procedure from the second item in list $\bm{L_{Bro}}$. In lines 20-21, we delete the last successful matching pair in $\bm{MatchT}$ and keep searching from the next item located behind the deleted item in $\bm{L_{Bro}}$, if the matching procedure is blocked by an infeasible buyer-seller pair (e.g, a buyer-seller pair that cannot meet the structure-preservation constraint). Lines 22-23 correspond to the failure of matching. In stage 2, lines 26-27 calculate the payment for every winning seller according to the proposed payment rule shown in (18). 

The proposed algorithm enables low computation complexity when solving the auction-based graph job allocation problem. For buyers' preference list generations, bubble sort is applied that brings a computation complexity of $\mathcal{O}\left({|\widetilde{\bm{\mathcal{R}_{b_{n,x}}}}|}^2\right)$ for each buyer. As for $\bm{L_{Bro}}$, merge sort algorithm is utilized with computation complexity of $\mathcal{O} \left(\mathrm{L}\times log \left(\mathrm{L}\right)\right)$, where $\mathrm{L}=|\bm{L_{Bro}}|$ denotes the number of items in $\bm{L_{Bro}}$. The proposed structure-preserved matching provides the best computation complexity performance of $\mathcal{O}(\mathrm{b})$ and the worst case complexity of $\mathcal{O} ((\mathrm{L}-\mathrm{b})\times \mathrm{b})$.

\subsection{Analysis of truthfulness and individual rationality}

\noindent
\textbf{Proposition 5 (Truthfulness):}
The proposed MaxUoSG algorithm makes each seller, i.e., $s_{i, j, k}\in {\bm{V}}^{\bm{s}}$ set its bid equal to the true valuation, i.e., $p_{i, j, k}=q_{i, j, k}$.

\begin{proof}
Due to the structure-preservation factor, any changes to the broker's preference list may have a great impact on the final matching solution. Therefore, two cases are considered: \textbf{Case 1}) the misreporting behavior of a seller causes no change to the orders of items in $\bm{L_{Bro}}$; and \textbf{Case 2}) the misreporting behavior of a seller causes changes to the orders of items in $\bm{L_{Bro}}$. The utility $u_{i, j, k}\ $ of a winning seller $s_{i, j, k}$ when $p_{i, j, k}=q_{i, j, k}$, is given by:
\begin{align}
\label{eq19}
u_{i, j, k}={\mathcal{P}}^{\#}_{i, j, k}-q_{i, j, k}\times{{\kappa}^{n, x~\#}_{i, j, k}}\tag{19}
\end{align}

A virtual seller $s'_{i, j, k}$ is considered owning the same properties with seller $s_{i, j, k}$, except for the untruthful bid $p'_{i, j, k}\neq q_{i, j, k}$, which denotes the case where seller $s_{i, j, k}$ misreports in the auction. The utility of $s'_{i, j, k}$ is given by:
\begin{align}
\label{eq20}
u'_{i, j, k}={\mathcal{P}}^{\#\prime}_{i, j, k}-q_{i, j, k}\times{{\kappa}^{n, x~\#\prime}_{i, j, k}},\tag{20}
\end{align}
where ${{\kappa}^{n, x~\#\prime}_{i, j, k}}$ is the graph job allocation solution via the proposed algorithm when $s_{i, j, k}$ misreports its bid. Based on the aforementioned utilities, the proof of truthfulness is presented based on the above mentioned two cases. For~\textbf{Case 1}, consider the following conditions:

\textbf{Case 1.1} ($p'_{i, j, k}<q_{i, j, k}$). If $s'_{i, j, k}$ wins the auction, the utility can be calculated as:
\begin{align}
\label{eq21}
u'_{i, j, k}=\left({\alpha}_n{g^{n, x}_{i, j, k}}'- \left({\alpha}_n{\widetilde{g^{n, x~\prime}_{i, j, k}}}-\widetilde{p'_{i, j, k}}\right)-q_{i, j, k}\right)\times{{\kappa}^{n, x~\#\prime}_{i, j, k}},\tag{21}
\end{align}
where ${g^{n, x~\prime}_{i, j, k}}$ and ${\widetilde{g^{n, x~\prime}_{i, j, k}}}$ denote the buyer $b_{n, x}$'s benefit obtained from virtual seller $s'_{i, j, k}$ and the first seller located after $s'_{i, j, k}$ in list $\bm{L_{n,x}}$, respectively. Similarly, if $s_{i, j, k}$ wins the auction, we get the utility:
\begin{align}
\label{eq22}
u_{i, j, k}= \left({\alpha}_ng^{n, x}_{i, j, k}- \left({\alpha}_n\widetilde{g^{n, x}_{i, j, k}}-\widetilde{p_{i, j, k}}\right)-q_{i, j, k}\right)\times{{\kappa}^{n, x~\#}_{i, j, k}},\tag{22}
\end{align}
When $p'_{i, j, k}<q_{i, j, k}$, $s'_{i, j, k}$  is always located before $s_{i, j, k}$ in each buyer's preference list, which lead to $\left({\alpha}_n{\widetilde{g^{n, x~\prime}_{i, j, k}}}-\widetilde{p'_{i, j, k}}\right)\geq \left({\alpha}_n\widetilde{g^{n, x}_{i, j, k}}-\widetilde{p_{i, j, k}}\right)$ and $u_{i, j, k}{\geq u}'_{i, j, k}$. Consequently, $s'_{i, j, k}$ will never be mapped to $b_{n, x}$ owing to the same properties with $s_{i, j, k}$ except for the bid. Thus, bidding truthfully will always lead to more utility.

\textbf{Case 1.2} ($p'_{i, j, k}>q_{i, j, k}$). It is obvious that $s_{i, j, k}$ will be located before $s'_{i, j, k}$ in each buyer's preference list.

\noindent \textbf{Case 1.2.1:} $s_{i, j, k}$ and $s'_{i, j, k}$ are located adjacently with each other in list $\bm{L_{n,x}}$. Here, we have $u'_{i, j, k}=u_{i, j, k}$ since $\widetilde{s'_{i, j, k}}$ is the only critical indicator. Thus, $s_{i, j, k}$ does not have to misreport his bid in the auction.

\noindent
\textbf{Case 1.2.2:} There is a seller $\widetilde{s_{i, j, k}}$ located between $s_{i, j, k}$ and $s'_{i, j, k}$ in list $\bm{L_{n,x}}$. Here, if $s'_{i, j, k}$ wins the auction, $\widetilde{s_{i, j, k}}$ will also win the auction. Due to the one-to-one mapping rule between buyers and sellers in the proposed auction, we have $u'_{i, j, k}=0$.

To meet the requirement of structure-preservation, a seller who misreports bid faces the following risks in \textbf{Case 2.} 

\noindent
\textbf{Risk 2.1:} the structure-preservation constraint may bring a seller $s'_{i, j, k}$ with failure in this auction even when $p'_{i, j, k}>q_{i, j, k}$, which leads to $u'_{i, j, k}=0$.

\noindent
\textbf{Risk 2.2:} the structure-preservation constraint leads to an equal utility for an untruthful seller $s'_{i,j,k}$, i.e., $u'_{i, j, k}=u_{i, j, k}$, when $\left({\alpha}_n{\widetilde{g^{n, x~\prime}_{i, j, k}}}-\widetilde{p'_{i, j, k}}\right)= \left({\alpha}_n\widetilde{g^{n, x}_{i, j, k}}-\widetilde{p_{i, j, k}}\right)$.

\noindent
\textbf{Risk 2.3:} the structure-preservation constraint leads to an non-positive utility for an untruthful seller $s'_{i,j,k}$, if ${\alpha}_n{g^{n, x~\prime}_{i, j, k}}-\left({\alpha}_n{{\widetilde{g^{n, x~\prime}_{i, j, k}}}}-\widetilde{p'_{i, j, k}}\right)<0$, which leads to $u'_{i, j, k}<0$.

Notably, misreporting behavior may bring an untruthful seller a higher utility in \textbf{Case 2} (examples are detailed in simulation). However, a seller has no idea how to adjust its bid to aviod the above mentioned risks while insuring a higher utility, owing to the non-transparent information in the proposed marketplace (e.g., the opportunistic V2V communication duration among different SPs, and the required weights 
${\omega}^{O_n}_{xx'}$ among buyers of each graph job). 

Consequently, in \textbf{Case 2}, being truthful stands for a risk-free option for each seller. In general, the proposed MaxUoSG algorithm offers truthfulness in this auction marketplace.

\end{proof}


\noindent
\textbf{Proposition 6 (Individual rationality of the sellers):}
All the sellers in the proposed auction scheme are individual rational via the proposed MaxUoSG algorithm.

\begin{proof}
\noindent
Given that seller $s_{i, j, k}$ bids truthfully and $s_{\widetilde{i, j, k}}$ is the seller located behind $s_{i, j, k}$ in the preference list of $b_{n, x}$, we have:
\begin{align}
\label{eq23}
{\alpha}_ng^{n, x}_{i, j, k}-p_{i, j, k}\geq {\alpha}_n\widetilde{g^{n, x}_{i, j, k}}-\widetilde{p_{i, j, k}},\tag{23}
\end{align}
which leads to
\begin{align}
\label{eq24}
{\alpha}_ng^{n, x}_{i, j, k}-{\alpha}_n\widetilde{g^{n, x}_{i, j, k}}+{\widetilde{p_{i, j, k}}}\geq p_{i, j, k}\tag{24}
\end{align}

Combining (22) with (24), we have $u_{i, j, k}\geq 0$. Thus, all the sellers in the proposed auction scheme have the property of individual rationality.
\end{proof}

\noindent
\textbf{Proposition 7 (Individual rationality of VMs and SPs):}
All SPs and VMs are willing to take part in the proposed auction scheme via the proposed MaxUoSG algorithm.
\begin{proof}
\noindent
For each VM $v_{i, j}\in {\bm{VM}}_{\bm{i}}$ and \textbf{SP} $S_i\in \bm{S}$, we calculate the utility of which as $u_{i, j}=\sum^{{\widetilde{r_{i,j}}}}_{k=1}{u_{i, j, k}}$, and $u_i=\sum^{|{\bm{VM}}_{\bm{i}}|}_{j=1}{u_{i, j}}$, respectively. According to \textbf{Proposition 6}, we have $u_{i, j}\geq 0$ and $u_i\geq 0$. In conclusion, all SPs and VMs are willing to participate in the proposed auction scheme.
\end{proof}

\section{Numerical Results and Performance Evaluation}
\begin{figure}[h!t]
\centerline{\includegraphics[width=\linewidth]{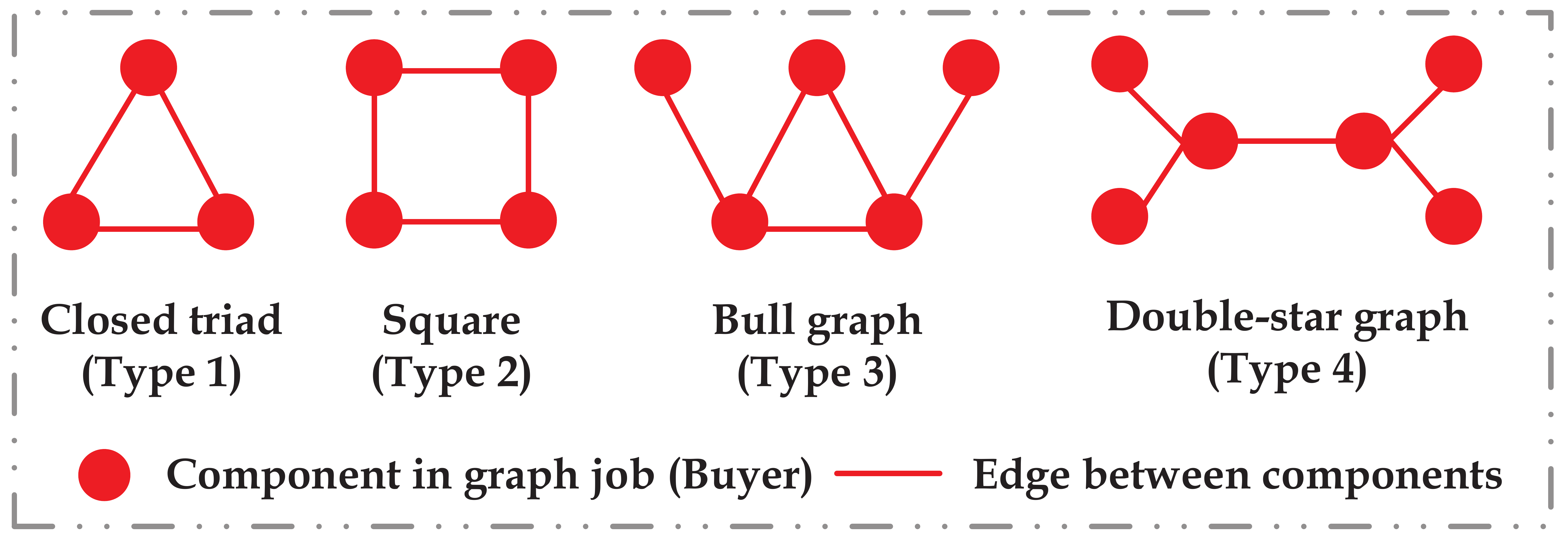}}
\caption{Graph job types considered in simulation~\cite{3,29}.}
\end{figure}

\begin{figure}[h!t]
\centerline{\includegraphics[width=1.0\linewidth]{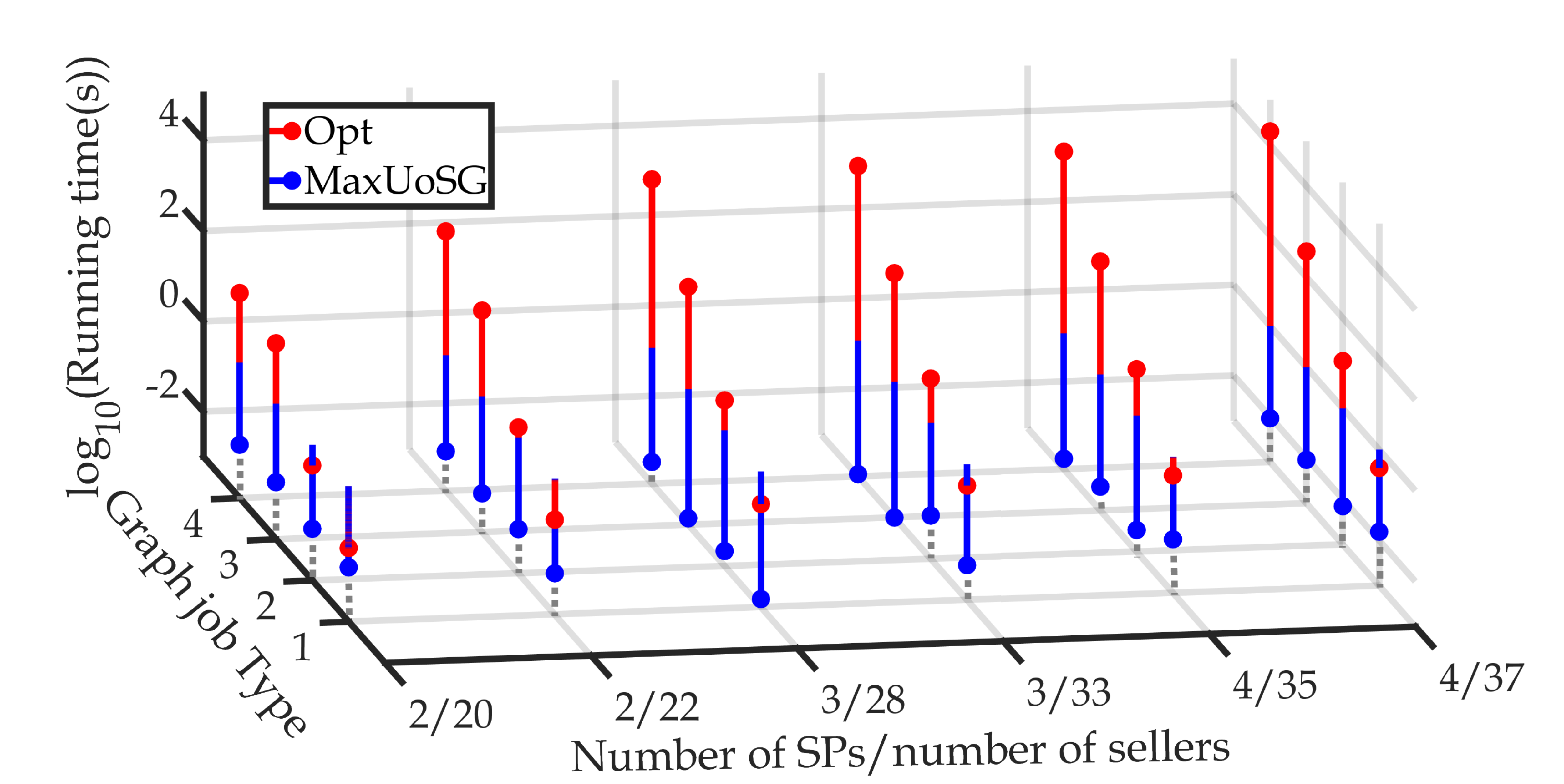}}
\caption{Running time performance of the optimal and the proposed MaxUoSG algorithm considering various number of buyers and sellers.}
\end{figure}
This section presents numerical results, illustrating the validity of the proposed algorithms. In the following, the performance of the optimal algorithm (``Opt'') and the structure-preserved matching algorithm based on UoSG maximization (``MaxUoSG'') comparing with the baseline methods, are analyzed in detail. Moreover, various problem sizes are investigated with different numbers of buyers and sellers. Baseline methods and the related procedures considered in the simulation are given below:

\begin{figure*}[h!t]
\centering
\subfigure[]{\includegraphics[width=.245\linewidth]{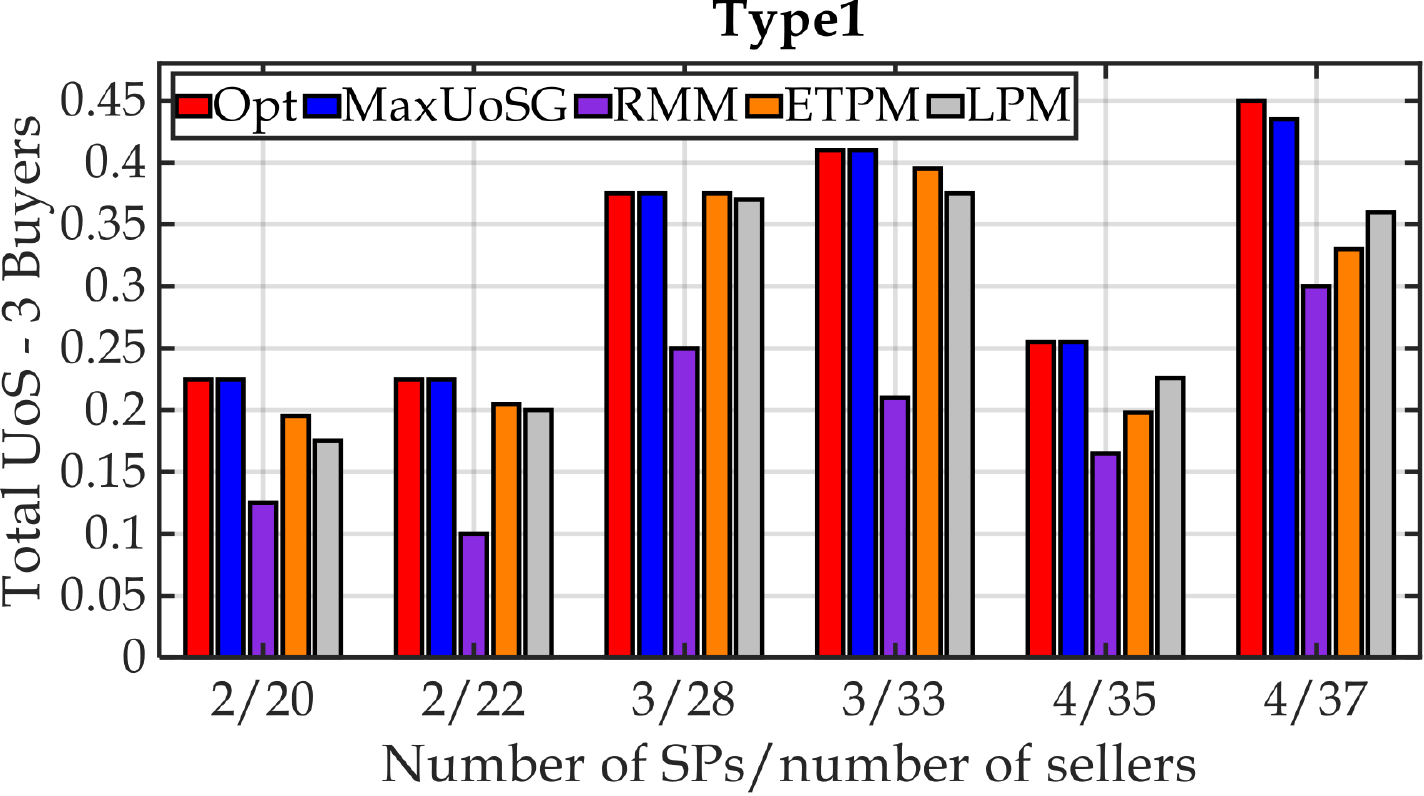}}
\subfigure[]{\includegraphics[width=.245\linewidth]{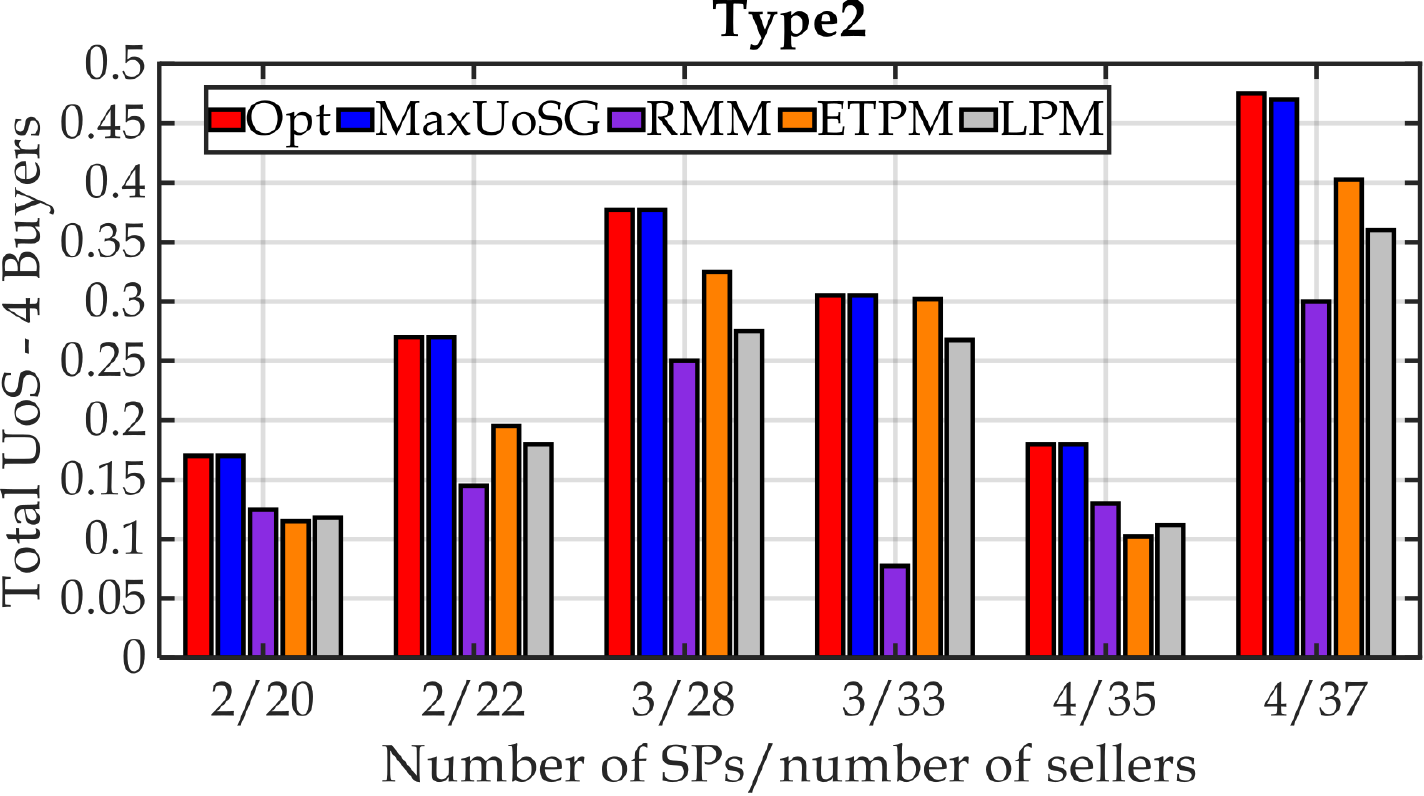}}
\subfigure[]{\includegraphics[width=.245\linewidth]{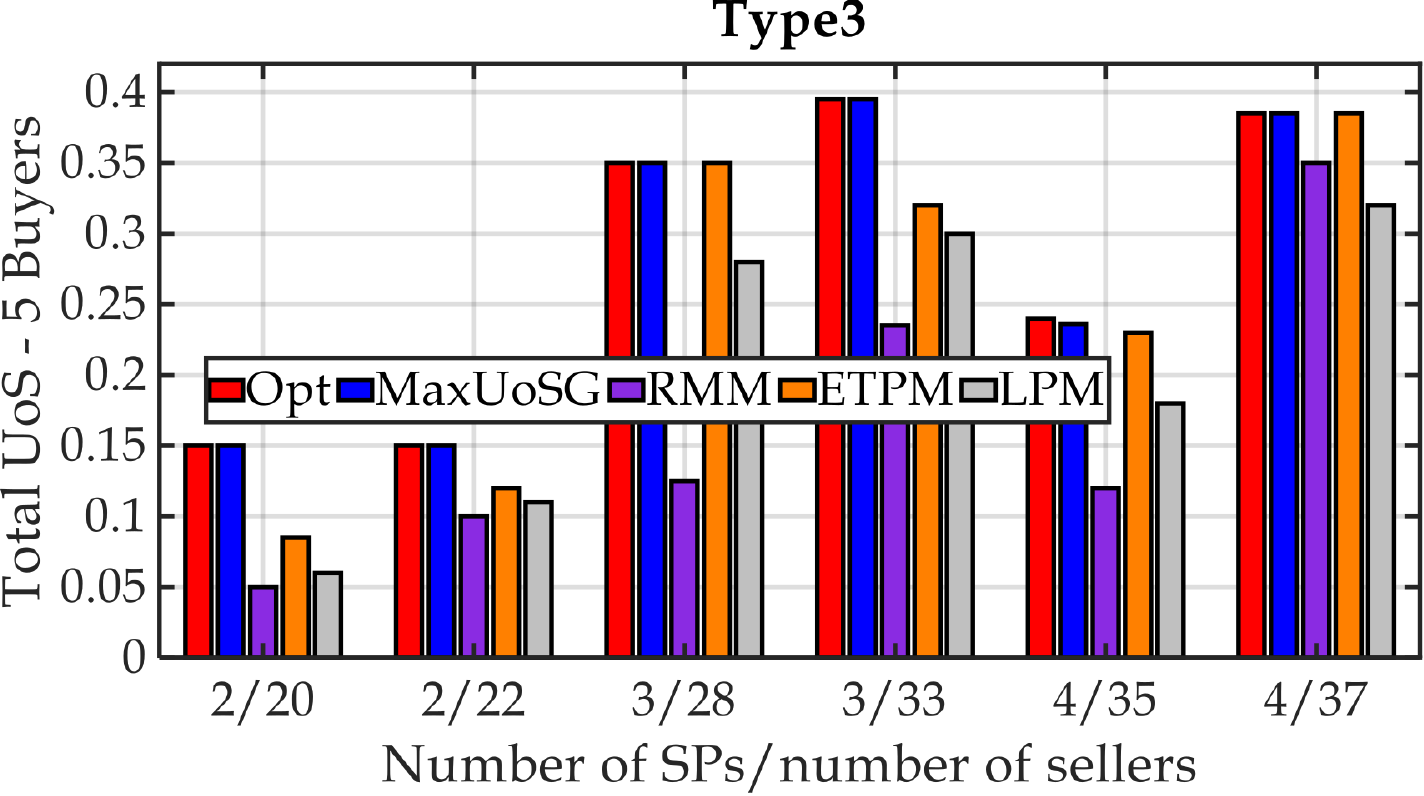}}
\subfigure[]{\includegraphics[width=.245\linewidth, height=.137\linewidth]{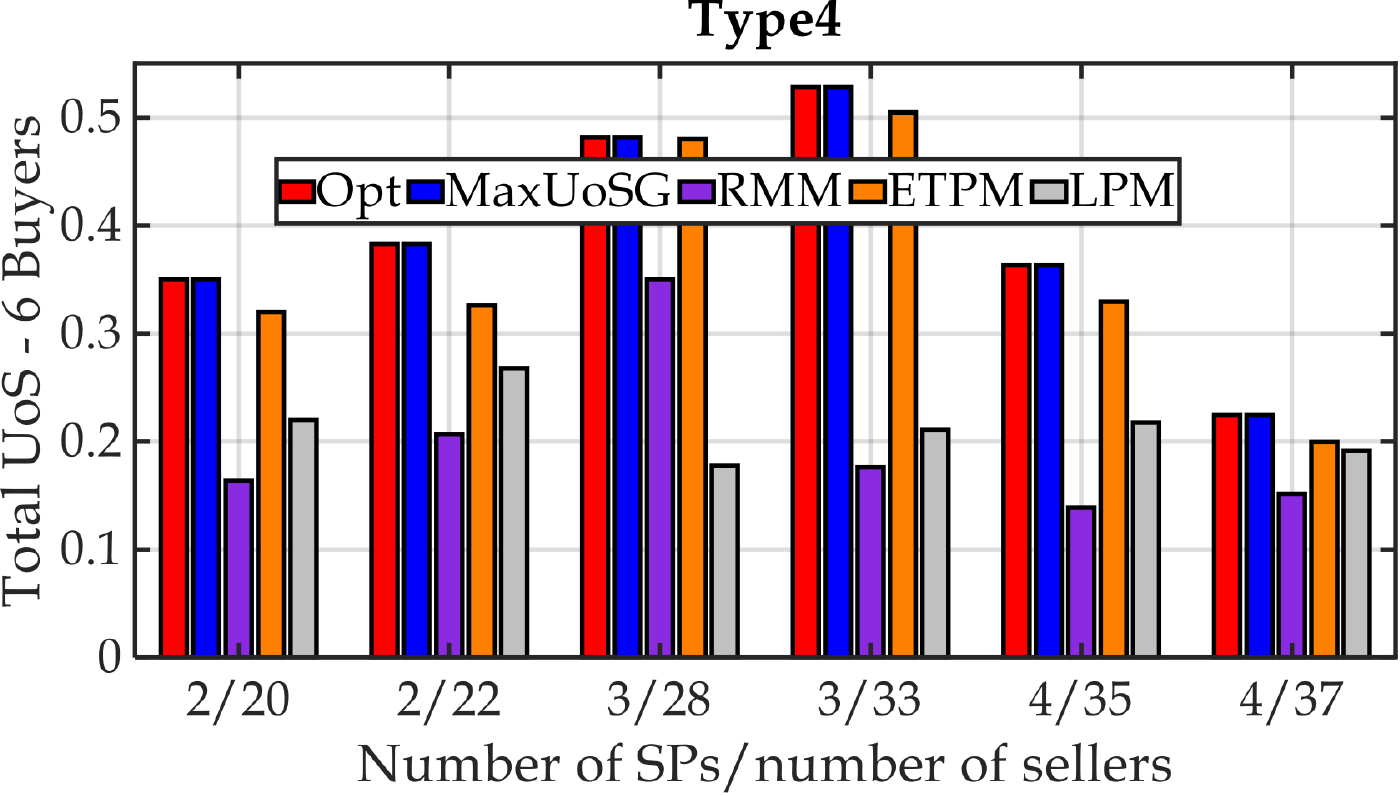}}

\subfigure[]{\includegraphics[width=.245\linewidth]{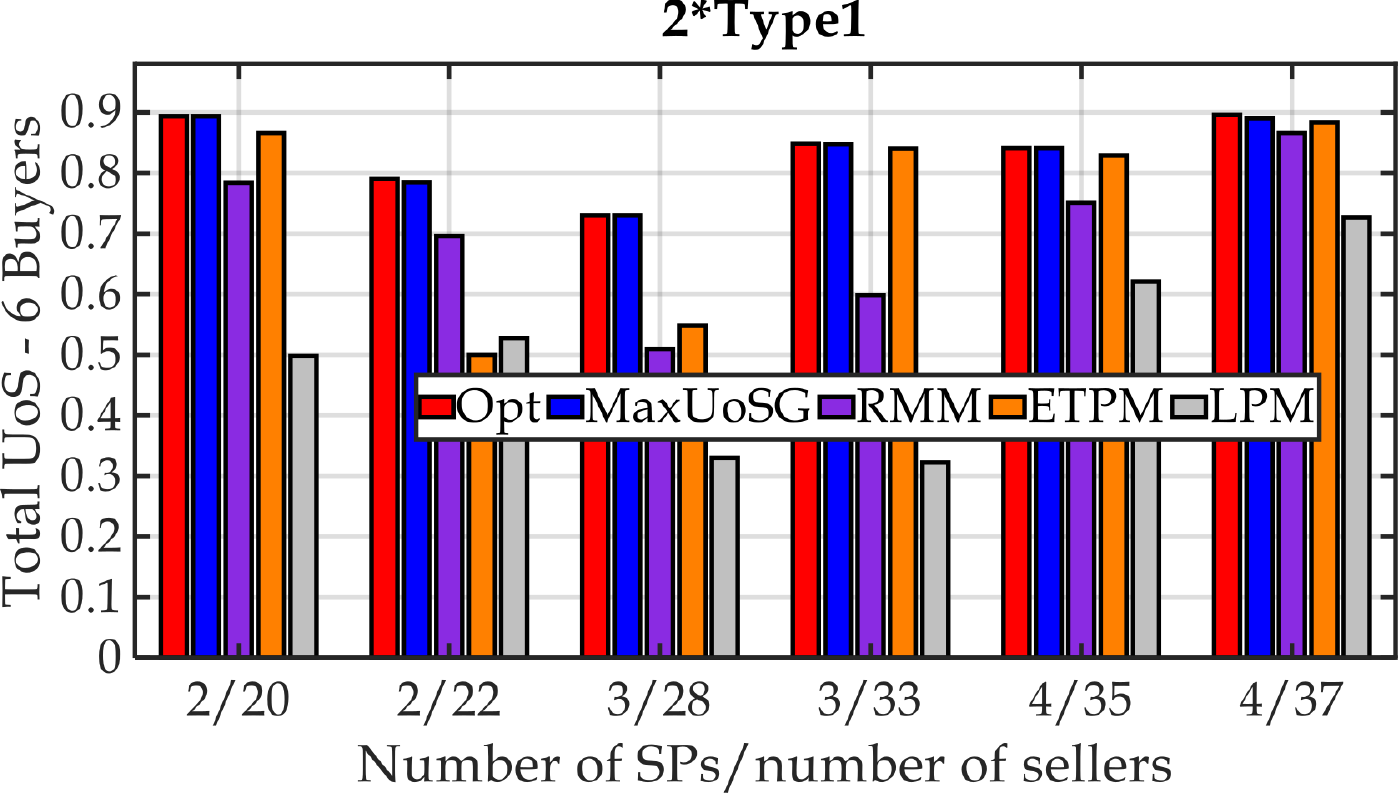}}
\subfigure[]{\includegraphics[width=.245\linewidth]{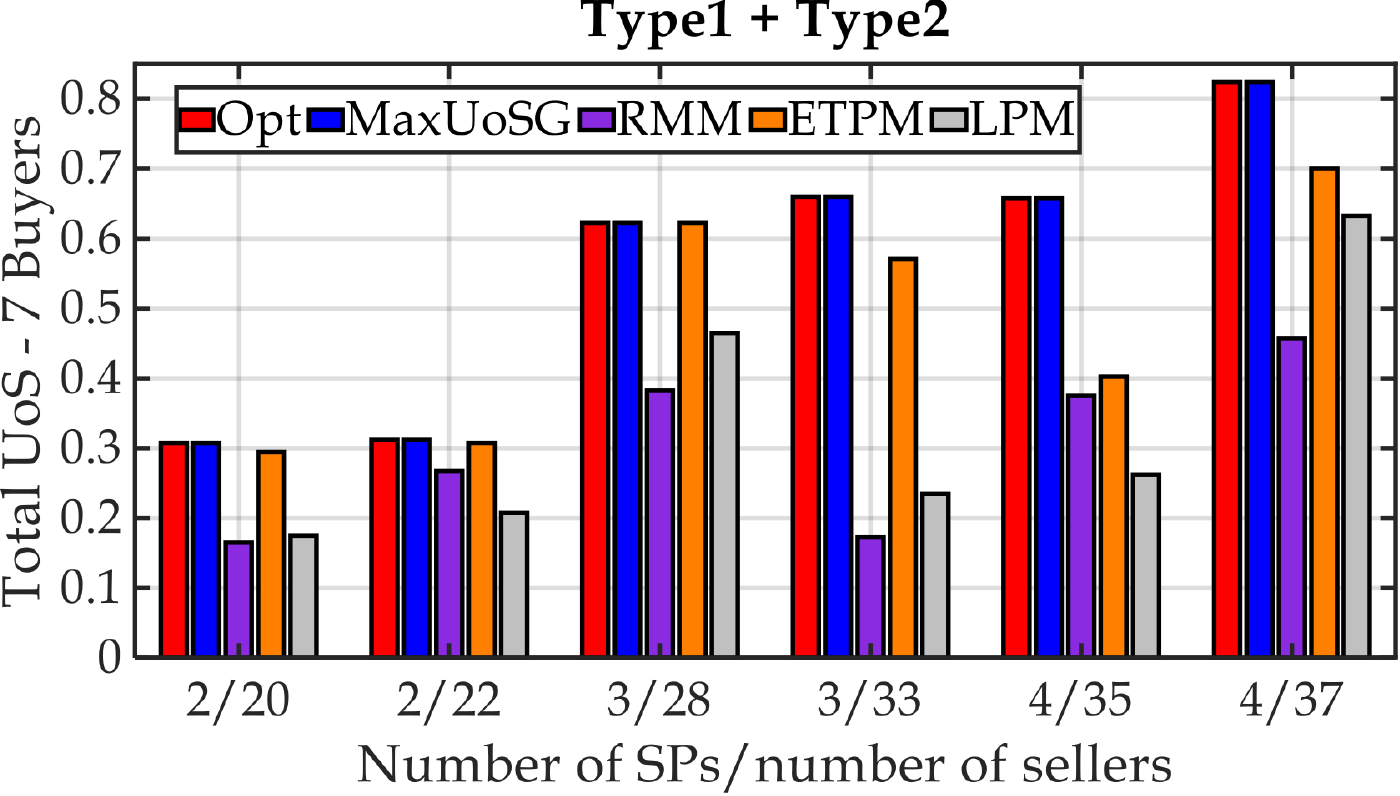}}
\subfigure[]{\includegraphics[width=.245\linewidth]{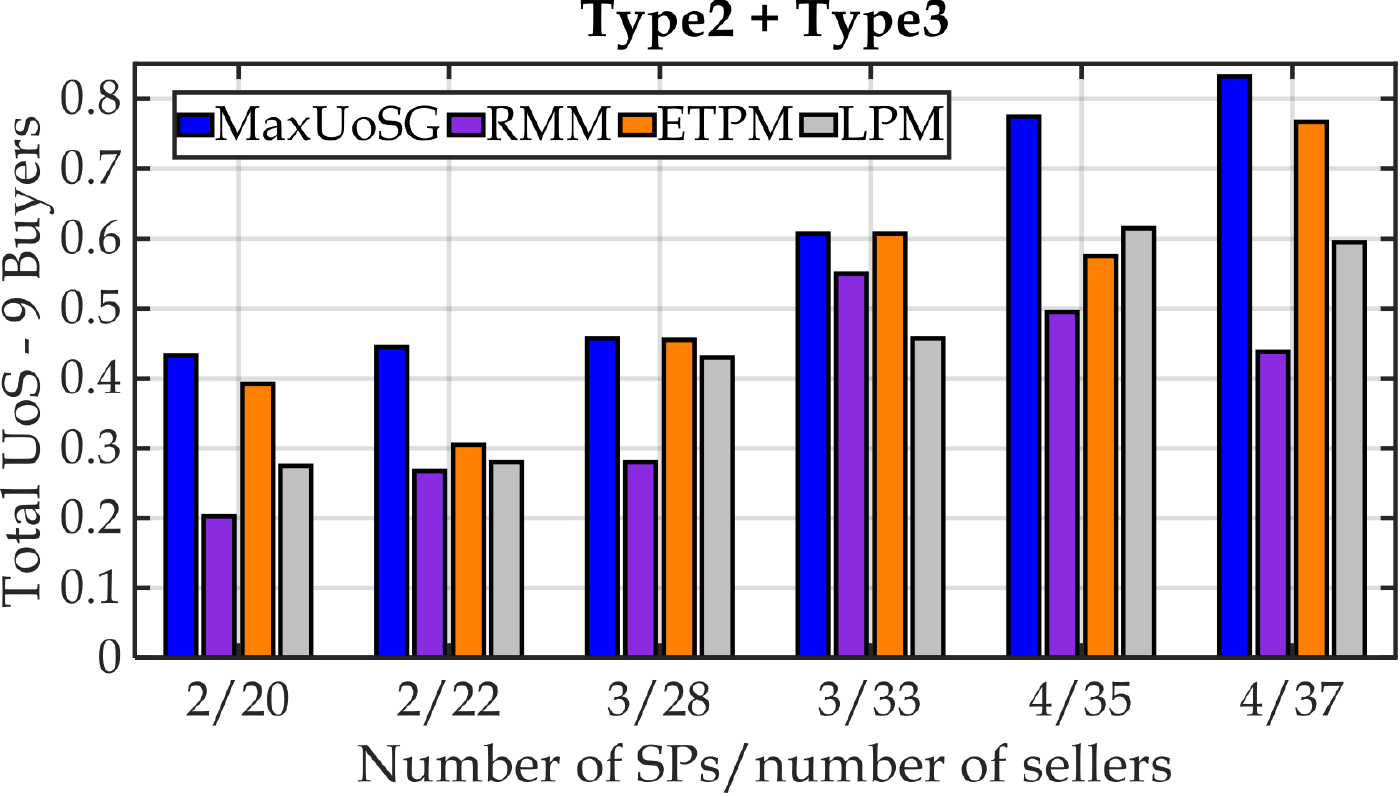}}~
\subfigure[]{\includegraphics[width=.245\linewidth]{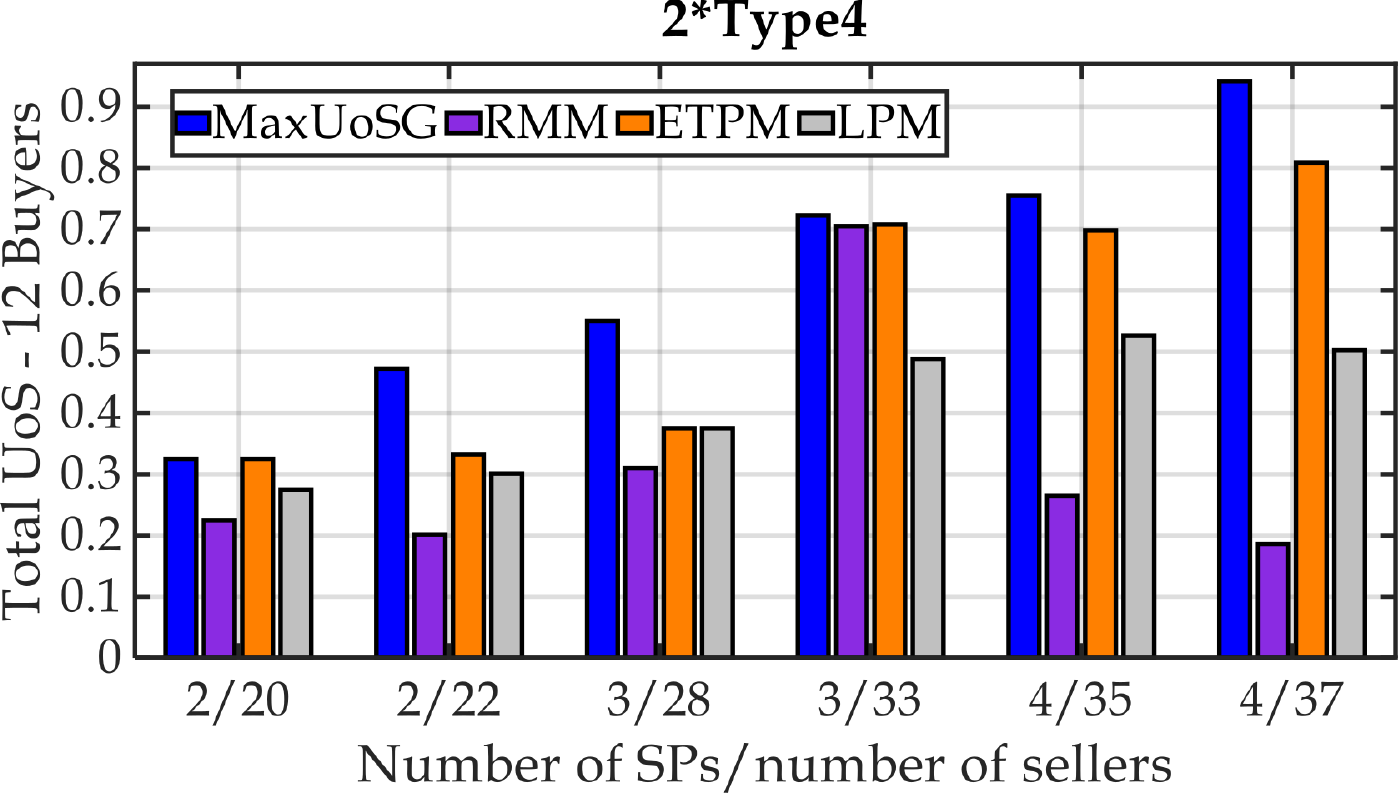}}
\caption{Performance evaluation of the total UoS in small problem size scenarios, where the title of each sub-figure denotes the related graph job type(s).}
\end{figure*}

\begin{figure*}[h!t]
\centering
\subfigure[]{\includegraphics[width=.245\linewidth]{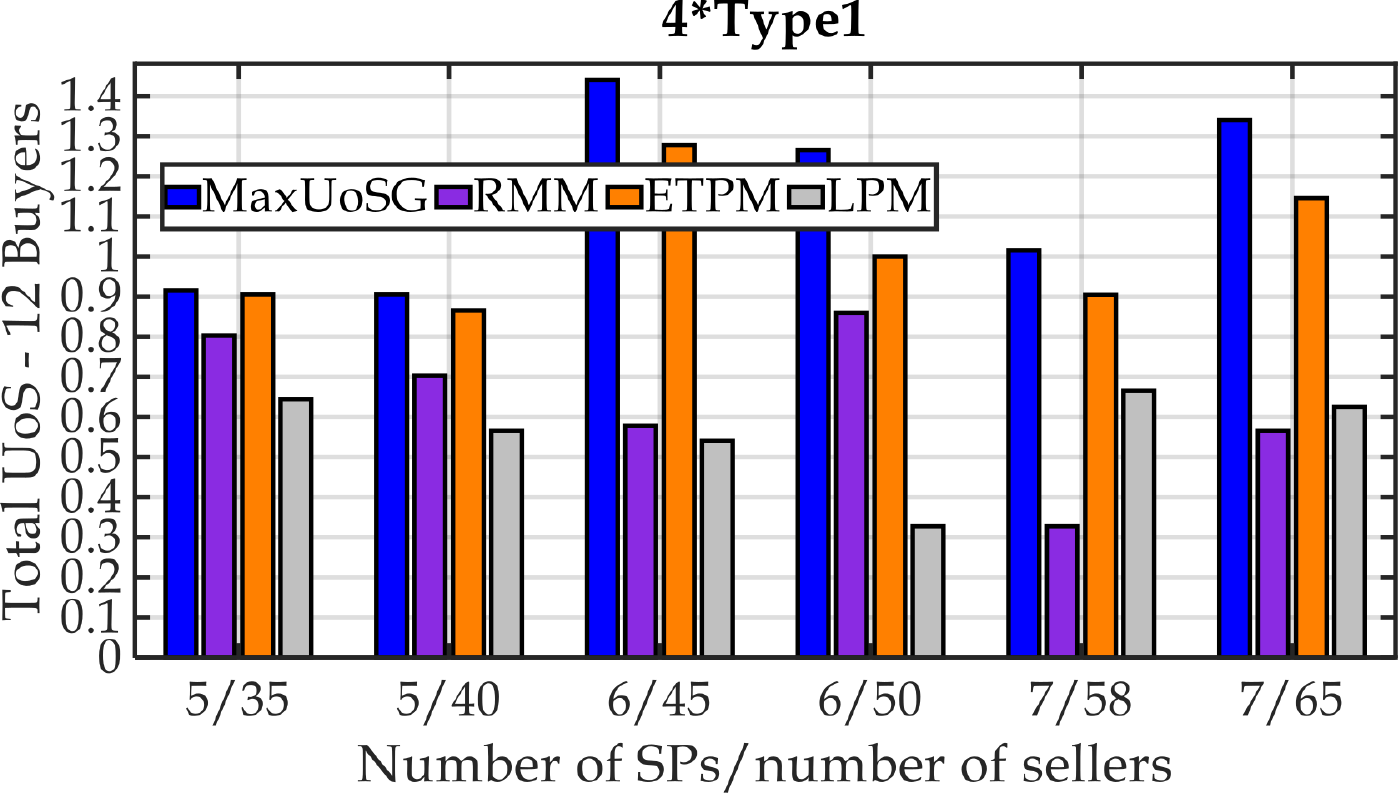}}
\subfigure[]{\includegraphics[width=.245\linewidth,height=.137\linewidth]{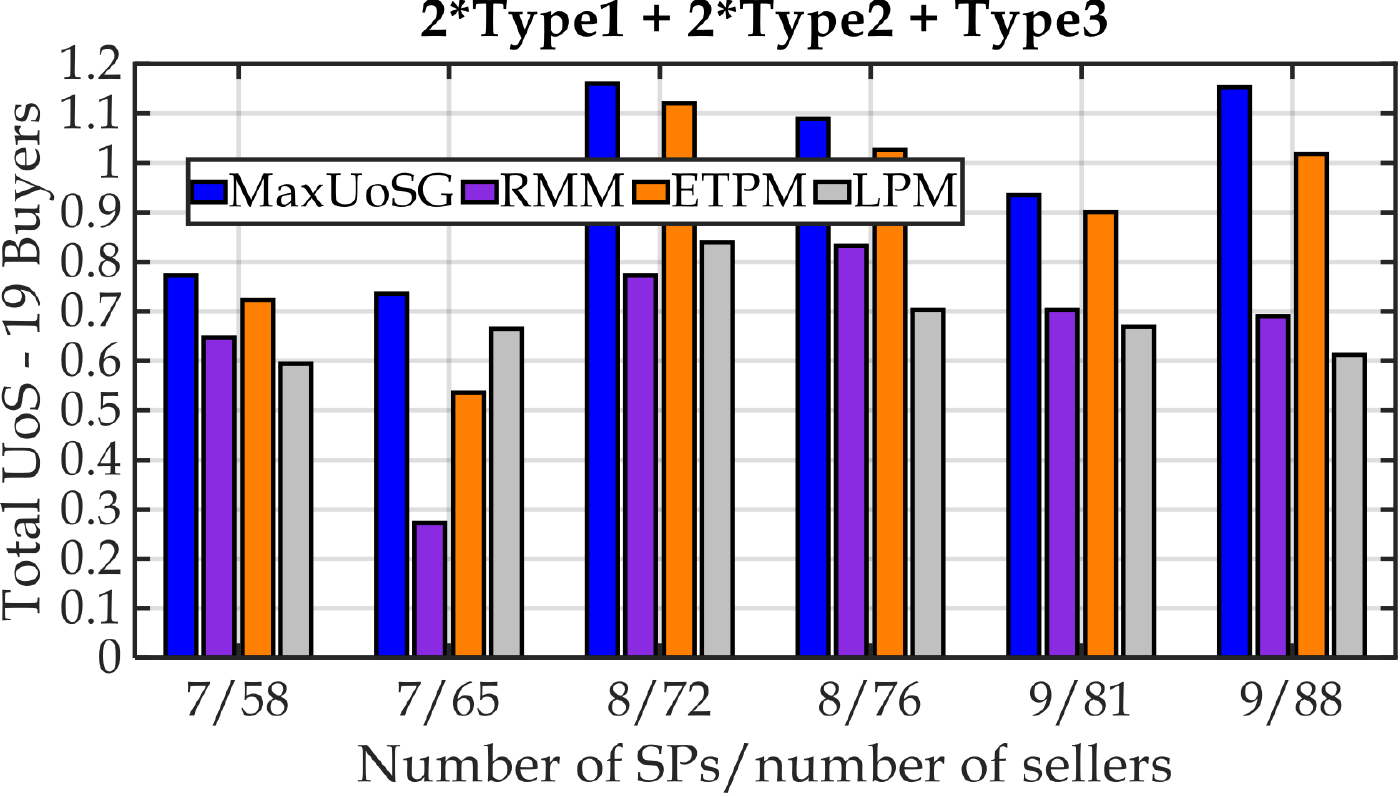}}
\subfigure[]{\includegraphics[width=.245\linewidth,height=.137\linewidth]{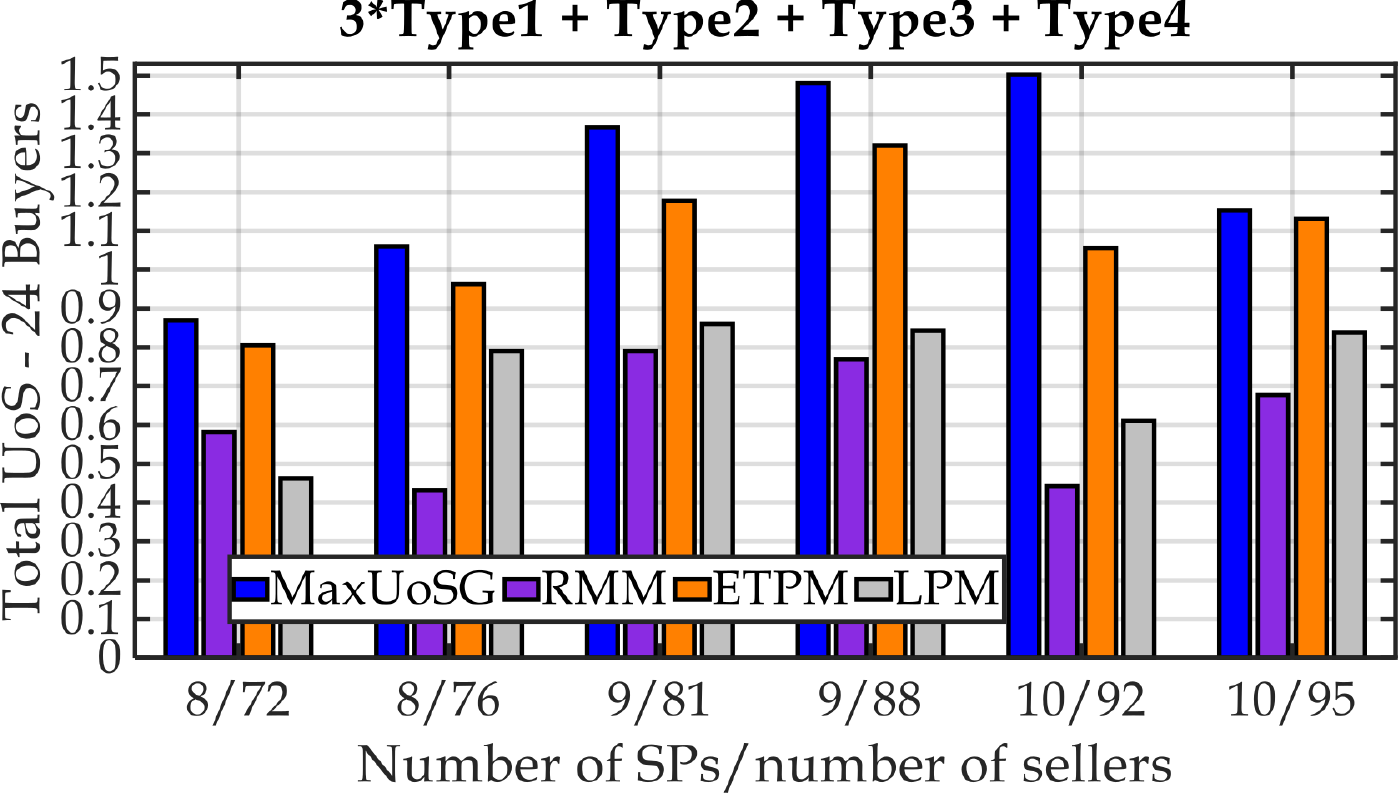}}
\subfigure[]{\includegraphics[width=.245\linewidth,height=.138\linewidth]{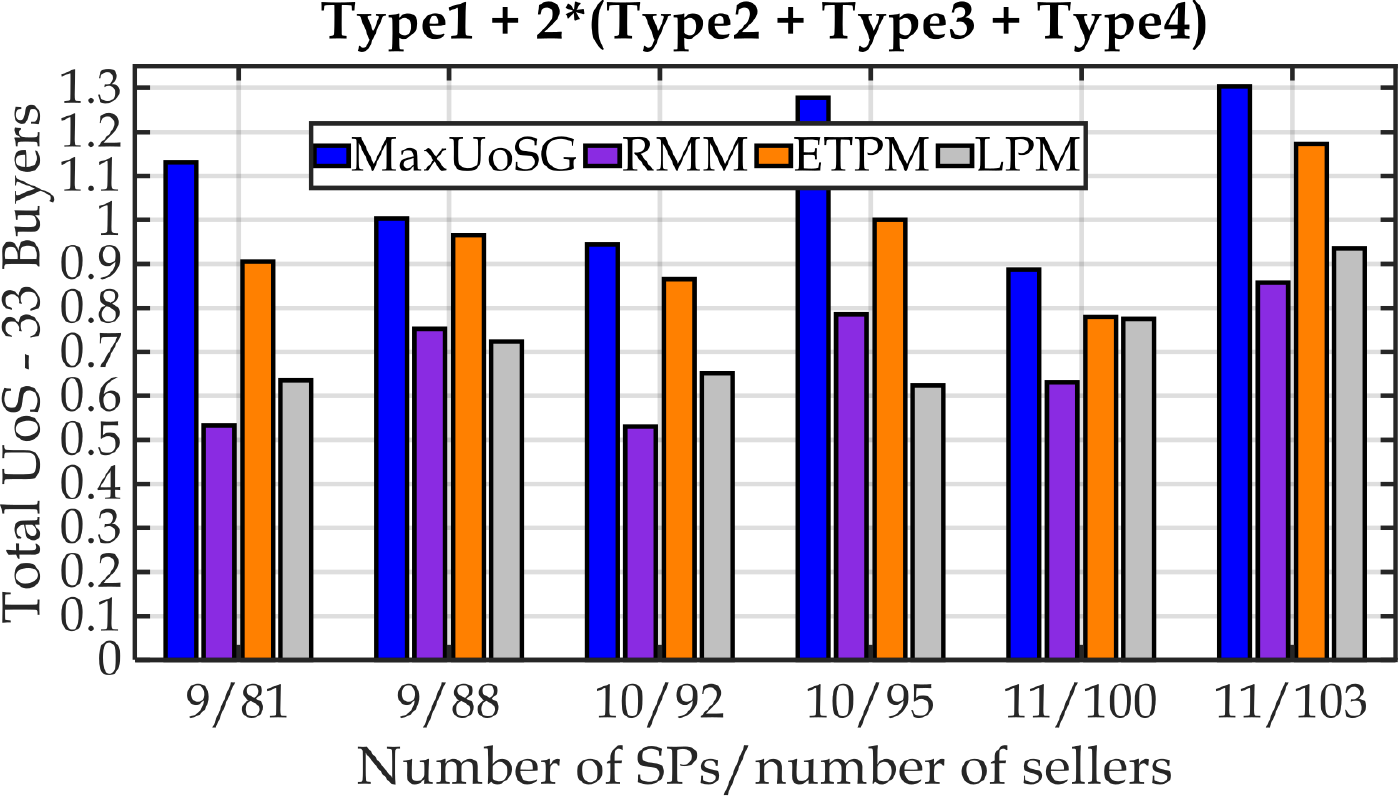}}
\caption{Performance evaluation of the total UoS in large problem size scenarios, where the title of each sub-figure denotes the related graph job type(s).}
\end{figure*}
\noindent
\textbf{1) Execution time preferred mechanism (ETPM)}: Randomly select a buyer and match it to the seller with the current lowest execution time, while satisfying constraints (C1) and (C2), until all the buyers are mapped to the sellers successfully.

\noindent
\textbf{2) Low price preferred mechanism (LPM)}: Randomly select a buyer and match it to the seller with the current lowest price, while satisfying constraints (C1) and (C2), until all the buyers are mapped to the sellers successfully.

\noindent
\textbf{3) Random matching mechanism (RMM)}: Randomly select a buyer and randomly match it to an available seller, while satisfying constraints (C1) and (C2), until all the buyers are mapped to the sellers successfully.

\subsection{Simulation setup}
Graph job structures considered in the simulation are depicted in Fig.~3. A monotone decreasing function formed by ${\mathcal{U}}^S (c_{m, y, r})={\beta}_2-{\beta}_{1}c_{m, y, r}$ is applied to determine the true valuation $q_{m, y, r}$ of each seller, where ${\beta}_{1}$ and ${\beta}_{2}$ are positive constants $({\beta}_{1}\in [0.7,0.9], {\beta}_{2}\in [0.9,1]$) which enables a higher true valuation of a more powerful seller. The simulation parameters are randomly obtained from the following intervals: $\varepsilon \in [0.9,0.95]$, ${\alpha}_n\in [1,1.5]$, $c_{m, y}\in [0.2,0.3]$, $t_{n, x}\in [0.6,0.7]$, ${\omega}^{o_n}_{xx'}\in [0.1,0.7]$, and ${\lambda}_{mm'}\in [0.05,0.06]$ for small problem sizes with a couple of JOs and SPs and ${\lambda} _{mm'}\in [0.01,0.02]$ for large problem sizes with more participants.

\subsection{Running time performance}

The running time performance comparison between the optimal algorithm and the proposed MaxUoSG is shown in Fig.~4, for various numbers of SPs/sellers and different job types depicted in Fig.~3. Notably, 10-based logarithm representation is applied since the gap between the running time of the two algorithms becomes too large as the graph job and VC structures become more complicated (e.g., by increasing the number of buyers and sellers). Compared with the proposed MaxUoSG, Fig.~4 indicates that as the number of buyers and sellers in a VC grows, the running time of obtaining the optimal solutions rises sharply, while that of the proposed MaxUoSG remains at a certain order of magnitude of ${10}^{-2}$ seconds, which makes the optimal algorithm unsuitable for fast-changing and large-scale networks. Moreover, different topological complexity of VC configurations (e.g., existence of more sellers and edges in VC graph) may also lead to a dramatic change in the running time. Specifically, obtaining the optimal solution takes more than ${2\times 10}^4$~seconds upon allocating a job with type 4 and four SPs (37 sellers), as shown in Fig.~4.

\subsection{The obtained UoS in various problem sizes}

\begin{table*}[!t]
\centering
\begin{center}\renewcommand\arraystretch{1.3}
     \caption{The average performance improvement of the proposed MaxUoSG compared to the baseline methods obtained from Fig. 5 and Fig. 6.}
     \setlength{\tabcolsep}{1.2mm}{
\begin{tabular}{|c|c|c|c|c|c|c|c|c|c|c|c|c|}
\hline
\diagbox{Baselines}{Figures}& 
Fig. 5(a)& 
Fig. 5(b)& 
Fig. 5(c)& 
Fig. 5(d)&
Fig. 5(e)&
Fig. 5(f)&
Fig. 5(g)&
Fig. 5(h)&
Fig. 6(a)&
Fig. 6(b)&
Fig. 6(c)&
Fig. 6(d)\\
\hline
RMM& 74.96$\%$ & 90.18$\%$ & 101.95$\%$ & 90.85$\%$ & 21.38$\%$ & 95.27$\%$ &68.11$\%$ & 141.67$\%$ & 97.73$\%$ &61.62$\%$ & 111.70$\%$ & 64.07$\%$ \\
\hline
ETPM& 12.24$\%$ & 30.15$\%$ & 20.42$\%$ & 9.07$\%$ & 16.07$\%$ & 17.11$\%$ &16.65$\%$ & 12.67$\%$ & 12.39$\%$ & 10.67$\%$ & 14.81$\%$ & 20.49$\%$ \\
\hline
LPM&  14.23$\%$ & 37.41$\%$ & 49.08$\%$ & 80.91$\%$ & 69.80$\%$ & 86.96$\%$ &37.35$\%$ & 50.11$\%$ & 118.2$\%$ & 43.57$\%$ & 62.88$\%$ & 36.76$\%$ \\
\hline
\end{tabular}}
\end{center}
\end{table*}
\begin{figure*}[h]
\centering
\subfigure[]{\includegraphics[width=.248\linewidth]{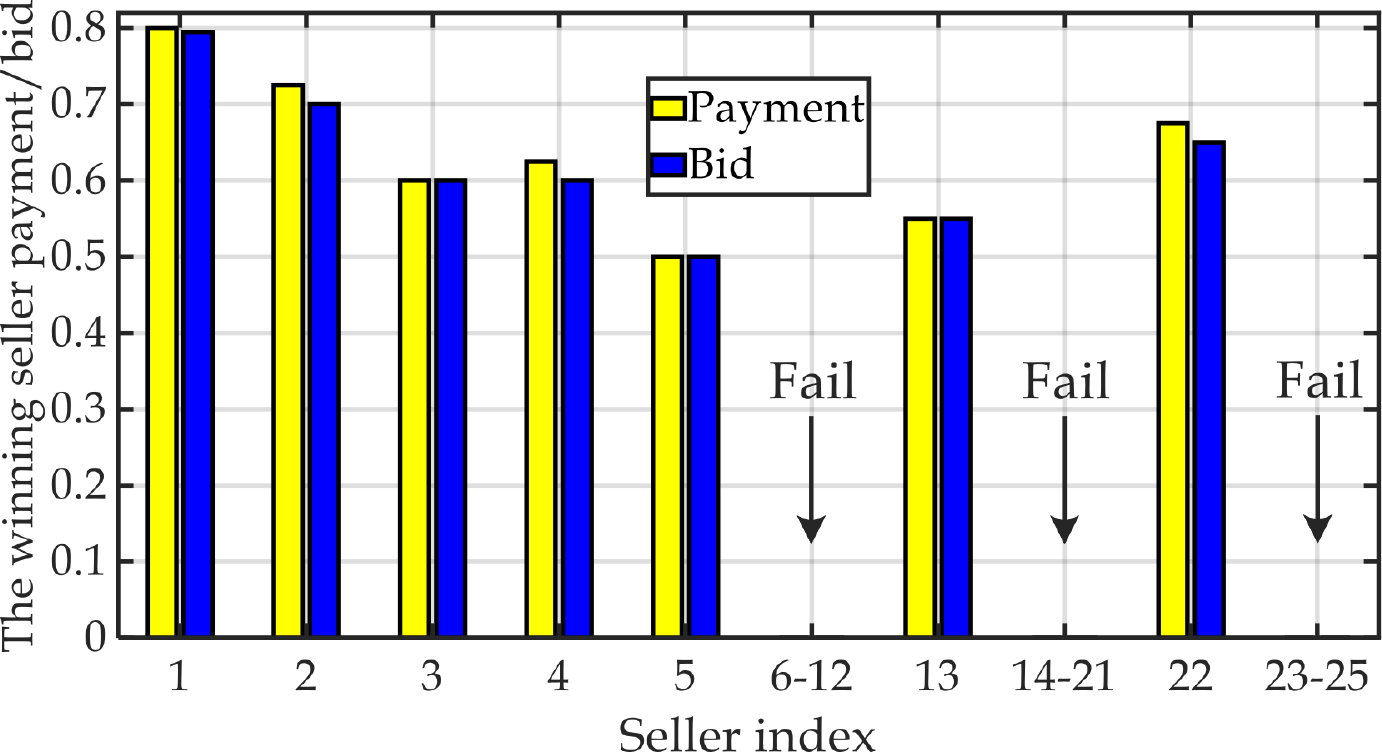}}
\subfigure[]{\includegraphics[width=.245\linewidth]{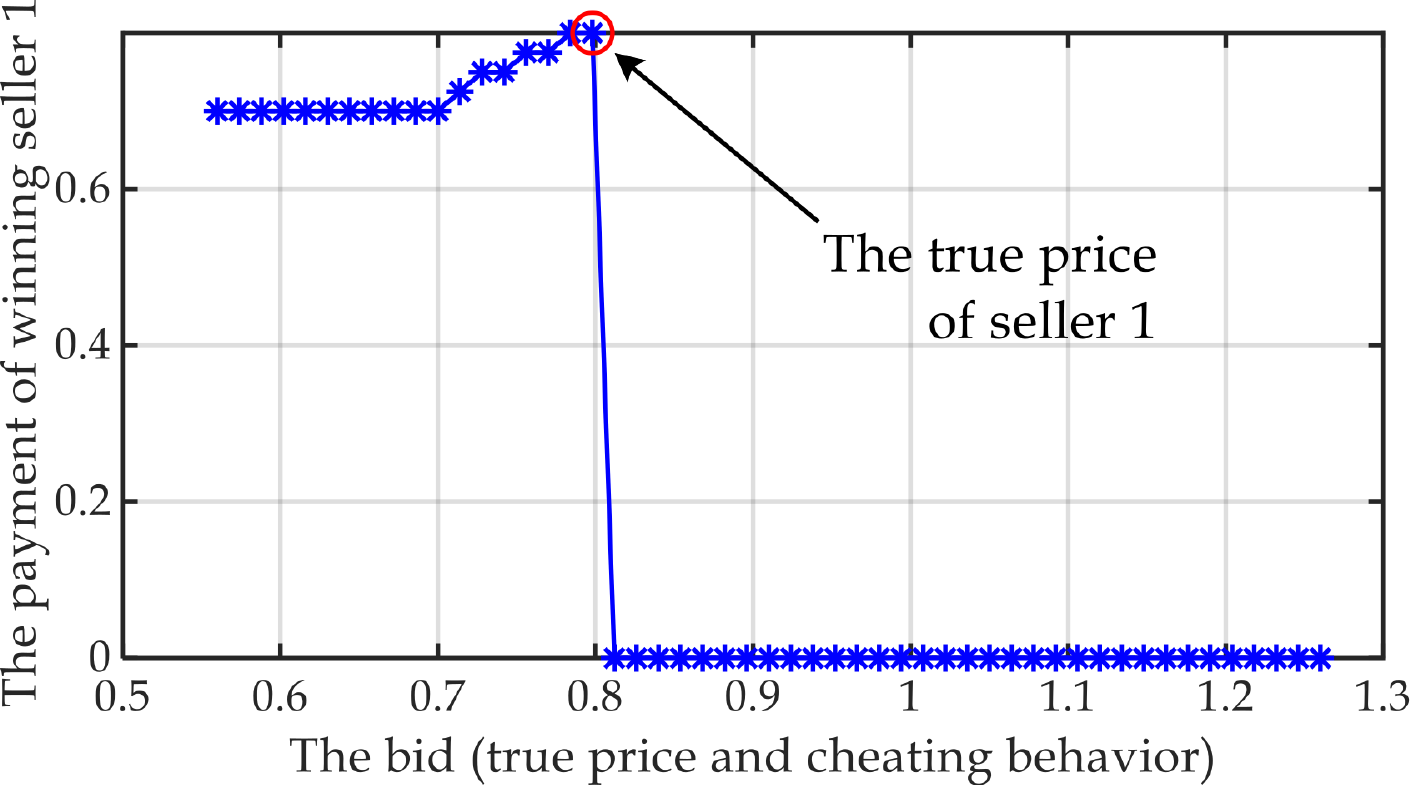}}
\subfigure[]{\includegraphics[width=.245\linewidth]{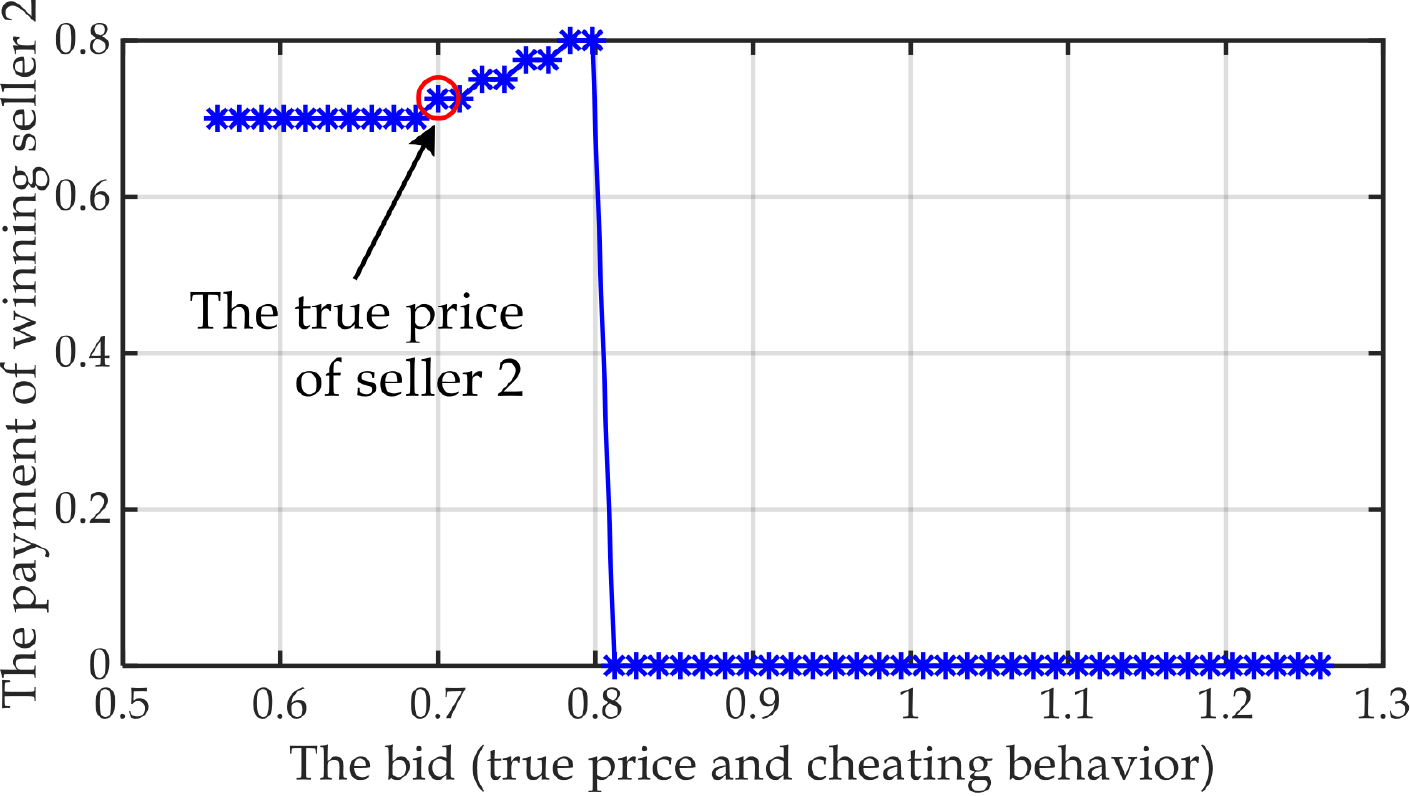}}
\subfigure[]{\includegraphics[width=.245\linewidth]{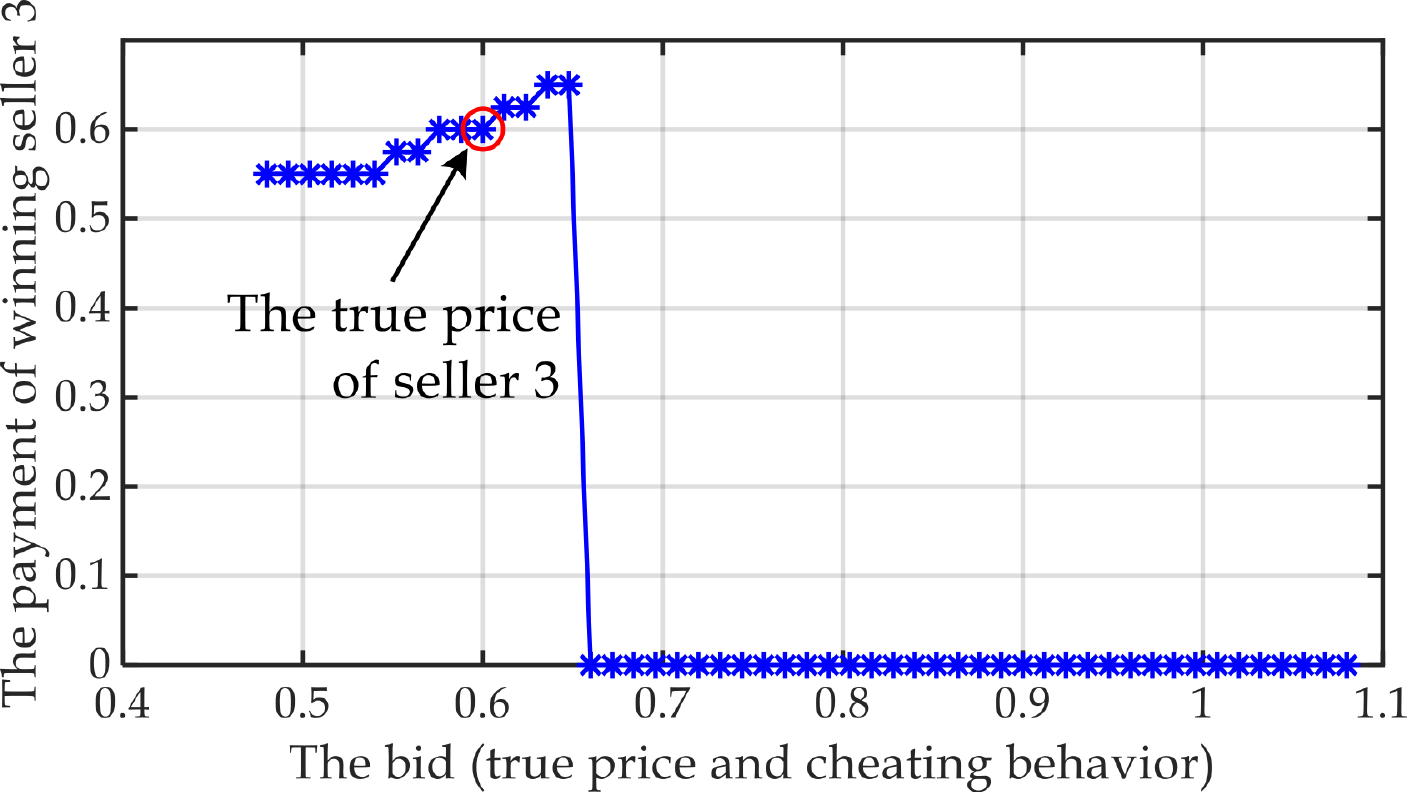}}

\subfigure[]{\includegraphics[width=.245\linewidth]{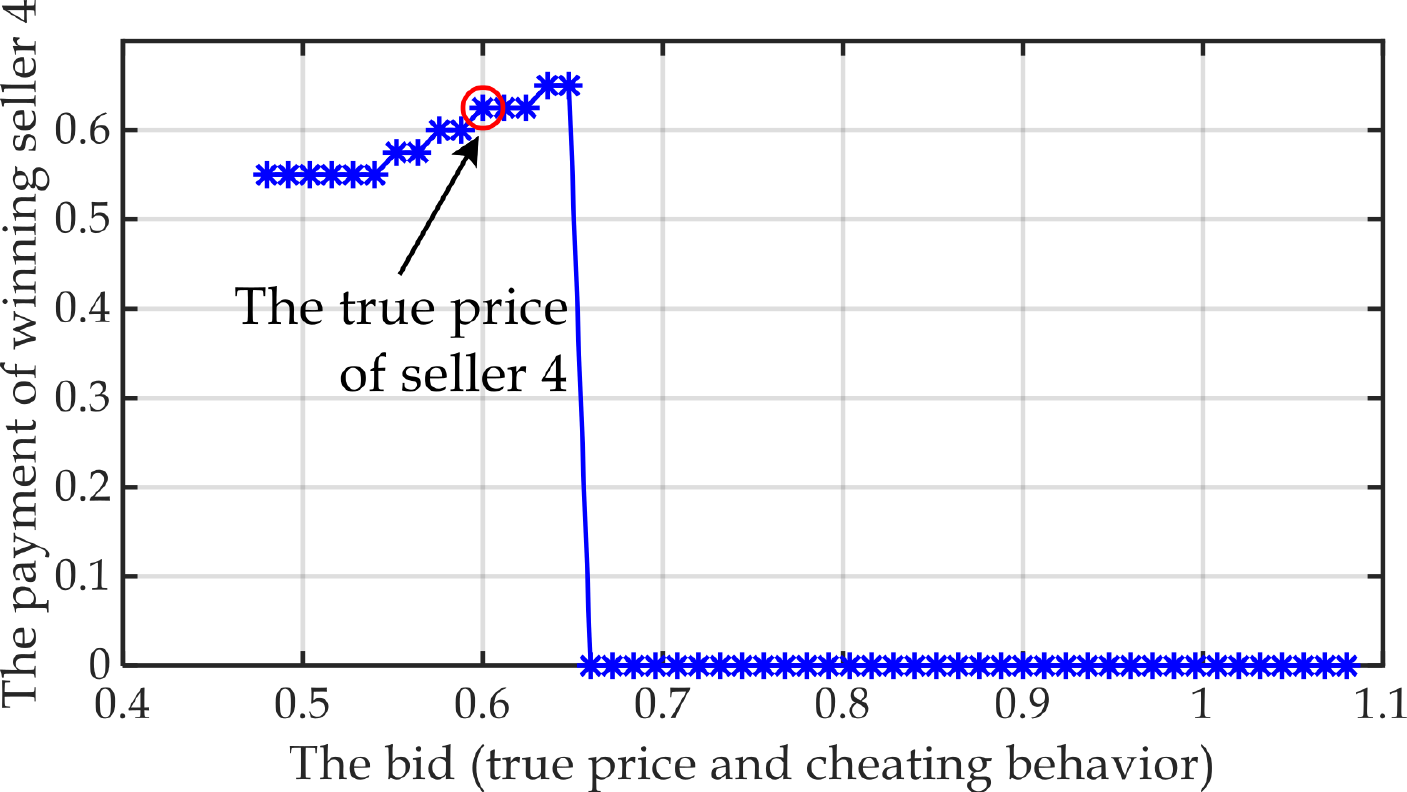}}
\subfigure[]{\includegraphics[width=.245\linewidth]{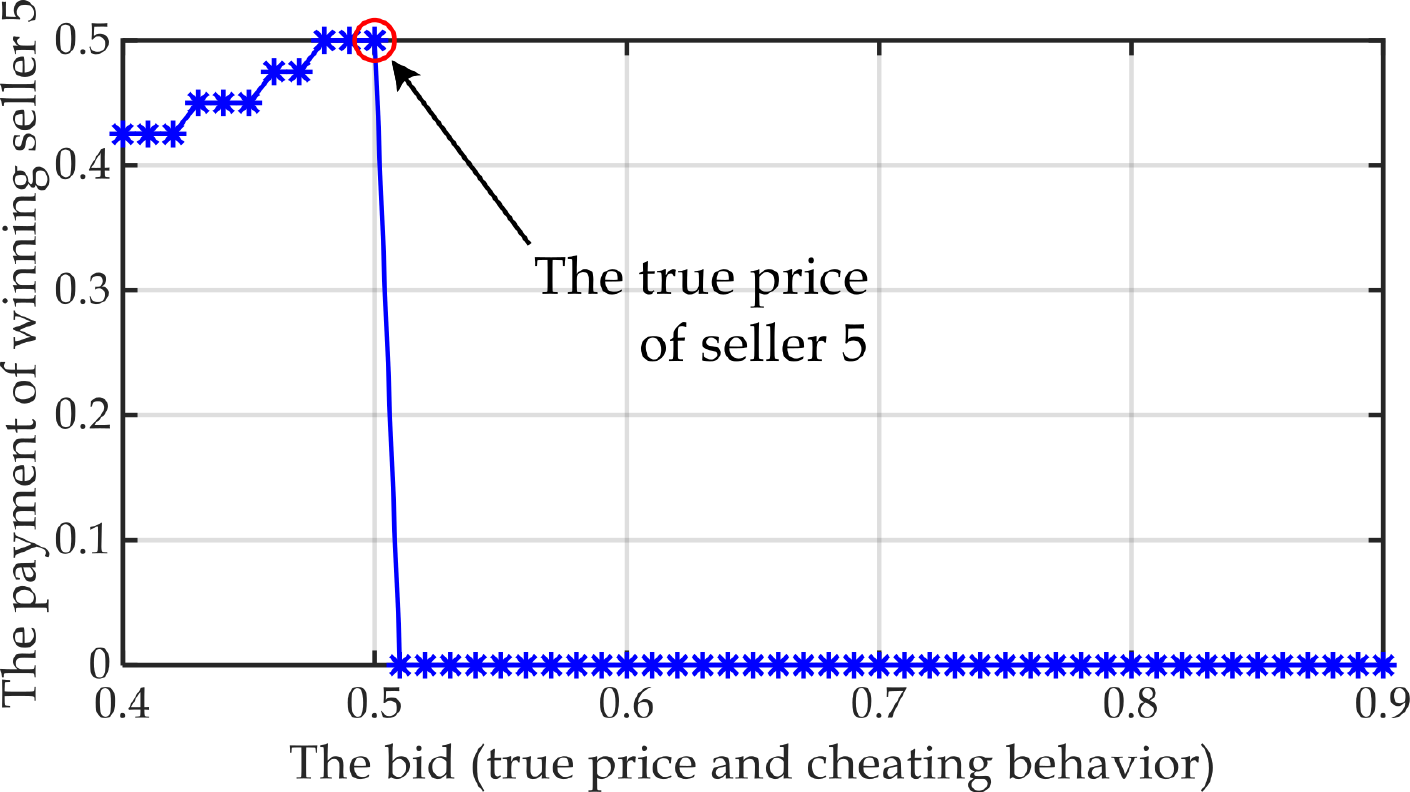}}
\subfigure[]{\includegraphics[width=.245\linewidth]{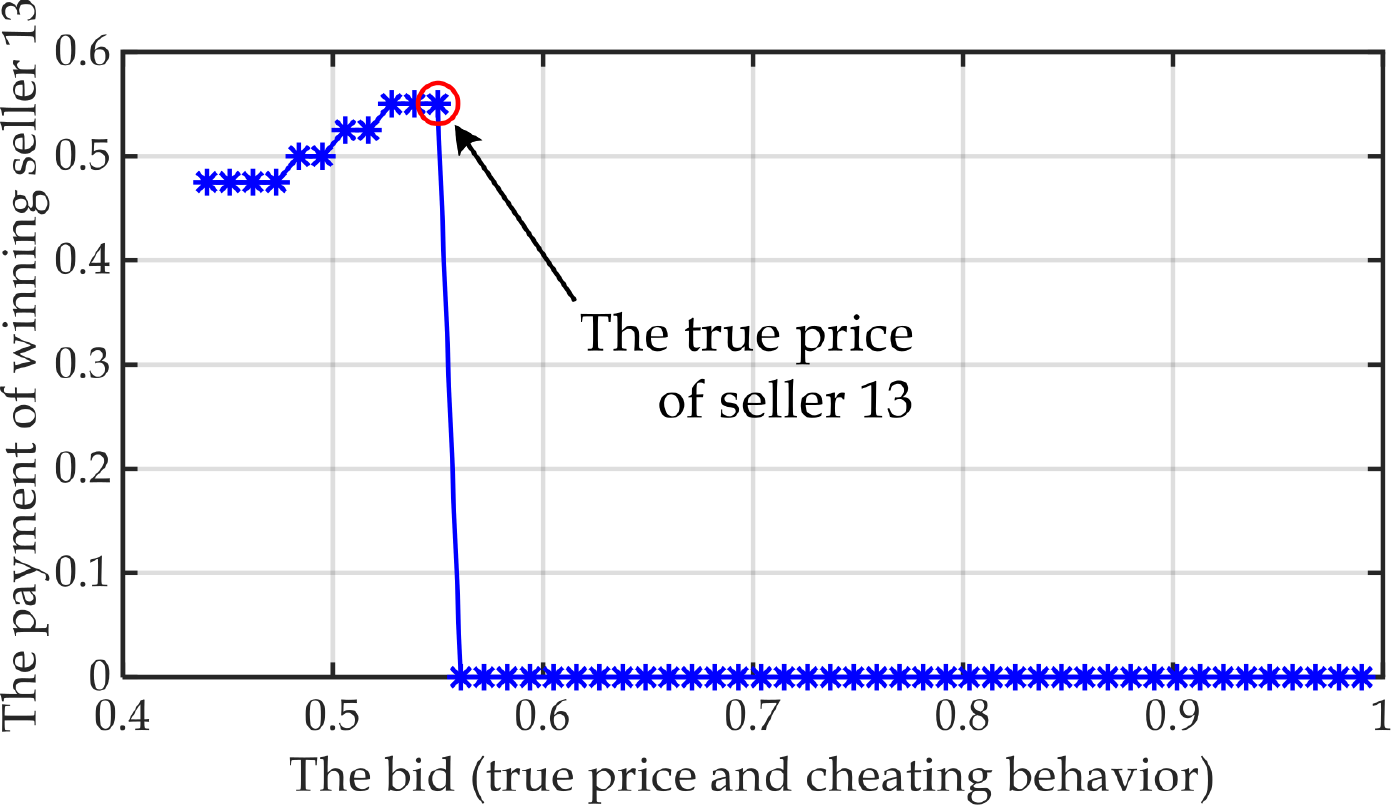}}
\subfigure[]{\includegraphics[width=.245\linewidth]{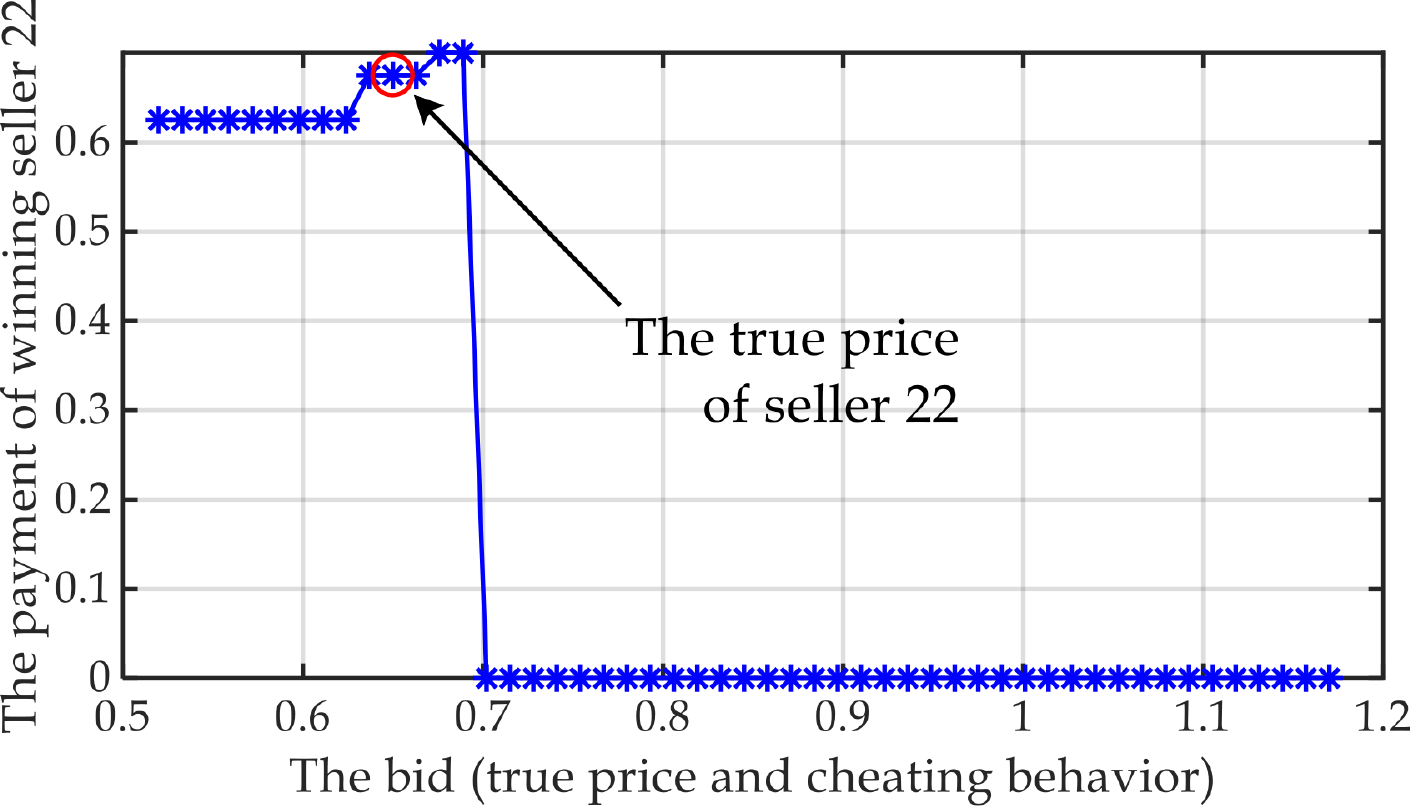}}

\caption{Performance evaluation of winning seller payments and bids in small problem size scenario.}
\end{figure*}

\begin{figure*}[h]
\centering
\subfigure[]{\includegraphics[width=.240\linewidth]{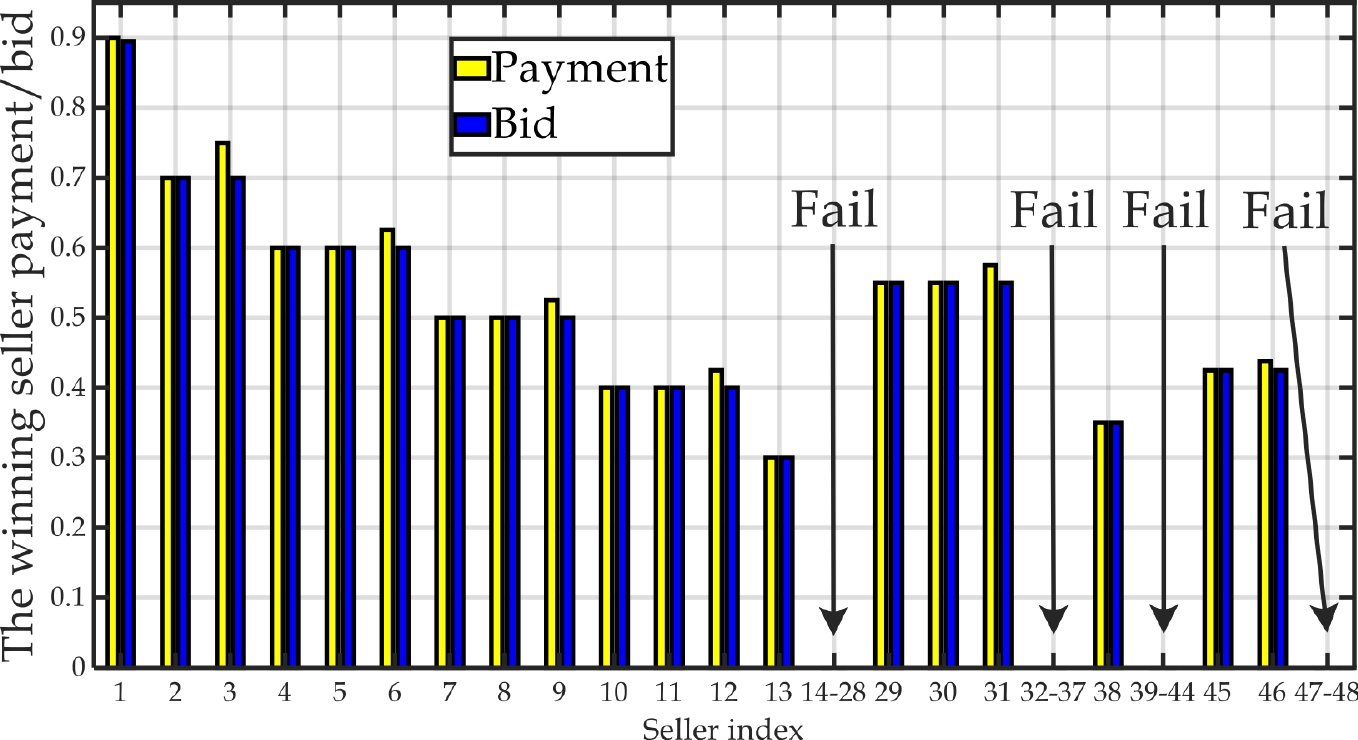}}
\subfigure[]{\includegraphics[width=.245\linewidth]{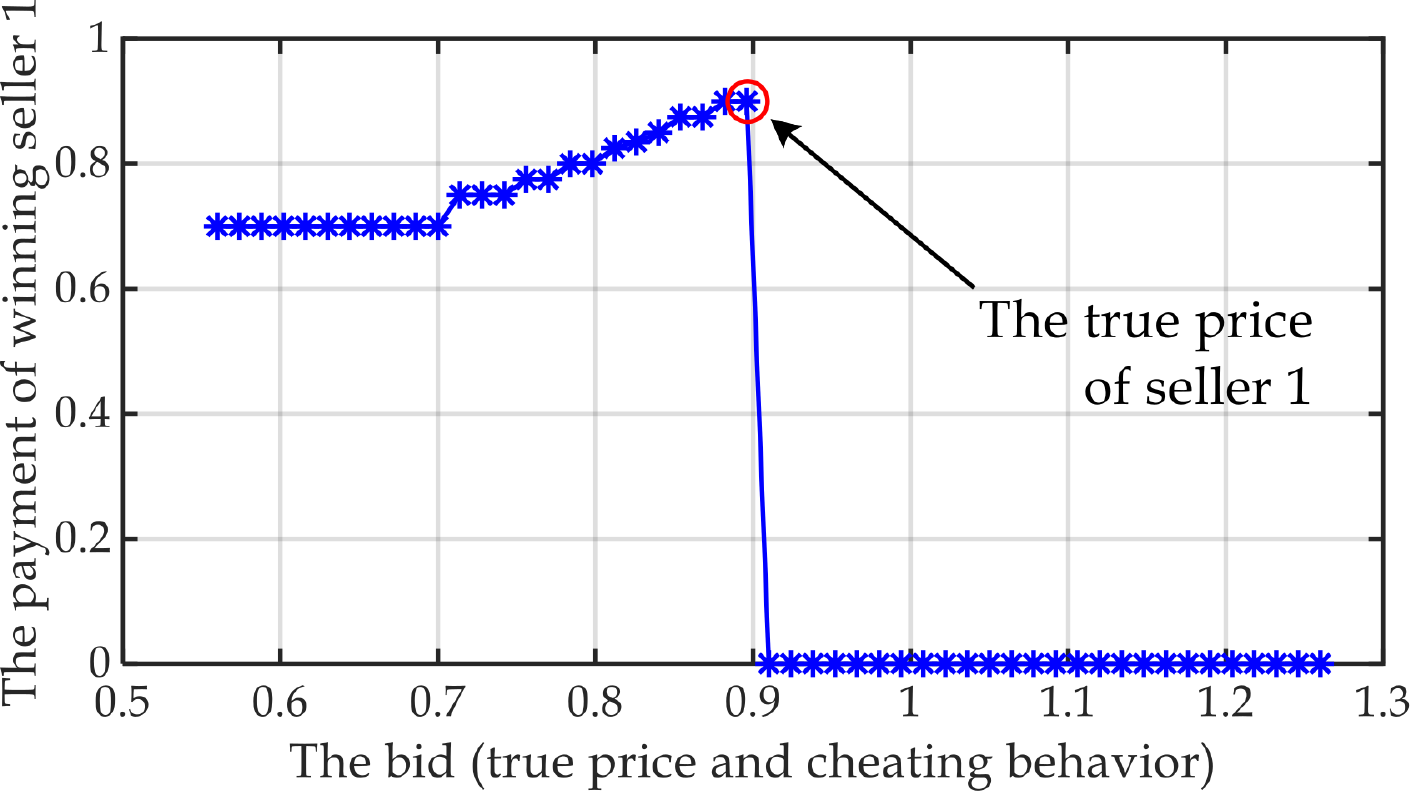}}
\subfigure[]{\includegraphics[width=.248\linewidth]{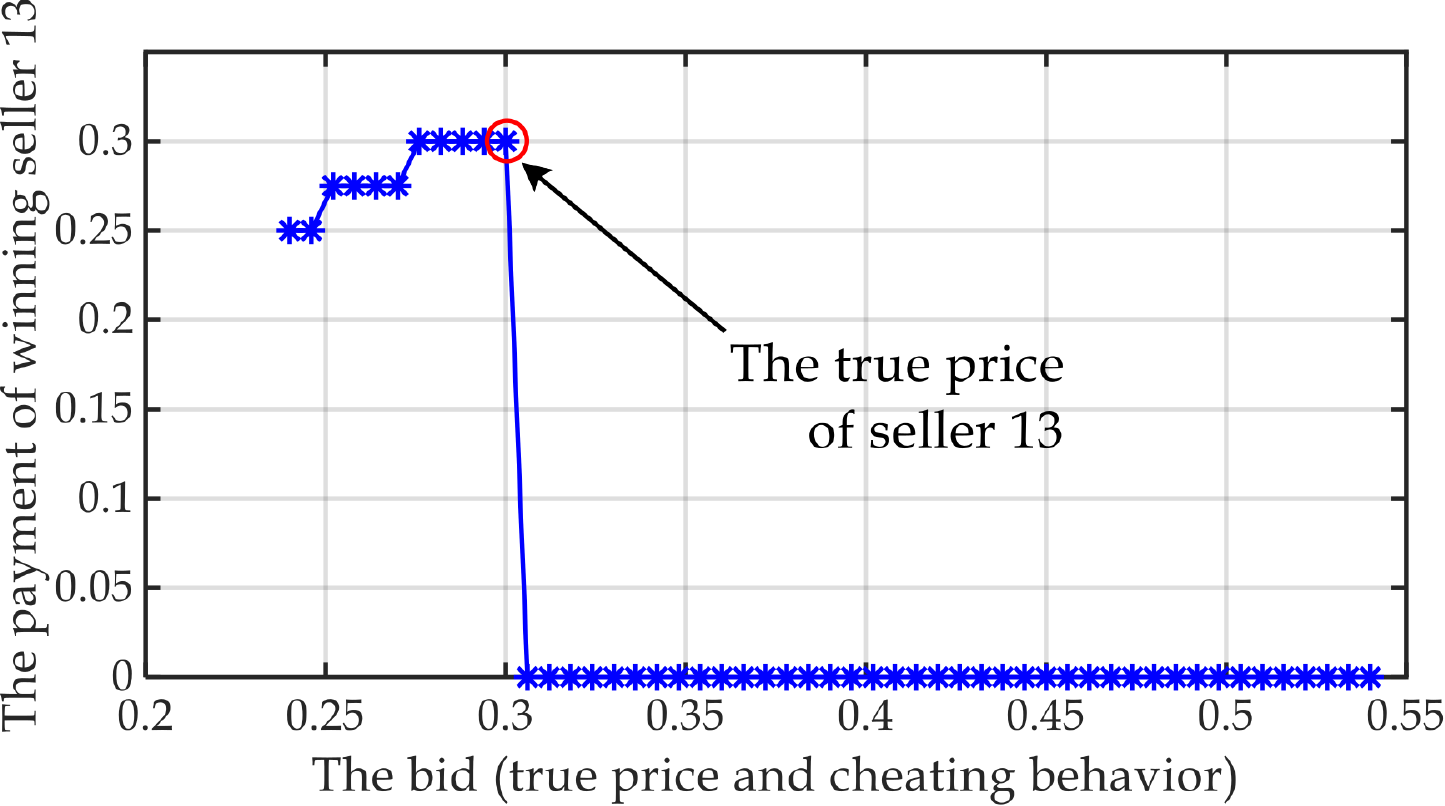}}
\subfigure[]{\includegraphics[width=.248\linewidth]{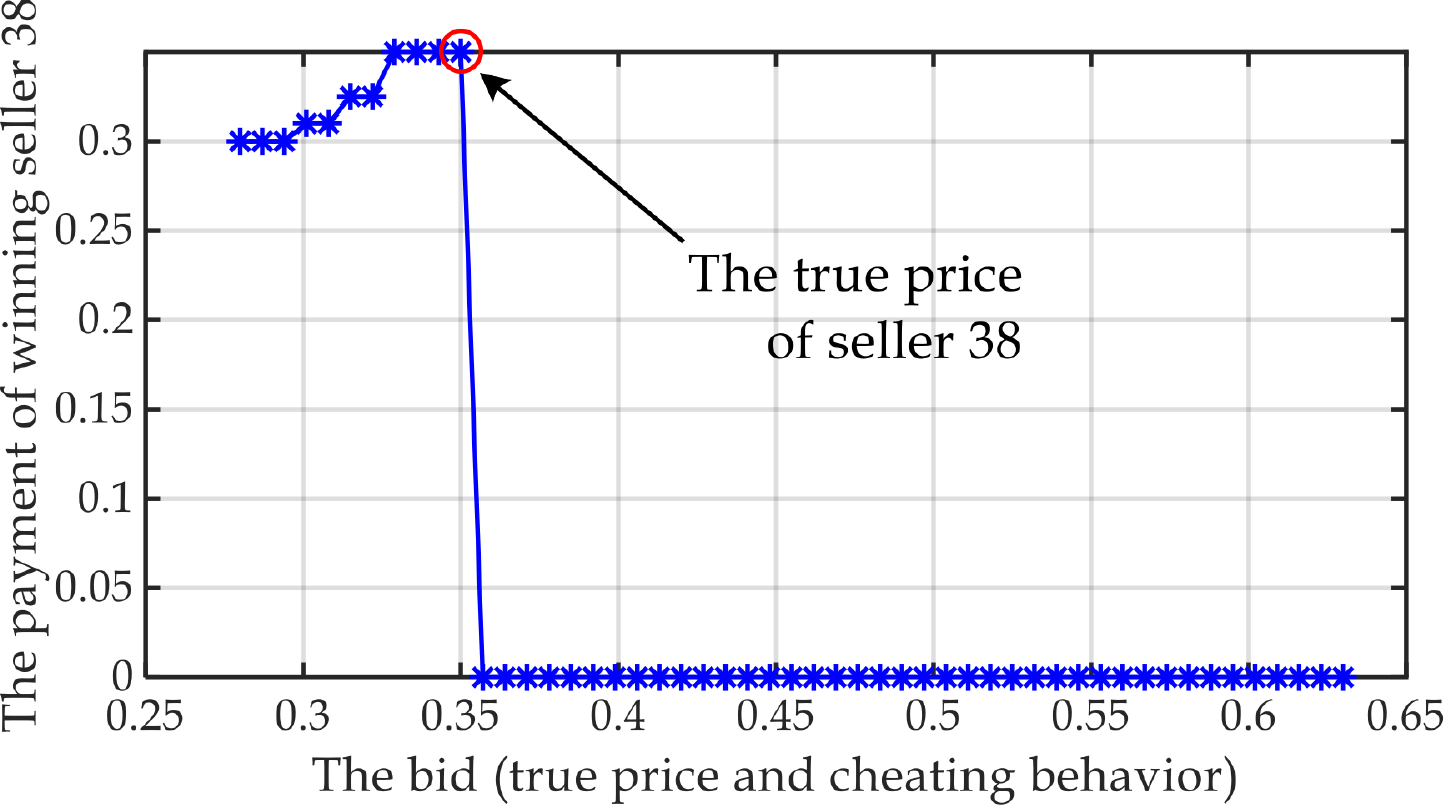}}

\subfigure[]{\includegraphics[width=.245\linewidth]{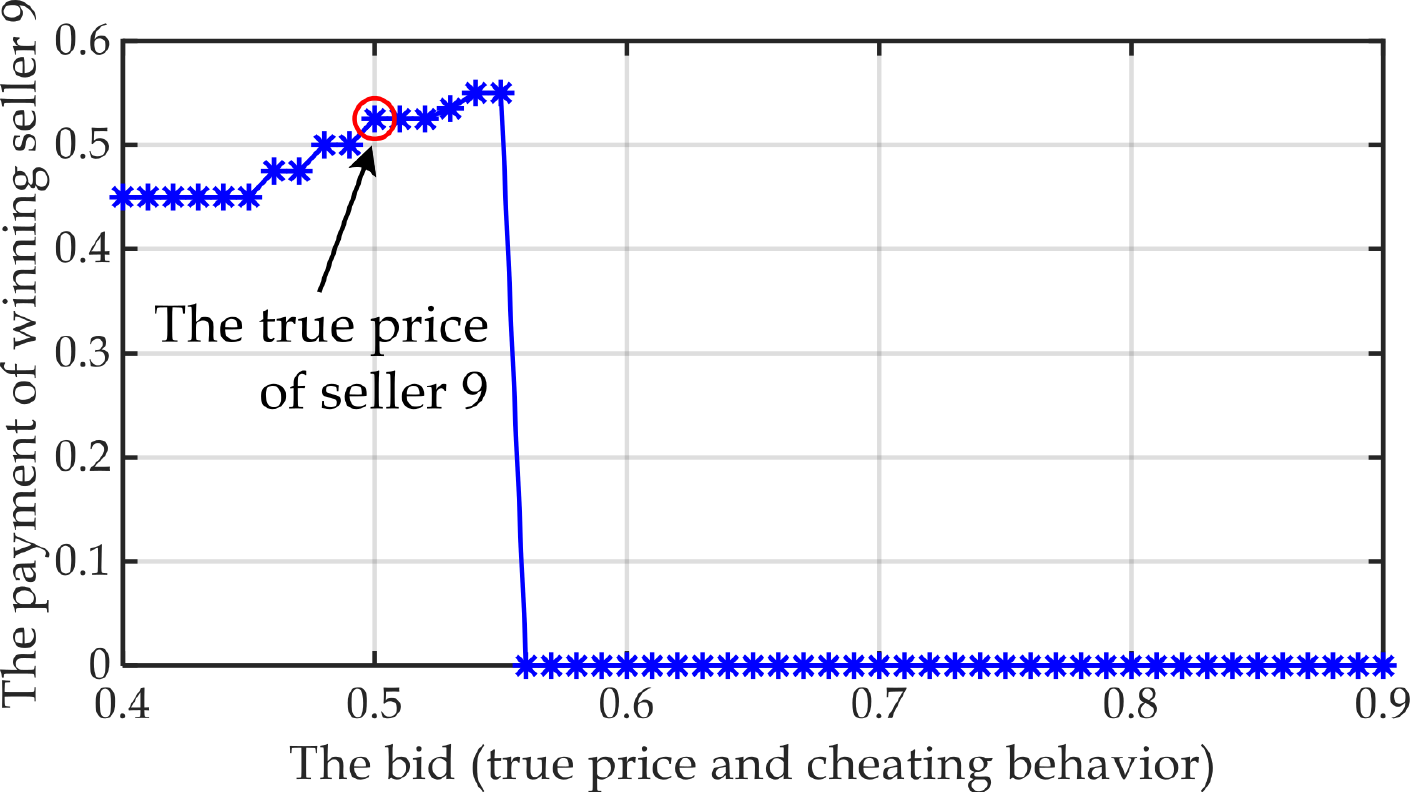}}
\subfigure[]{\includegraphics[width=.246\linewidth]{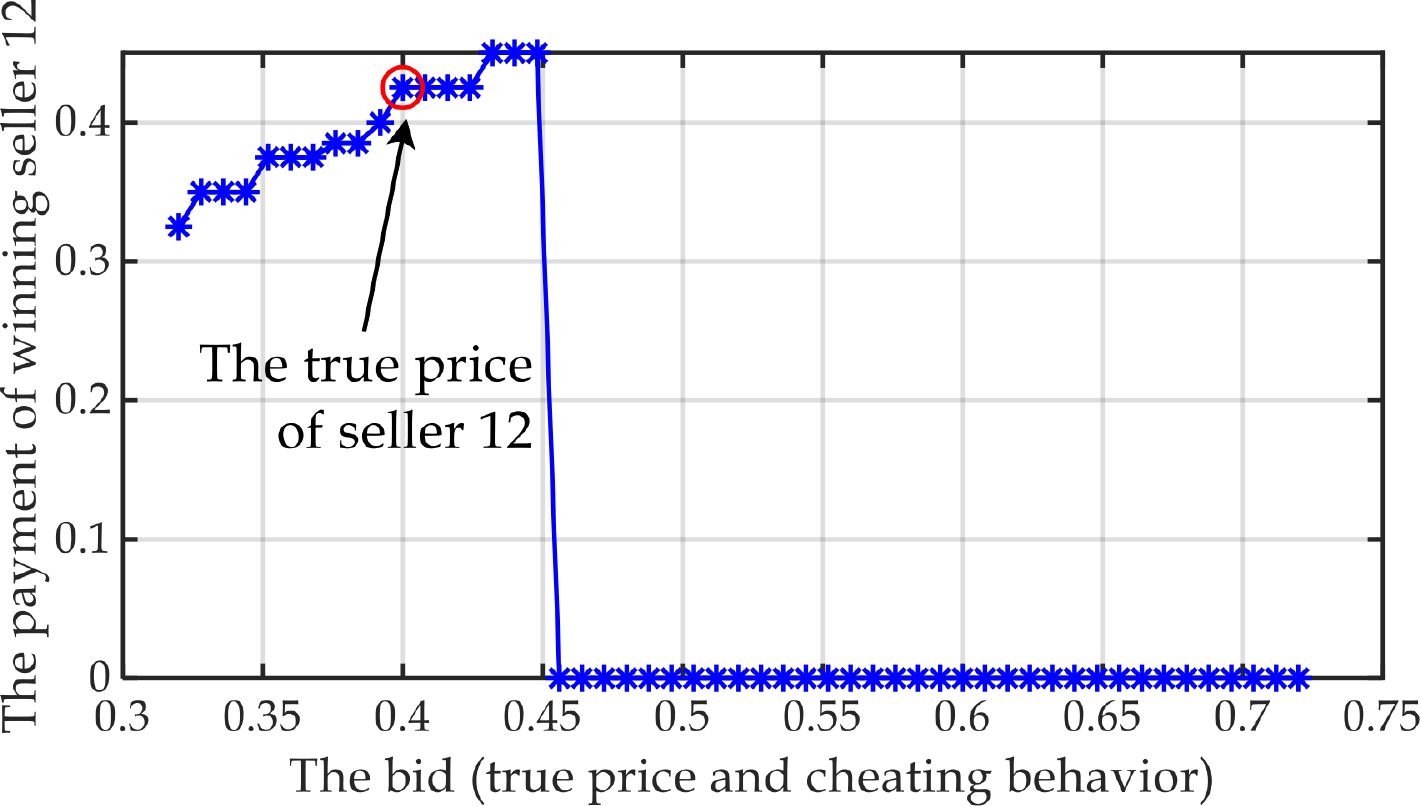}}
\subfigure[]{\includegraphics[width=.245\linewidth]{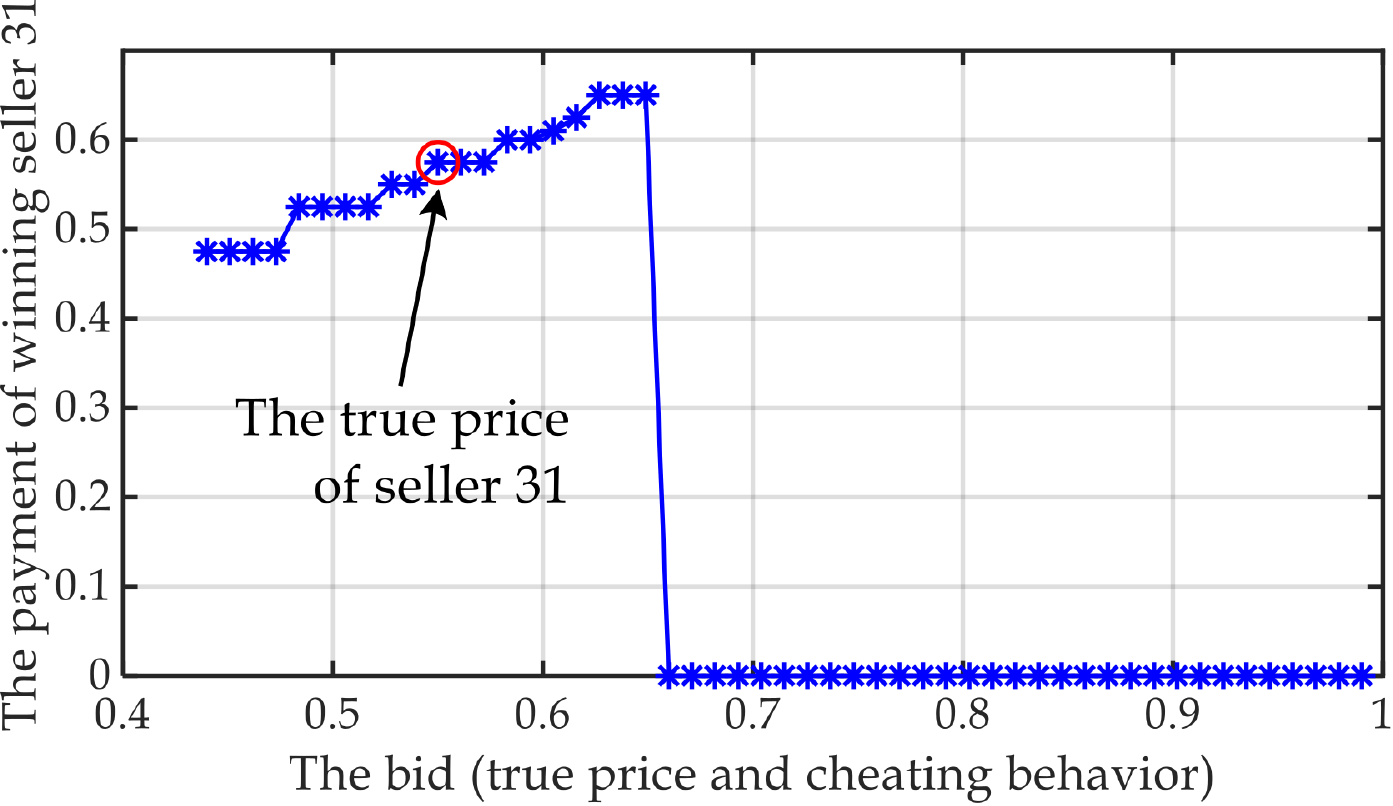}}
\subfigure[]{\includegraphics[width=.245\linewidth]{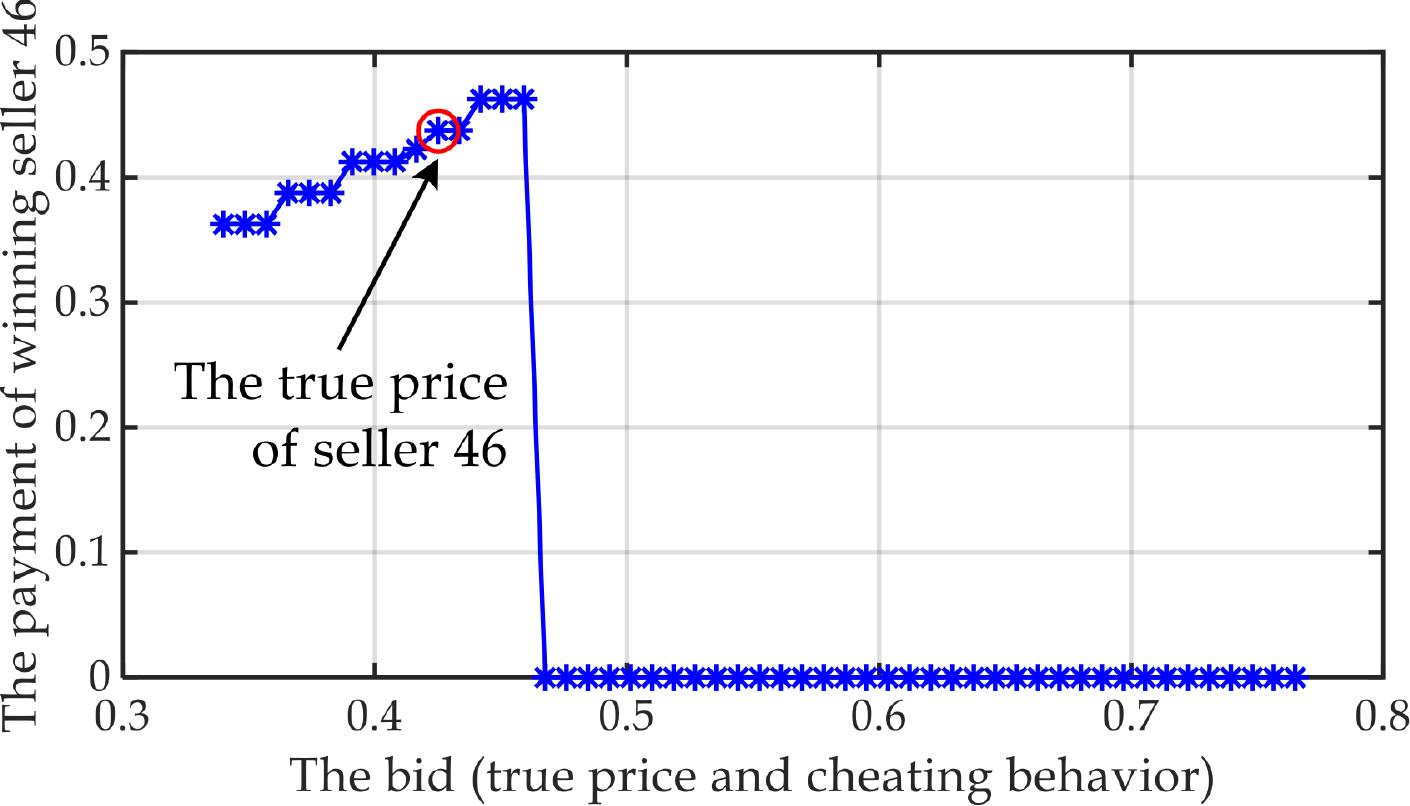}}
\caption{Performance evaluation of winning seller payments and bids in large problem size scenario.}
\end{figure*}

The comparisons of the sum of the UoS (the value of the objective function (3)) between the baseline methods, the optimal, and the MaxUoSG algorithms for various graph job types are shown in Fig.~5 for small problem size scenarios with a couple of buyers and sellers. Fig.~5(a), Fig.~5(b), Fig.~5(c) and Fig.~5(d) reveal that the proposed MaxUoSG exhibits the same or similar performance with the optimal algorithm, and outperforms the baseline methods RMM, ETPM and LPM upon allocating one graph job. Similarly, upon allocating two graph jobs, Fig.~5(e), Fig.~5(f), Fig.~5(g) and Fig.~5(h) demonstrate that the proposed MaxUoSG achieves better UoS than the baseline methods that is close to that of the optimal solutions. Notably, as an excessively long time is needed to obtain the optimal solutions with an increase in the number of buyers and sellers, performance evaluation of the optimal algorithm is ignored in Fig.~5(g) and Fig.~5(h).

For large problem sizes containing more buyers and sellers, the comparisons of the sum of the buyers' UoS between the baseline methods, the optimal and the MaxUoSG algorithms are shown in Fig.~6. Note that the performance evaluation of the optimal algorithm is not considered in Fig.~6 since the running time becomes prohibitively large when the number of participants in the marketplace grows. As can be seen from Fig.~6, the proposed MaxUoSG algorithm always results in a better value of the objective function (3) rather than that of RMM, ETPM and LPM, in various scenarios considering different numbers of buyers and sellers. For a better illustration, \textbf{TABLE 2} is presented that describes the average performance improvement revealed in Fig. 5 and Fig. 6 upon utilizing the proposed MaxUoSG compared with the baseline methods in various problem sizes. 

\subsection{The desired economical properties}

Simulations related to the economical properties of truthfulness and individual rationality for the proposed MaxUoSG algorithm considering different problem sizes are depicted in Fig. 7 and Fig. 8. Considering a small problem size case with one graph job type 1 and one type 2 (totally 7 buyers) and 3 SPs (totally 25 sellers), the comparison between the winning sellers’ payments and bids is depicted in Fig. 7(a). As can be seen from this figure, each winning seller will get a final payment no less than his bid, which is related to the individual rationality property. Fig. 7(b)-Fig. 7(h) exhibit both the untruthful and truthful bidding behaviors of each winning seller, where the value of $x$-axis is defined as the submitted bids and the red solid circles indicate the truthful prices and the related payments. Specifically, Fig. 7(b), Fig. 7(f) and Fig. 7(g) show that misreporting behavior in bidding will not lead to higher payments to the sellers, which encourages them to behave truthfully. Sellers 2, 3, 4 and 22 in Fig. 7(c), Fig. 7(d), Fig. 7(e) and Fig. 7(h) are facing with the risk of getting worse payments when misreporting in the auction. According to \textbf{Case 2} in \textbf{Proposition 5}, misreporting behaviors may cause some changes to the preference lists of the buyers and the broker, which may result in a different buyer-seller pair selection owing to the structure preservation constraint. Thus, sellers 2, 3, 4 and 22 in these figures may get chances to obtain better payments when they are untruthful. However, non-transparent information in the marketplace makes sellers unconsciousness of arranging their bids to avoid potential risks. Thus, bidding truthfully stands for a risk-free behavior. Take Fig. 7(e) as an example, seller 4 will definitely get a worse payment when bidding lower than his true price 0.6. Also, seller 4 is insensible of how to raise his bid for getting a better payment before encountering the risk of non-payment after bidding larger than 0.648. As a conclusion, sellers observe the truthfulness property in our proposed framework.

Simulation results for large problem sizes considering 4 JOs (totally 19 buyers, including two graph job type 2, one type 3 and one type 4) and 5 SPs (totally 48 sellers) are presented in Fig. 8. Similar to Fig. 7(a), Fig. 8(a) shows that the payment of each winning seller stays no less than his bid, which makes all sellers individual rational. Subsequently, seven winning sellers as repesentatives are randomly chosen from 19 winning sellers to conduct payment evaluation under untruthful and truthful behaviors. Fig. 8(b), Fig. 8(c) and Fig. 8(d) indicate that the sellers do not get better payments when misreporting their bids. Moreover, Fig. 8(e), Fig. 8(f), Fig. 8(g) and Fig. 8(h) reveal the risks that the sellers may undergo when misreporting their bids in the marketplace. Consequently, sellers are motivated to bid truthfully in our proposed framework.

\section{Conclusion}
This paper studies a novel truthful auction-based graph job allocation scheme for VC-assisted IoV, which is formulated as an IP problem. For small problem sizes with a couple of buyers and sellers, an optimal algorithm is introduced along with a VCG-based payment rule. For large problem sizes containing more participants and complicated job and VC structures, a low computation complexity structure-preserved matching algorithm based on UoSG maximization, and the corresponding payment role are proposed. The effectiveness of the proposed algorithms is revealed through comprehensive simulations.  Several potential future directions can be concerned such as considering the cooperation among the SPs, and designing new payment rules to avoid risks if possible.

\section*{Acknowledgement}

This work is supported in part by the National Natural Science Foundation of China (grant nos. 61971365, 61871339, 61901403), Digital Fujian Province Key Laboratory of IoT Communication, Architecture and Safety Technology (grant no. 2010499), the State Key Program of the National Natural Science Foundation of China (grant no. 61731012), the Major Research Plan of the National Natural Science Foundation of China (grant no. 91638204) and the US National Science Foundation (grant nos. ECCS-1444009, CNS-1824518).

%

\begin{IEEEbiography}[{\includegraphics[width=1in,height=1.25in,clip,keepaspectratio]{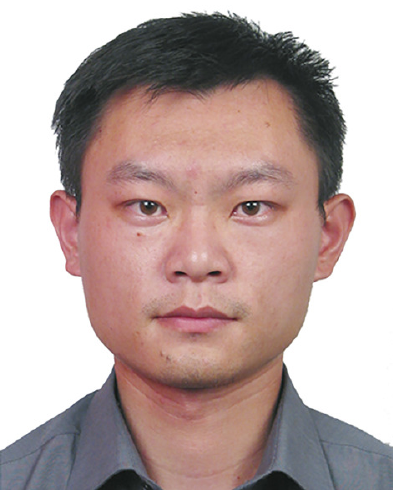}}]{Zhibin Gao}
(gaozhibin@xmu.edu.cn) received his B.S. degree in Communication Engineering in 2003, M.S. degree in Radio Physics in 2006, and Ph.D. in Communication Engineering in 2011 from Xiamen University, where he is a senior engineer of communication engineering. His current research interests include wireless communication, mobile network resource management, and signal processing.
\end{IEEEbiography}

\begin{IEEEbiography}[{\includegraphics[width=1in,height=1.25in,clip,keepaspectratio]{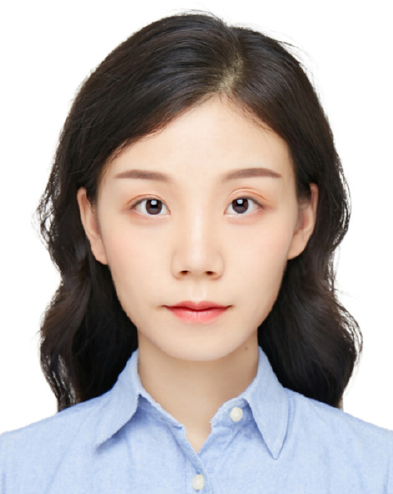}}]{Minghui LiWang}
[M'19] (mliwang@uwo.ca) received her B.S. degree in Computer Science and Technology in 2013 and Ph. D degree in Information {\&} Communication Engineering, Xiamen University, China in 2019. She is now a postdoc fellow in the Department of Electrical {\&} Computer Engineering, University of Western Ontario, Ontario, Canada. She was a visiting scholar at NC State University, NC, USA during 2017 to 2018. Her research interests are wireless communication systems, mobile edge computing, resource optimization {\&} management and Internet of Vehicles.
\end{IEEEbiography}

\begin{IEEEbiography}[{\includegraphics[width=1in,height=1.25in,clip,keepaspectratio]{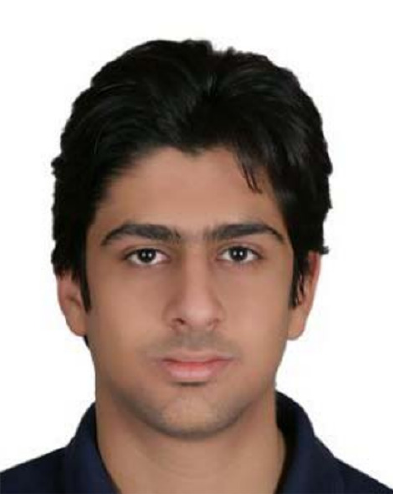}}]{Seyyedali Hosseinalipour}[S’18]
(shossei3@ncsu.edu) received his B.S. degree in Electrical Engineering from Amirkabir University of Technology (Tehran Polytechnic), Tehran, Iran in 2015. He is pursuing a Ph.D. degree in the Department of Electrical and Computer Engineering at North Carolina State University, Raleigh, NC, USA. His research interests include analysis of wireless networks, resource allocation and load balancing for cloud networks, and resource allocation and task scheduling for vehicular ad-hoc networks.
\end{IEEEbiography}
\vspace{-0.1cm}
\begin{IEEEbiography}[{\includegraphics[width=1in,height=1.25in,clip,keepaspectratio]{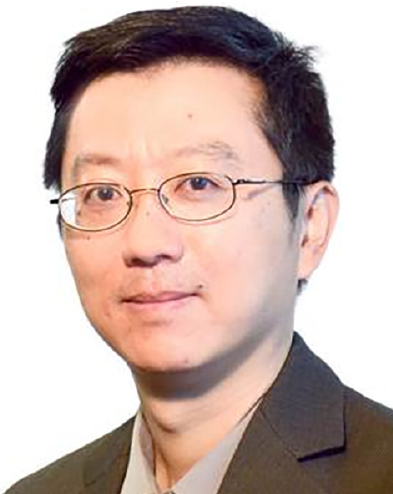}}]{Huaiyu Dai}
[F'17] (hdai@ncsu.edu) received the B. E. and M.S. degrees in Electrical Engineering from Tsinghua University, Beijing, China, in 1996 and 1998, respectively, and the Ph.D. degree in electrical engineering from Princeton University, Princeton, NJ in 2002. He was with Bell Labs, Lucent Technologies, Holmdel, NJ, in summer 2000, and with AT {\&} T Labs-Research, Middletown, NJ, in summer 2001. He is currently a Professor of Electrical and Computer Engineering with NC State University, Raleigh. His research interests are in the general areas of communication systems and networks, advanced signal processing for digital communications, and communication theory and information theory. His current research focuses on networked information processing and crosslayer design in wireless networks, cognitive radio networks, network security, and associated information-theoretic and computation-theoretic analysis. He has served as an editor of IEEE Transactions on Communications, IEEE Transactions on Signal Processing, and IEEE Transactions on Wireless Communications. Currently he is an Area Editor in charge of wireless communications for IEEE Transactions on Communications. He co-edited two special issues of EURASIP journals on distributed signal processing techniques for wireless sensor networks, and on multiuser information theory and related applications, respectively. He co-chaired the Signal Processing for Communications Symposium of IEEE Globecom 2013, the Communications Theory Symposium of IEEE ICC 2014, and the Wireless Communications Symposium of IEEE Globecom 2014. He was a co-recipient of best paper awards at 2010 IEEE International Conference on Mobile Ad-hoc and Sensor Systems (MASS 2010), 2016 IEEE INFOCOM BIGSECURITY Workshop, and 2017 IEEE International Conference on Communications (ICC 2017).
\end{IEEEbiography}
\vfill
\begin{IEEEbiography}[{\includegraphics[width=1in,height=1.25in,clip,keepaspectratio]{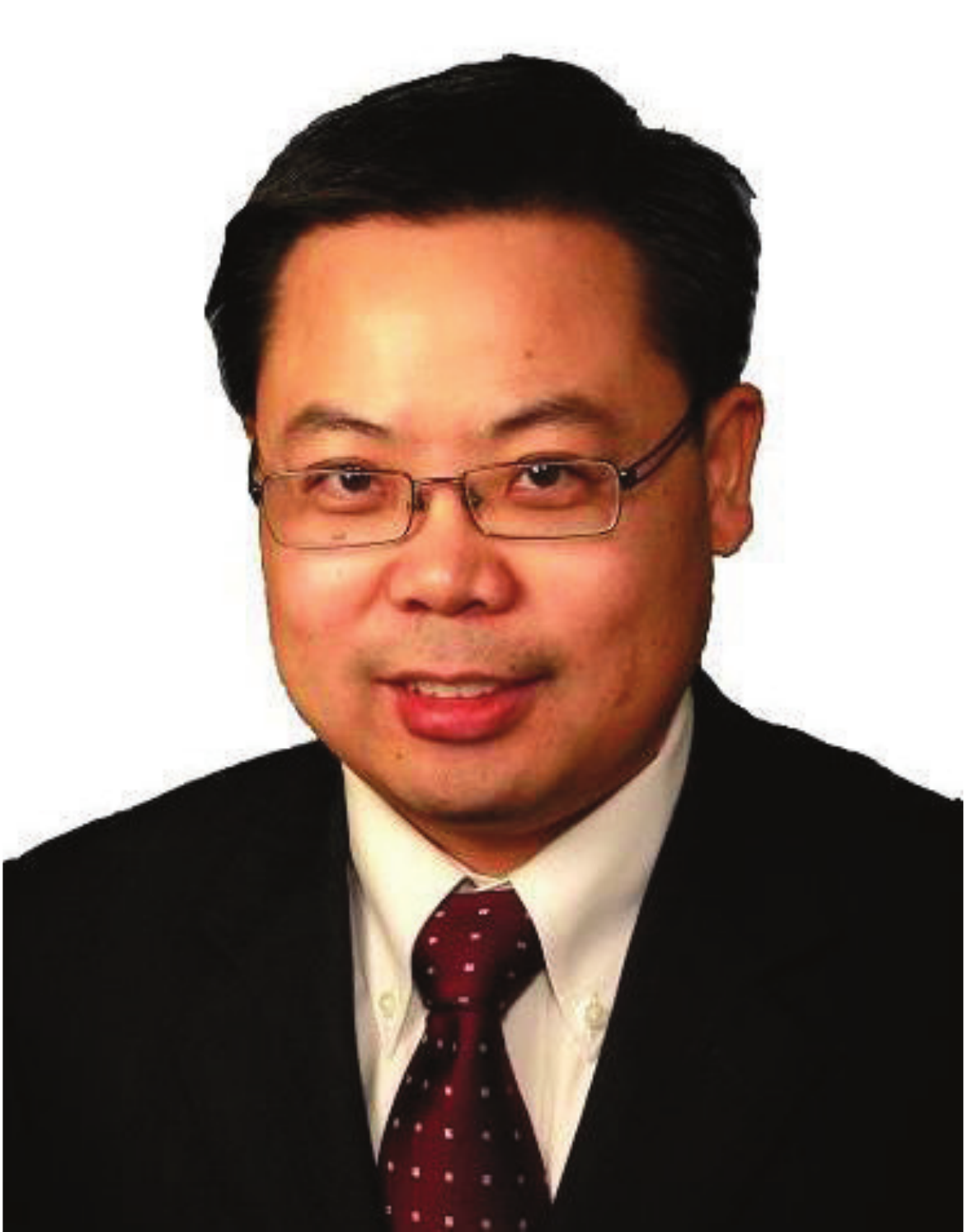}}]{Xianbin Wang}[F'17] (xianbin.wang@uwo.ca) is a Professor and Tier 1 Canada Research Chair at Western University, Canada. He received his Ph.D. degree in electrical and computer engineering from National University of Singapore in 2001. Prior to joining Western, he was with Communications Research Centre Canada (CRC) as a Research Scientist/Senior Research Scientist between July 2002 and Dec. 2007. From Jan. 2001 to July 2002, he was a system designer at STMicroelectronics. His current research interests include 5G and beyond, Internet-of-Things, communications security, machine learning and intelligent communications. Dr. Wang has over 400 peer-reviewed journal and conference papers, in addition to 30 granted and pending patents and several standard contributions. Dr. Wang is a Fellow of Canadian Academy of Engineering, a Fellow of Engineering Institute of Canada, a Fellow of IEEE and an IEEE Distinguished Lecturer. He has received many awards and recognitions, including Canada Research Chair, CRC President’s Excellence Award, Canadian Federal Government Public Service Award, Ontario Early Researcher Award and six IEEE Best Paper Awards. He currently serves as an Editor/Associate Editor for IEEE Transactions on Communications, IEEE Transactions on Broadcasting, and IEEE Transactions on Vehicular Technology. He was also an Associate Editor for IEEE Transactions on Wireless Communications between 2007 and 2011, and IEEE Wireless Communications Letters between 2011 and 2016. He was involved in many IEEE conferences including GLOBECOM, ICC, VTC, PIMRC, WCNC and CWIT, in different roles such as symposium chair, tutorial instructor, track chair, session chair and TPC co-chair. Dr. Wang is currently serving as the Chair of ComSoc SPCE Technical Committee.
\end{IEEEbiography}

\vfill

\end{document}